\documentclass[fleqn,usenatbib]{mnras}

\usepackage{newtxtext,newtxmath}
\usepackage[T1]{fontenc}
\usepackage{ae,aecompl}
\usepackage{graphicx}	% Including figure files
\usepackage{amsmath}	% Advanced maths commands

\usepackage[figure,figure*]{hypcap}
\usepackage{float}
\usepackage{mathcomp}
\usepackage{textcomp}
\usepackage{latexsym}
\usepackage{gensymb}

\usepackage[dvipsnames]{xcolor}

\title[Quantified diffuse light in compact groups]{Quantified diffuse light in compact groups of galaxies}

\author[D. Poliakov et al.]{
Denis Poliakov,$^{1}$
Aleksandr V. Mosenkov,$^{2}$
Noah Brosch,$^{3}$
Shuki Koriski$^{3}$
\newauthor
and R. Michael Rich$^{4}$
\newauthor
\\
$^{1}$Saint Petersburg State University, Department of Astrophysics, St. Petersburg 198504, Russia\\
$^{2}$Pulkovo Observatory of the Russian Academy of Sciences, St. Petersburg 196140, Russia\\
$^{3}$The Wise Observatory and the Raymond and Beverly Sackler School of Physics and Astronomy, The Faculty of Exact Sciences,\\ 
~Tel Aviv University, Tel Aviv 69978, Israel\\
$^{4}$Division of Astronomy, Department of Physics and Astronomy, UCLA, 430 Portola Plaza, Box 951547,\\ 
~Los Angeles, CA 90095-1547, USA\\\\
}

\date{Accepted XXX. Received YYY; in original form ZZZ}

\pubyear{2020}

\begin{document}
\label{firstpage}
\pagerange{\pageref{firstpage}--\pageref{lastpage}}
\maketitle

% Abstract of the paper
\begin{abstract}
The vast majority of stars in galaxy groups are contained within their constituent galaxies. Some small fraction of stars is expected, however, to follow the global dark matter potential of the group. In compact groups, interactions between the galaxies should be frequent. This leads to a more intensive material stripping from the group members, which finally forms an intra-group light component (IGL). Therefore, the distribution of the IGL should be related to the distribution of the total mass in the compact group and its dynamical status. In this study we consider the distribution and fraction of the IGL in a sample of 36 Hickson compact groups (HCGs). 
We use deep observations of these compact groups (down to surface brightness $\sim 28$ mag\,arcsec$^{-2}$ in the $r$ band) obtained with the WISE $28$-inch telescope. 
For five HCGs with a bright symmetric IGL component, we carry out multicomponent photometric decomposition to simultaneously fit the galaxy profiles and the IGL. For the remaining groups, we only fit the profiles of their constituent galaxies.
We find that the mean surface brightness of the IGL correlates with the mean morphology of the group: it becomes brighter in the groups with a larger fraction of early-type galaxies. On the other hand, the IGL brightness depends on the total luminosity of the group.
The IGL profile tends to have a S\'ersic index $n\sim0.5-1$, which is generally consistent with the mass density profile of dark matter haloes in compact groups obtained from cosmological simulations.
\end{abstract}

% Select between one and six entries from the list of approved keywords.
% Don't make up new ones.

%%%%%%%%%%%%%%%%%%%%%%%%%%%%%%%%%%%%%%%%%%%%%%%%%%

%%%%%%%%%%%%%%%%% BODY OF PAPER %%%%%%%%%%%%%%%%%%
\section{Introduction}
\label{sec:intro}
As shown in many studies \citep[see e.g.][]{1974ApJ...193L...1O,1974Natur.250..309E,1979ARA&A..17..135F,2016A&A...594A..13P}, the collisionless dark matter (DM) makes up most of the mass-energy in the Universe after the dark energy and interacts with the visible matter only gravitationally. This fact allows us to detect the DM only indirectly.

Over decades, different approaches have been proposed to detect the DM: using gravitational lensing \citep{2004ApJ...604..596C,2004ApJ...606..819M,2011A&ARv..19...47K,2013SSRv..177...75H}, X-ray observations \citep{1993Natur.363...51P,1994ApJ...436...44E,1995A&A...301..348S,1996MNRAS.283..690P,2001Natur.409...39B}, a combination of optical and X-ray observations \citep{1998ApJ...496...73M,2011sf2a.conf..147M}, and by means of the intra-cluster light (ICL) detection \citep{1998ApJ...502..141D,2007AJ....134..466K,2015MNRAS.449.2353B,2018MNRAS.474..917M,2018ApJ...857...79J}. Gravitational lensing is a powerful tool to obtain the detailed mass distribution in galaxy clusters and groups, but its reconstruction procedure is very complicated. It requires having deep images and corresponding redshifts of the target object as well as the lensed one. The highly ionised gas, bound by the gravitational potential of a cluster or a group, follows the mass distribution only in relaxed systems \citep{2004ApJ...604..596C,2004ApJ...606..819M}. Thus, the X-ray emission produced by hot gas, cannot always be used as a tracer of the mass distribution. 

Recently, \citet{2019MNRAS.482.2838M} proposed a promising method for reconstructing the distribution of the DM in clusters. They compared the bi-dimensional distributions of the DM in massive galaxy clusters with the distribution of the faint ICL in them. They used NIR images of six Hubble Frontier Field (HFF) clusters, X-ray images from the \textit{Chandra} \citep{2000SPIE.4012....2W} Data Archive and mass maps derived from gravitational lensing data provided by the HST Frontier Fields Initiative \citep{2017ApJ...837...97L}. To quantify the similarity between three bi-dimensional distributions, the ICL distribution, the mass maps and the X-ray distribution, they obtained isocontours for each of the different components. Then, they computed the Modified Hausdorff distance for each pair of the isocontours. Relying on this quantitative analysis, they demonstrate the suitability of the ICL to trace the shape of the total mass distribution in the galaxy clusters. They argue that in most cases the ICL traces better the mass distribution for the DM than the X-ray emission of the hot gas. 
The relation between the diffuse intracluster light and the galaxy cluster matter distribution measured through weak lensing was studied for 528 clusters in \citet{2020MNRAS.tmp.3482S}. They also found the similarity of these distributions. The results of \citet{2019MNRAS.482.2838M} and \citet{2020MNRAS.tmp.3482S} are confirmed by recent N-body simulations \citep{2020MNRAS.494.1859A}. 

Many other observations \citep{1997MNRAS.284L..11T,2002ApJ...570..119D,2002AJ....123..760A,2003AJ....125..514A,2004ApJ...614L..33A,2004ApJ...615..196F,2005AJ....129.2585A,2005ApJ...618..195G,2005MNRAS.358..949Z,2007MNRAS.378.1575S,2011ApJS..195...15D,2015MNRAS.448.1162D,2017ApJ...834...16M,2017ApJ...846..139M,2018MNRAS.474.3009D,2019lssc.confE...9K,2019ApJ...874..165Z,2020ApJS..247...43K,2020A&A...639A..14S,2021ApJS..252...27K} and simulations \citep{2007MNRAS.377....2M,2007ApJ...666...20P,2010MNRAS.406..936P,2011ApJ...732...48R,2014MNRAS.437.3787C,2014MNRAS.437..816C,2015MNRAS.451.2703C,2018MNRAS.479..932C,2018MNRAS.475..648P,2020ApJ...901..128C} in the last decades were aimed at exploring the diffuse light in galaxy clusters.
The diffuse light was mostly detected around the brightest cluster galaxies (BCGs). Naturally, the formation of the ICL was associated with the formation and evolution of the BCG. According to the above mentioned studies, the diffuse light in galaxy clusters and groups is related to stellar stripping and galaxy mergers. During a partial tidal stellar stripping in these systems, dwarf galaxies supply stars to the diffuse light. Thus, the formation of the diffuse light is linked to the assembly of a cluster or a group. 

Compact groups are of particular interest for studying galaxy merging and the properties of interacting galaxies.  According to \citet{1992ApJ...399..353H}, compact groups represent groups of galaxies, the densities of which are equivalent to what is observed in the cores of galaxy clusters, but with much lower, modest velocity dispersions of order 200 km\,s$^{-1}$, comparable to the velocity dispersion in elliptical galaxies (see also \citealt{1997ARA&A..35..357H} and references therein). The classical definition of a compact group implies that it is an isolated group of a few (up to a few tens) of galaxies within a 3 magnitude range. 
The very dense environment in compact groups favours intensive interactions and mergers, which suggests that the group galaxies must display or must have shown in the past various signatures of interactions: tidal stellar streams and tails, plumes, shells, diffuse envelopes, fans, and bridges. Obviously, in such a compact configuration, interactions of galaxies must affect their morphology and effectively strip matter from them. The dispersed stars should settle onto the common gravitational potential of the group and form the IGL \citep{2007ApJ...666...20P,2007AJ....133.2630C}. 
Therefore, it is expected that the properties of the IGL should be related to the evolutionary status and properties of the group galaxies because in such dense environments the evolution of galaxies should be greatly affected by external processes \citep[see e.g.][]{ 2008MNRAS.388.1537M,2012A&A...539A..46A}.

A simple visual inspection of compact groups from the Hickson Compact Group catalogue \citep[HCG][]{1982ApJ...255..382H} reveals that some groups indeed show the presence of the IGL.  
Evidence for diffuse components in compact groups was also found in subsequent studies \citep{1995AJ....110.1498P,2000AJ....120.2355N,2003ApJ...585..739W,2005MNRAS.364.1069D,2006A&A...457..771A,2008MNRAS.388.1433D}. 

In this paper, we aim to study deep observations of 39 compact groups from the HCG catalogue to distinguish the diffuse IGL from the light of the galaxies. We also analyse the relations between several dynamical and photometric characteristics of the groups to link their dynamical status with their morphology.
For the first time, we simultaneously quantify the IGL profile and profiles of the individual group members in five HCGs. In contrast to \citet{2008MNRAS.388.1433D}, who used a wavelet analysis to separate the IGL, we not only determine its fraction to the total luminosity of the group, but derive its parameters using a generalized elliptical 2D S\'ersic function. In addition, we decompose the remaining compact groups and derive the structural parameters of their constituent galaxies. This provides more robust information on the structural properties of galaxies in compact groups than previously done in the literature \citep{2008A&A...484..355D,2012A&A...543A.119C}.

The paper is organised as follows. Section~\ref{sec:sample} presents a description of the dataset used. In Section~\ref{sec:data}, we describe the observations and their preparation for a subsequent analysis. In Section~\ref{sec:method}, we describe our method for photometric fitting of the galaxy and IGL profiles. In Section~\ref{sec:results}, we present our main results. A discussion and conclusions are given in Sections~\ref{sec:discussion} and~\ref{sec:conclusion}, respectively.

\section{The sample}
\label{sec:sample}
Our initial sample consisted of 39 objects from the HCG catalogue \citep{1982ApJ...255..382H}. We inspected these groups using the NASA/IPAC Extragalactic Database (NED) database\footnote{\url{https://ned.ipac.caltech.edu/}} to examine whether the visually close galaxies in these groups truly belong to the groups. Based on the results of this inspection, three groups (HCG\,41, HCG\,73 and HCG\,77) were rejected because less than three galaxies in these groups have concordant radial velocities. \citet{ 2013MNRAS.433.3547D} show that triplets are a natural extension of compact groups based on their global properties. Therefore, we do not exclude the triplets from our sample and consider them as compact groups. Thus, our final sample comprises 36 galaxy groups. Also, based on our imaging (see below) and the NED cross-identifications, we found that some groups consist of more galaxies than listed in the original Hickson catalogue.  

A visual analysis of our images showed that 5 compact groups (HCG\,8, 17, 35, 37, 74) from the final sample may contain an IGL of a rather elliptical shape. In two cases, HCG\,94 and 98 also contain an IGL but it demonstrates a relatively less symmetric shape. Therefore, we decided not to consider these two groups in our subsequent photometric decomposition since the aim of this study is to fit a radially-symmetric 2D model profile to the IGL.

In Sect.~\ref{sec:method}, we carry out multicomponent photometric decomposition of the mentioned five groups with an IGL.
The general properties of the decomposed groups are summarised in the top five lines of Table~\ref{tab:gen_prop_1}. The bottom part of the table lists the properties for the remaining 31 groups. The intrinsic 3D velocity standard deviation was calculated using formula (1) presented in \citet{1992ApJ...399..353H}:
\begin{equation} \label{eq:Dsigma}
	D_{\sigma} = \left[3 \left(\left<v^{2}\right> - \left<v\right>^{2} - \left<\sigma_{v}^{2}\right>\right)\right]^{1/2}\,,
\end{equation}
where $v$ is the observed radial velocity and $\sigma_{v}$ is the estimated velocity error. If formula~(\ref{eq:Dsigma}) returns a complex value, we use bootstrapping to estimate $D_{\sigma}$.
The crossing times were calculated using formula~(2) from \citet{1992ApJ...399..353H}:
\begin{equation} \label{eq:t_c}
	t_\mathrm{c} = \frac{4 R}{\pi D_{\sigma}}\,,
\end{equation}
where $R$ is the median length of the two-dimensional galaxy-galaxy separation vector in the group, hereafter median separation.

To obtain physical sizes, we use the standard spatially-flat 6-parameter $\Lambda$CDM cosmology model that includes a Hubble constant $H_{0} =  67.74$ km\,s$^{-1}$ Mpc$^{-1}$, matter density parameter $\Omega_{m} = 0.3089$, dark energy density parameter $\Omega_{\Lambda} = 0.6911$ \citep[see ][p.32, Table 4, last column]{2016A&A...594A..13P}.
For 33 out of 36 groups, we estimate an average surface brightness in the geometric centre of each group (see Sect.~\ref{sec:results}). The remaining three groups (HCG\,1, HCG\,2 and HCG\,97) show significant contamination near the geometric centre from foreground bright stars or galaxies which do not belong to the group. Therefore we do not estimate the average surface brightness for them.

\begin{table*}
	\caption{General properties of the selected compact groups: equatorial coordinates (from NED); mean redshifts (from NED); median separation; intrinsic 3D velocity standard deviations, defined in formula (1) in \citep{1992ApJ...399..353H}, or using bootstrapping when formula (1) gives a complex value; crossing time; number of galaxies in each group; mean surface brightness of the IGL and its standard deviation (see Sect.~\ref{sec:results}); mean numerical morphological galaxy type (see Sect.~\ref{sec:relation}).}
	\centering
	\begin{tabular}{l c c c c c c c c c c}
        \hline \hline
        Group & RA (J2000) & Dec. (J2000) & $z$ & $R$ & $D_{\sigma}$ & $t_\mathrm{c} $ & Num. Gal. & $\mu_{\mathrm{IGL},r}$ & std($\mu_{\mathrm{IGL},r}$) & $\langle T\rangle$ \\
              & deg & deg &    & kpc & km\,s$^{-1}$ & Gyr           &           & mag\,arcsec$^{-2}$ & mag\,arcsec$^{-2}$ & \\
\hline
HCG\,8  &$12.40341$  &$23.58082$  &$0.054$ &$60.68$ &$860.0$ &$0.09$  &$4/8^{a}$   &$26.1$  &$0.6$   &$-2.6$   \\
HCG\,17 &$33.51875$  &$13.31508$  &$0.0601$&$33.07$ &$446.0$ &$0.094$ &$5$     &$26.0$  &$0.6$   &$-3.5$   \\
HCG\,35 &$131.33147$ &$44.52162$  &$0.0543$&$68.15$ &$481.0$ &$0.18$  &$6$     &$25.7$  &$0.5$   &$-1.8$   \\
HCG\,37 &$138.3986$  &$30.01417$  &$0.0225$&$41.95$ &$697.0$ &$0.077$ &$5$     &$26.0$  &$0.5$   &$2.0$    \\
HCG\,74 &$229.86776$ &$20.89372$  &$0.0399$&$59.98$ &$527.0$ &$0.145$ &$5$     &$25.8$  &$0.5$   &$-3.8$   \\
\hline
HCG\,1  &$6.50083$   &$25.71813$  &$0.034$ &$75.19$ &$131.0$ &$0.733$ &$4$     &\texttwelveudash   &\texttwelveudash   &$-0.2$   \\
HCG\,2  &$7.87517$   &$8.43125$   &$0.0145$&$80.32$ &$70.0$  &$1.46$  &$4/3^{a}$   &$27.6$  &$1.0$   &$4.3$    \\
HCG\,3  &$8.6144$&$-7.5931$   &$0.026$ &$119.05$&$428.0$ &$0.354$ &$4/3^{a}$   &\texttwelveudash   &\texttwelveudash   &$1.0$    \\
HCG\,5  &$9.72625$   &$7.06$  &$0.0408$&$38.83$ &$226.0$ &$0.219$ &$4/3^{a}$   &$27.0$  &$0.8$   &$-0.7$   \\
HCG\,7  &$9.8496$&$0.87817$   &$0.014$ &$67.67$ &$184.0$ &$0.468$ &$4/5^{a}$   &$28.3$  &$1.1$   &$-0.6$   \\
HCG\,12 &$21.8904$   &$-4.67053$  &$0.0483$&$91.63$ &$417.0$ &$0.28$  &$5$     &$26.7$  &$0.7$   &$-0.8$   \\
HCG\,13 &$23.09216$  &$-7.88107$  &$0.0416$&$71.3$  &$297.0$ &$0.306$ &$5$     &$26.5$  &$0.7$   &$-1.6$   \\
HCG\,18 &$39.77818$  &$18.38308$  &$0.0137$&$7.85$  &$72.0$  &$0.139$ &$4/3^{a}$   &$26.5$  &$0.6$   &\texttwelveudash  \\
HCG\,20 &$41.06248$  &$26.10298$  &$0.0481$&$46.48$ &$465.0$ &$0.127$ &$6/5^{a}$   &$26.4$  &$0.6$   &$-3.4$   \\
HCG\,25 &$50.18221$  &$-1.05192$  &$0.0211$&$92.78$ &$100.0$ &$1.177$ &$7/5^{a}$   &$27.2$  &$1.0$   &$0.0$    \\
HCG\,44 &$154.50198$ &$21.81225$  &$0.0046$&$60.04$ &$492.0$ &$0.155$ &$4/5^{a}$   &$26.9$  &$0.9$   &$1.4$    \\
HCG\,59 &$177.11086$ &$12.71123$  &$0.0137$&$38.18$ &$424.0$ &$0.115$ &$5$     &$26.9$  &$1.0$   &$4.2$    \\
HCG\,69 &$208.87805$ &$25.0628$   &$0.0293$&$45.33$ &$209.0$ &$0.277$ &$4$     &$26.4$  &$0.7$   &$1.0$    \\
HCG\,71 &$212.76906$ &$25.48492$  &$0.0308$&$77.45$ &$465.0$ &$0.212$ &$4/3^{a}$   &$27.6$  &$1.0$   &$4.7$    \\
HCG\,72 &$221.98021$ &$19.05747$  &$0.0433$&$68.37$ &$855.0$ &$0.102$ &$6/7^{a}$   &$26.4$  &$0.8$   &$-0.4$   \\
HCG\,76 &$232.92453$ &$7.30793$   &$0.034$ &$111.09$&$408.0$ &$0.346$ &$7$     &$26.9$  &$0.8$   &$-0.9$   \\
HCG\,78 &$237.11658$ &$68.20768$  &$0.0318$&$69.14$ &$992.0$ &$0.089$ &$4/3^{a}$   &$27.3$  &$0.9$   &$3.7$    \\
HCG\,80 &$239.80149$ &$65.22588$  &$0.0311$&$38.64$ &$458.0$ &$0.107$ &$4$     &$27.0$  &$0.8$   &$6.2$    \\
HCG\,81 &$244.55958$ &$12.79522$  &$0.0498$&$28.26$ &$262.0$ &$0.138$ &$4$     &$26.7$  &$0.6$   &$0.2$    \\
HCG\,82 &$247.09196$ &$32.82366$  &$0.0361$&$108.45$&$1071.0$&$0.129$ &$4$     &$26.6$  &$0.7$   &$0.5$    \\
HCG\,84 &$251.03369$ &$77.83618$  &$0.0556$&$92.54$ &$329.0$ &$0.358$ &$6/5^{a}$   &$25.8$  &$0.5$   &$-2.6$   \\
HCG\,86 &$297.9967$  &$-30.82598$ &$0.0197$&$70.2$  &$470.0$ &$0.19$  &$4$     &$25.3$  &$0.6$   &$-3.5$   \\
HCG\,88 &$313.09504$ &$-5.75791$  &$0.0201$&$79.73$ &$204.0$ &$0.497$ &$4/6^{a}$   &$27.4$  &$1.0$   &$4.5$    \\
HCG\,89 &$320.04505$ &$-3.909$&$0.0296$&$90.76$ &$68.0$  &$1.699$ &$4$     &$28.0$  &$1.0$   &$6.2$    \\
HCG\,93 &$348.85098$ &$18.98311$  &$0.0164$&$108.75$&$324.0$ &$0.427$ &$5/4^{a}$   &$27.5$  &$1.0$   &$0.2$    \\
HCG\,94 &$349.31868$ &$18.71966$  &$0.0422$&$50.07$ &$914.0$ &$0.07$  &$7/4^{a}$   &$25.8$  &$0.4$   &$0.2$    \\
HCG\,95 &$349.88239$ &$9.49184$   &$0.0396$&$46.56$ &$550.0$ &$0.108$ &$4/3^{a}$   &$27.3$  &$0.8$   &$3.0$    \\
HCG\,96 &$351.9929$  &$8.77406$   &$0.0292$&$46.22$ &$240.0$ &$0.245$ &$4$     &$27.0$  &$1.0$   &$3.0$    \\
HCG\,97 &$356.84559$ &$-2.32599$  &$0.0222$&$88.54$ &$727.0$ &$0.155$ &$5/9^{a}$   &\texttwelveudash   &\texttwelveudash   &$2.2$    \\
HCG\,98 &$358.55316$ &$0.37327$   &$0.0265$&$43.65$ &$150.0$ &$0.37$  &$4/3^{a}$   &$26.1$  &$0.6$   &$-3.3$   \\
HCG\,100&$0.33653$   &$13.13255$  &$0.0181$&$51.08$ &$206.0$ &$0.316$ &$4$     &$27.4$  &$1.1$   &$2.8$    \\
\hline
        \multicolumn{10}{l}{$^{a}$The first value is the published number of galaxies in the Hickson catalogue, the second value is the number of galaxies after}\\ 
        \multicolumn{10}{l}{our revision based on our observations and the NED database.}\\
    \end{tabular}
    \label{tab:gen_prop_1}
\end{table*} 

\section{The data}
\label{sec:data}
Our dataset consists of 41 stacked images, which were obtained with the $28-$inch 'Jay Baum Rich telescope' (JBRT) at the Wise Observatory, Israel \citep{2015Ap&SS.359...49B}. This is a $28-$inch (0.7\,m) Centurion$-28$ prime-focus f/3.1 reflector, imaging an $\sim 1\degr$ wide field of view onto a CCD camera behind a doublet field-corrector lens. The camera is a Finger Lakes Instruments ProLine 16801 equipped with a five-position filter wheel and thermoelectrically cooled to approximately $-30^{\circ}~C$ using water assist. The plate scale is 0.83 arcsec\,pixel$^{-1}$, the images are digitized to $16$ bits, the readout is done at $8$ MHz, and the dark counts are $\sim 0.05$\,s$^{-1}$\,pixel$^{-1}$. The 4k $\times$ 4k chip covers a bit less than 1 deg$^2$ of the sky and has a peak quantum efficiency of $67\%$ at 661~nm.

Each HCG was mostly observed during one or several nights a week around the New Moon. For each object we obtained between 30 to 76 individual images. These images were exposed for 300\,s through the luminance (L) filter and were dithered by as much as 20~arcsec. Bias, dark and flat field exposures were collected in each observing night, the flats were taken at dusk and/or at dawn. The reduction procedure were conducted using the \texttt{THELI} package \citep{2005AN....326..432E,2013ApJS..209...21S} This procedure included bias and dark subtraction, flat fielding, registration and median-combining while rejecting outlier pixels to eliminate hot pixels, meteors and passing planes, cosmic ray tracks etc. The final stacked images were also astrometrically solved.
 
\begin{figure}
	\label{fig:hist_1}
    \begin{center}
        \includegraphics[width = 0.5\textwidth]{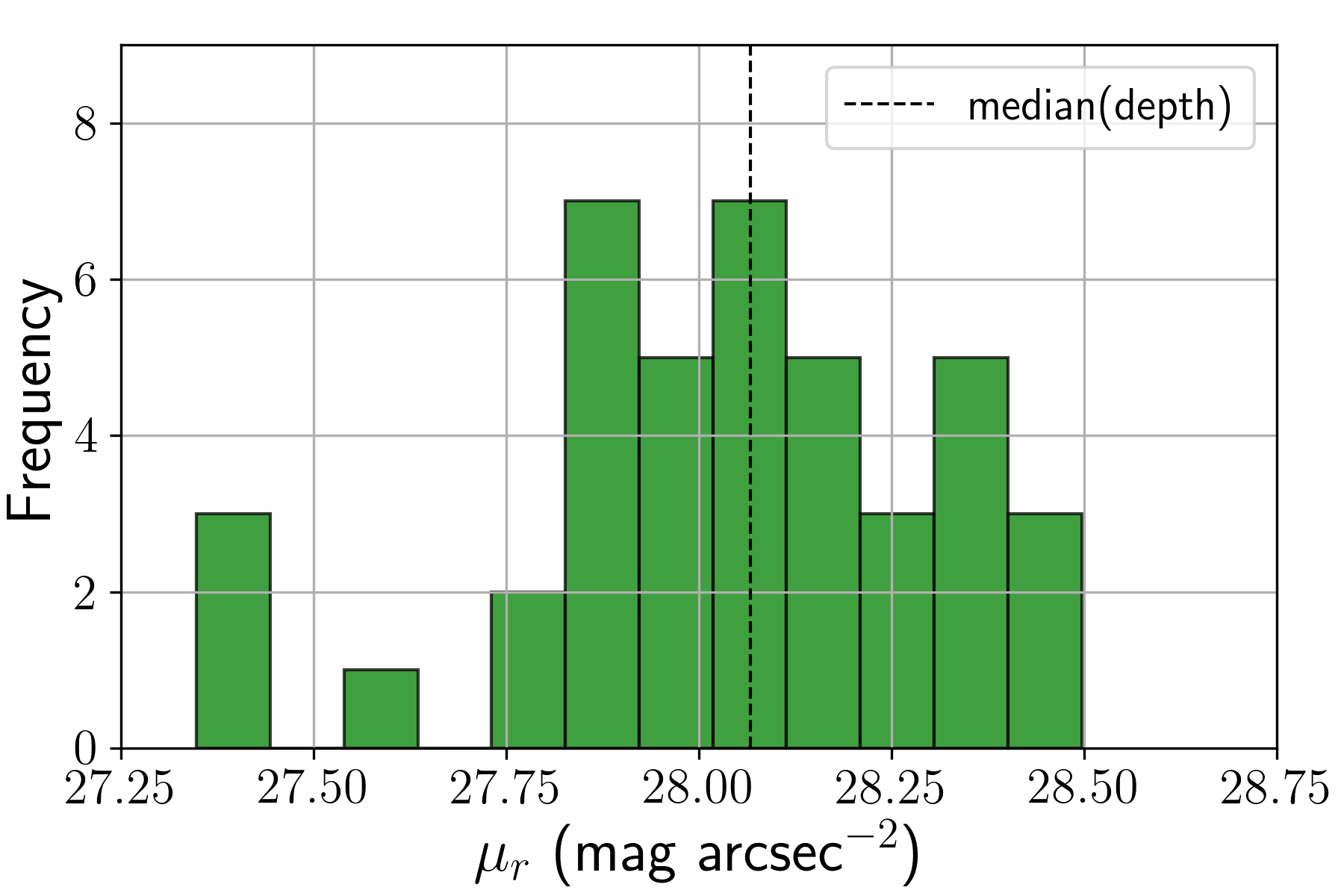}
    \end{center}
    \caption{Photometric depth for all dataset used.}
\end{figure}

The total exposure of the composed images reaches several hours. On average, the depth of the final images reaches the surface brightness $\sim 28.1$~mag\,arcsec$^{-2}$ in the $r$ band,  calculated at the $3\sigma$ level in a box of $10\times10$~arcsec$^2$. The histogram in Fig.~\ref{fig:hist_1} depicts the distribution of our deep images by the photometric depth.

%Pipeline
All images were processed in a semi-automated regime to prepare them for a subsequent analysis in Sect.~\ref{sec:method}. The algorithm, described below, is realised in the Python package \textbf{\texttt{IMAN}}\footnote{\url{https://bitbucket.org/mosenkov/iman_new/src/master/}}. 

First, we performed photometric calibration using multiple non-saturated stars with a high signal-to-noise ratio selected in each frame. We use these stars not only to do photometric calibration, but also to create a core of the point spread function (PSF) for each image (see below). Then, we cross-correlated the selected stars with the Gaia DR2 \citep{2016A&A...595A...1G,2018A&A...616A...1G} 
the Sloan Digital Sky Survey DR16 \citep[SDSS, ][]{2000AJ....120.1579Y,2020ApJS..249....3A} 
and PanSTARRS \citep[][if the frame is not covered by the SDSS fields]{2016ApJ...822...66F} 
photometric database  to estimate the zero-point in the $r$ band for each image. The mean of the standard deviation of the zero-point for the calibration stars for 36 groups listed in Table \ref{tab:gen_prop_1} is 0.05~mag. Adding a colour term has a negligible affect on our calibration and only slightly improves the error on the zero point by 0.01~mag, therefore we do not use it in our calibration. Also, to correct for Galactic extinction, we use the 3D dust map by \citet{2019ApJ...887...93G}.

After that, we cropped each image and masked out all sources using \texttt{SExtractor} \citep{1996A&AS..117..393B} for the area outside the group, with the detection threshold parameter \texttt{DETECT\_THRESH} = 1.8 for more than three contiguous pixels \texttt{DETECT\_MINAREA} = 3. For the region inside the galaxy group we employed the {\tt{mtobjects}} tool\footnote{\url{https://github.com/CarolineHaigh/mtobjects}} \citep{teeninga2015improved}. To preserve a sufficient background with no sources and, at the same time, decrease the size of the image for a photometric decomposition, we used a box with a side 4-5 times larger than the group diameter. To avoid a possible contamination by the scattered light from the masked objects, their linear size was increased by a factor of $1.5$. Then the sky level was fitted with a polynomial of the degree $\leq 3$ based on an iterative method: we started from a zero-order polynomial, increasing this value after each iteration to achieve a flat background in the sky-subtracted image.

\begin{figure}
	\label{fig:extend_psf_star}
    \begin{center}
        \includegraphics[width = 0.4\textwidth]{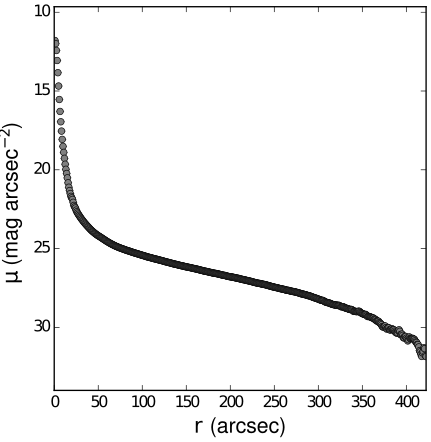}
    \end{center}
    \caption{Azimuthally-averaged radial surface-brightness profile of the extended PSF.}
\end{figure}

As the PSF can significantly affect the true profiles of galaxies and an IGL, this effect should be accounted for while carrying out photometric decomposition \citep{2016ApJ...823..123T}. To reproduce the wings of an extended PSF for all images under consideration, we used an observation of the isolated saturated star HD114946 in one of our deep images and the PSFs of the aforementioned non-saturated stars (used for our photometric calibration) to create the inner part (core) of the PSF individually for each image. We then merge the inner (core) and outer (extended) PSFs by normalising them in an annulus $5$\,arcsec in width where the core and the wings overlap (see \citealt{2017A&A...601A..86K} for details). Using the \texttt{IRAF/ELLIPSE} routine \citep{1993ASPC...52..173T} , we created its azimuthally averaged profile up to a radius of $\approx 400$\,arcsec (see Fig.~\ref{fig:extend_psf_star}). The mean of the full width at half maximum (FWHM) of the PSF for 36 groups listed in Table \ref{tab:gen_prop_1} is $2.5\pm0.4$\,arcsec.

% Mask
In the last step, for each image we created a final mask to exclude from further consideration the light of all sources which do not belong to the group. 

\section{The fitting method}
\label{sec:method}
The IGL is an extended and diffuse light source. Various astronomical and instrumental effects can influence on its detection. Therefore, the separation of the IGL from the light of the galaxies in a group is not always a simple task. 
Here we decided to use photometric decomposition which is a common approach to separate different components in an image. To our knowledge, this method is applied here for the first time to distinguish the IGL from the light of the galaxies. We use the \texttt{GALFIT} code \citep{2002AJ....124..266P,2010AJ....139.2097P} to model both the light of individual galaxies and an IGL. In most cases, we adopt a 2D S\'ersic function \citep{1968adga.book.....S} to describe the profile of individual galaxies. If a galaxy is too small to be resolved, we use a model of the PSF to fit it. The 2D S\'ersic surface-brightness profile with the major-axis intensity given by:
\begin{equation} \label{eq:ellipt_sersic_1}
	I(r) = I_\mathrm{e} \exp \left(-\nu_\mathrm{n}\left[\left(\frac{r}{r_\mathrm{e}}\right)^{1/n} - 1 \right]\right)\,,
\end{equation}
where $I_\mathrm{e}$ is the surface brightness at the effective radius $r_\mathrm{e}$ and $n$ is the S\'ersic index, whereas $\nu_\mathrm{n}$ is given by the solution of the equation
\begin{equation} \label{eq:ellipt_sersic_2}
	\Gamma(2\,n) = 2 \gamma(2n, \nu_\mathrm{n})\,,
\end{equation}
where $\Gamma(x)$ is the gamma function and $\gamma(s, x)$ is the incomplete gamma function.

Following \citet{1990MNRAS.245..130A}, the shape of generalised elliptical isophotes with the semi-major axis $a$ and semi-minor axis $b$ is described by 
\begin{equation} \label{eq:ellipt_sersic_3}
	\left(\frac{|x|}{a}\right)^{c_{0} + 2} + \left(\frac{|y|}{b}\right)^{c_{0} + 2} = 1\,,
\end{equation}
where $c_{0} = 0$ corresponds to a perfect ellipse, values $c_{0} < 0$ correspond to discy isophotes, while values $> 0$ describe boxy isophotes.
The other free parameters are the position angle and the ellipticity ($e = 1 - b/a$). For the IGL, we use a generalised elliptical 2D S\'ersic function, whereas for galaxies in the group pure elliptical isophotes are assumed for simplicity. The use of a generalised shape of the isophotes for the IGL allows us to better control their somewhat not elliptical shape. In principal, it is possible to apply Fourier modes in \texttt{GALFIT} for significantly asymmetric photometric profiles, but the risk of parameter degeneracies significantly increases in this case. Therefore, we decided not to use this possibility in our fitting.  

To obtain the residual IGL profile, we subtract the models of the galaxies in the group from the original image. The parameters of the galaxies (and of the IGL for the five compact groups) are listed in Sect.~\ref{sec:results}.

To compute the errors of the free parameters, we use a set of Monte Carlo simulations. Decomposition was repeated $15$ times for each of the five groups with a different flat sky level, $I_{\mathrm{sky}} \in N(0, \sigma^{2})$, where $\sigma$ is the standard deviation of the sky background in each image. 
We found that the errors of the apparent surface brightnesses (or total magnitudes) are less than $5\%$, whereas the relative errors on the remaining fit parameters of the galaxy models is less than $10\%$. The errors, estimated for the effective radius and the S\'ersic index of the IGL are less than $30\%$,  and less than $5\%$ for the apparent surface brightness errors. We note that the smaller errors correspond to groups with the largest angular size. A relatively big separation between the galaxies in these groups yielded a more robust estimation of the parameters. Due to the large angular size of an IGL component, the effect of the flatness of the sky on the IGL parameters is greater than its effect on the parameters of the constituent galaxies. This can be explained by the fact that the IGL has a very low surface brightness, therefore any significant gradient of the sky background may potentially lead to an erroneous model of the IGL.
In addition to the impact of the sky background uncertainty, the \texttt{GALFIT} optimization procedure itself does not ensure finding a global minimum of $\chi^2$. Also, this procedure is sensitive to the initial guess on the input parameters.

\section{Results}
\label{sec:results}
%%%%%%%%%%% 
In our sample, HCG\,44 has the lowest median redshift $z = 0.0046$. For peculiar velocities of $\sim 300$ km\,s$^{-1}$, the effects of the peculiar motions on the distance to this group can reach $20\%$ (resulting in the luminosity and physical size errors to be as large as $50\%$ and $20\%$, respectively). According to the NED database, the uncertainty in distance estimation for the member galaxies of HCG\,44 is not significantly less than $20\%$. Therefore, we decided to use the cosmological distance scale for this group, despite its significant uncertainty. The remaining groups in our sample have median redshifts larger than $0.0137$, hence the effects of the peculiar motions on the computed distances are less than $8\%$. Thus, we decided not to correct the radial velocities for the peculiar velocities of the galaxies. We compute the absolute magnitudes of the individual galaxies, assuming that their luminosity distances are all based on the median redshifts for the group galaxies and taking into account the correction for Galactic extinction and k-correction.
\begin{equation} \label{eq:k_correction}
	M_{r} = m_{r} - \mbox{ MD } - K_{r}\,,
\end{equation}
where MD is the distance modulus, $K_{r}$ is the k-correction. To compute the k-correction, we employed an analytical approach presented in \citet{2010MNRAS.405.1409C,2012MNRAS.419.1727C}. In their method, the k-correction is approximated by two-dimensional low-order polynomials of only two parameters: redshift and one observed colour. We used the source code provided by \citet{2010MNRAS.405.1409C}\footnote{\url{http://kcor.sai.msu.ru/}}, the group redshifts and dust reddening from the NED \citep{2011ApJ...737..103S}, and total magnitudes in the $g$ and $r$ photometric bands provided in the SDSS DR16. To estimate the physical sizes, we used the angular diameter distance. 

For the selected group members, we provide their names, coordinates, redshifts, and decomposition parameters: the apparent and absolute magnitudes, effective radii, and S\'ersic indices. In the case of point source profiles, we provide only their apparent and absolute magnitudes. Additionally, we provide the $c_{0}$ parameter, which controls the diskyness/boxyness of the IGL
isophotes, and its contribution to the total luminosity of the group $f_{\mathrm{IGL}}$: 
\begin{equation} \label{eq:f_IGL}
	f_{\mathrm{IGL}} = \frac{F_{\mathrm{IGL}}}{F_{\mathrm{IGL}} + F_{\mathrm{gal}}}\,,
\end{equation}
where $F_{\mathrm{IGL}}$ is the total flux of the IGL, and $F_{\mathrm{gal}}$ -- the total flux of the galaxies belonging to the group. All these results are summarised in Tables~\ref{tab:result_1}, \ref{tab:result_1_2}.

\begin{table*} 
	\caption{Results of multicomponent decomposition for the galaxies of the selected groups. It lists the labels of individual galaxies, their names (from NED), coordinates (from NED), redshifts (from NED), and the multicomponent \texttt{GALFIT} decomposition parameters: the apparent magnitude, effective radius, and S\'ersic index.}
	\centering
    \begin{tabular}{l l c c c c c c c}
        \hline \hline
     Group  & Component & NED Name          & RA, Dec. (J2000)        & $z$    &$m_r$&$M_r$& $r_\mathrm{e}$  & $n$ \\
            & label     &                   & deg, deg                &        &     mag        &   mag          & kpc      &     \\
\hline
HCG\,8      &galaxy a  &VV\,521 NED01      &$12.39229$, $23.57825$   &$0.0536$&$14.32$ &$-22.73$&$14.05$ &$10.7$   \\
            &galaxy b  &VV\,521 NED02      &$12.39683$, $23.59156$   &$0.0533$&$14.98$ &$-22.07$&$4.07$  &$4.86$   \\
            &galaxy c  &VV\,521 NED03      &$12.39893$, $23.58415$   &$0.0568$&$14.94$ &$-22.11$&$4.1$   &$4.32$   \\
            &galaxy d  &VV\,521 NED04      &$12.40263$, $23.57339$   &$0.0544$&$14.11$ &$-22.94$&$35.39$ &$12.32$  \\
            &galaxy a1 &MCG\,+04-03-009    &$12.38807$, $23.57364$   &$0.057$ &$17.0$  &$-20.05$&$3.63$  &$0.67$   \\
            &galaxy a2 &MCG\,+04-03-009    &$12.3855$, $23.57392$    &$0.057$ &$18.05$ &$-19.0$ &\texttwelveudash &\texttwelveudash \\
            &galaxy b1 &MCG\,+04-03-007    &$12.38086$, $23.59495$   &$0.0533$&$17.37$ &$-19.68$&\texttwelveudash &\texttwelveudash \\
            &galaxy b2 &MCG\,+04-03-007    &$12.38296$, $23.5959$    &$0.0533$&$18.35$ &$-18.7$ &\texttwelveudash &\texttwelveudash \\
\hline
HCG\,17     &galaxy a  &HCG\,017A          &$33.52135$, $13.31104$   &$0.0608$&$15.15$ &$-22.16$&$6.71$  &$6.57$   \\
            &galaxy b  &WISEA\,J021403.85+131847.2 &$33.51602$, $13.31313$   &$0.0601$&$15.27$ &$-22.04$&$7.54$  &$10.23$  \\
            &galaxy c  &WISEA\,J021405.04+131902.2 &$33.5209$, $13.3173$     &$0.0608$&$15.81$ &$-21.5$ &$5.07$  &$8.02$   \\
            &galaxy d  &WISEA\,J021407.58+131823.6 &$33.53163$, $13.30656$   &$0.0583$&$16.84$ &$-20.47$&$1.77$  &$7.4$    \\
            &galaxy e  &HCG\,017E          &$33.51754$, $13.31905$   &$0.06$  &$18.05$ &$-19.26$&\texttwelveudash &\texttwelveudash \\
\hline
HCG\,35     &galaxy a  &WISEA\,J084521.24+443114.0 &$131.33855$, $44.52057$  &$0.0531$&$14.84$ &$-22.21$&$5.66$  &$3.54$   \\
            &galaxy b  &WISEA\,J084520.59+443032.1 &$131.3358$, $44.50895$   &$0.0545$&$14.0$  &$-23.05$&$11.09$ &$5.4$    \\
            &galaxy c  &WISEA\,J084518.42+443139.5 &$131.32675$, $44.52768$  &$0.0542$&$14.41$ &$-22.64$&$8.6$   &$9.43$   \\
            &galaxy d  &2MASX\,J08452066+4432231 &  $131.33629$, $44.53975$  &$0.0527$&$15.67$ &$-21.38$&$5.98$  &$1.65$   \\
            &galaxy e  &WISEA\,J084520.75+443012.2 &$131.33648$, $44.50333$  &$0.0557$&$16.53$ &$-20.52$&$0.75$  &$4.49$   \\
            &galaxy f  &WISEA\,J084520.87+443159.5 &$131.33697$, $44.53324$  &$0.0545$&$17.19$ &$-19.86$&$1.65$  &$2.8$    \\
\hline
HCG\,37     &galaxy a  &NGC\,2783          &$138.41444$, $29.99297$  &$0.0225$&$11.81$ &$-23.31$&$18.56$ &$5.37$   \\
            &galaxy b  &NGC\,2783B         &$138.38813$, $30.00014$  &$0.0225$&$13.78$ &$-21.34$&$12.74$ &$1.01$   \\
            &galaxy c  &MCG\,+05-22-020    &$138.40563$, $29.99956$  &$0.0245$&$15.21$ &$-19.91$&$2.29$  &$1.61$   \\
            &galaxy d  &MCG\,+05-22-016    &$138.39082$, $30.01578$  &$0.0205$&$15.68$ &$-19.44$&$2.73$  &$1.22$   \\
            &galaxy e  &MCG\,+05-22-018    &$138.39174$, $30.03983$  &$0.0216$&$15.65$ &$-19.47$&$1.79$  &$2.02$   \\
\hline
HCG\,74     &galaxy a  &NGC\,5910 NED02    &$229.8531$, $20.89635$   &$0.0409$&$12.99$ &$-23.4$ &$17.23$ &$4.41$   \\
            &galaxy b  &NGC\,5910 NED01    &$229.85109$, $20.8908$   &$0.0399$&$14.52$ &$-21.87$&$5.2$   &$5.79$   \\
            &galaxy c  &NGC\,5910 NED03    &$229.85765$, $20.89954$  &$0.0409$&$15.84$ &$-20.55$&$2.43$  &$3.3$    \\
            &galaxy d  &WISEA\,J151931.81+205301.0 &$229.88244$, $20.88356$  &$0.039$ &$15.4$  &$-20.99$&$3.13$  &$2.41$   \\
            &galaxy e  &WISEA\,J151927.78+205431.8 &$229.86579$, $20.90888$  &$0.0383$&$17.1$  &$-19.29$&$1.79$  &$1.0$    \\
\hline
    \end{tabular} 
    \label{tab:result_1}
\end{table*}

\begin{table*} 
	\caption{Results of multicomponent decomposition for the IGL of the selected groups. It lists the coordinates, and the multicomponent \texttt{GALFIT} decomposition parameters: coordinates, the apparent and absolute magnitude, the IGL fraction, the central surface brightness, effective radius, S\'ersic index, generalized ellipse parameter $c_{0}$, separation between the geometric centre of the group and the IGL centre $\delta_{c}$, and the ratio between the $\delta_{c}$ and the median separation $R$.}
	\centering
    \begin{tabular}{l l c c c c c c c c c}
        \hline \hline
        Group  & RA, Dec. (J2000)     & $m_r$ & $M_r$ & $f_{\mathrm{IGL}}$& $\mu_{0,r}$ & $r_\mathrm{e}$  & $n$  & $c_{0}$ & $\delta_\mathrm{c}$ &	$\delta_\mathrm{c} / R$ \\
               & deg, deg             &     mag          &   mag            &                   & mag\,arcsec$^{-2}$ & kpc      &      &        & kpc          &  \\
\hline
HCG\,8  &$12.39116$, $23.57919$   &$14.09$ &$-22.91$&$0.251$ &$21.44$ &$33.33$ &$1.53$  &$-0.2$  &$15.86$ &$0.26$   \\
HCG\,17 &$33.51674$, $13.31384$   &$15.78$ &$-21.47$&$0.163$ &$24.39$ &$30.81$ &$0.54$  &$0.0$   &$20.26$ &$0.61$   \\
HCG\,35 &$131.33303$, $44.52$     &$15.03$ &$-21.97$&$0.128$ &$25.96$ &$100.0$ &$0.85$  &$0.8$   &$10.67$ &$0.16$   \\
HCG\,37 &$138.4079$, $29.9916$    &$13.63$ &$-21.47$&$0.127$ &$24.74$ &$72.0$  &$1.35$  &$1.0$   &$34.87$ &$0.83$   \\
HCG\,74 &$229.84939$, $20.89713$  &$15.28$ &$-21.07$&$0.075$ &$26.68$ &$70.0$  &$0.4$   &$0.2$   &$35.63$ &$0.59$   \\
\hline
    \end{tabular} 
    \label{tab:result_1_2}
\end{table*}      

\begin{figure}
	\label{fig:IGL_HCG74}
    \begin{center}
        \includegraphics[width = 0.4\textwidth]{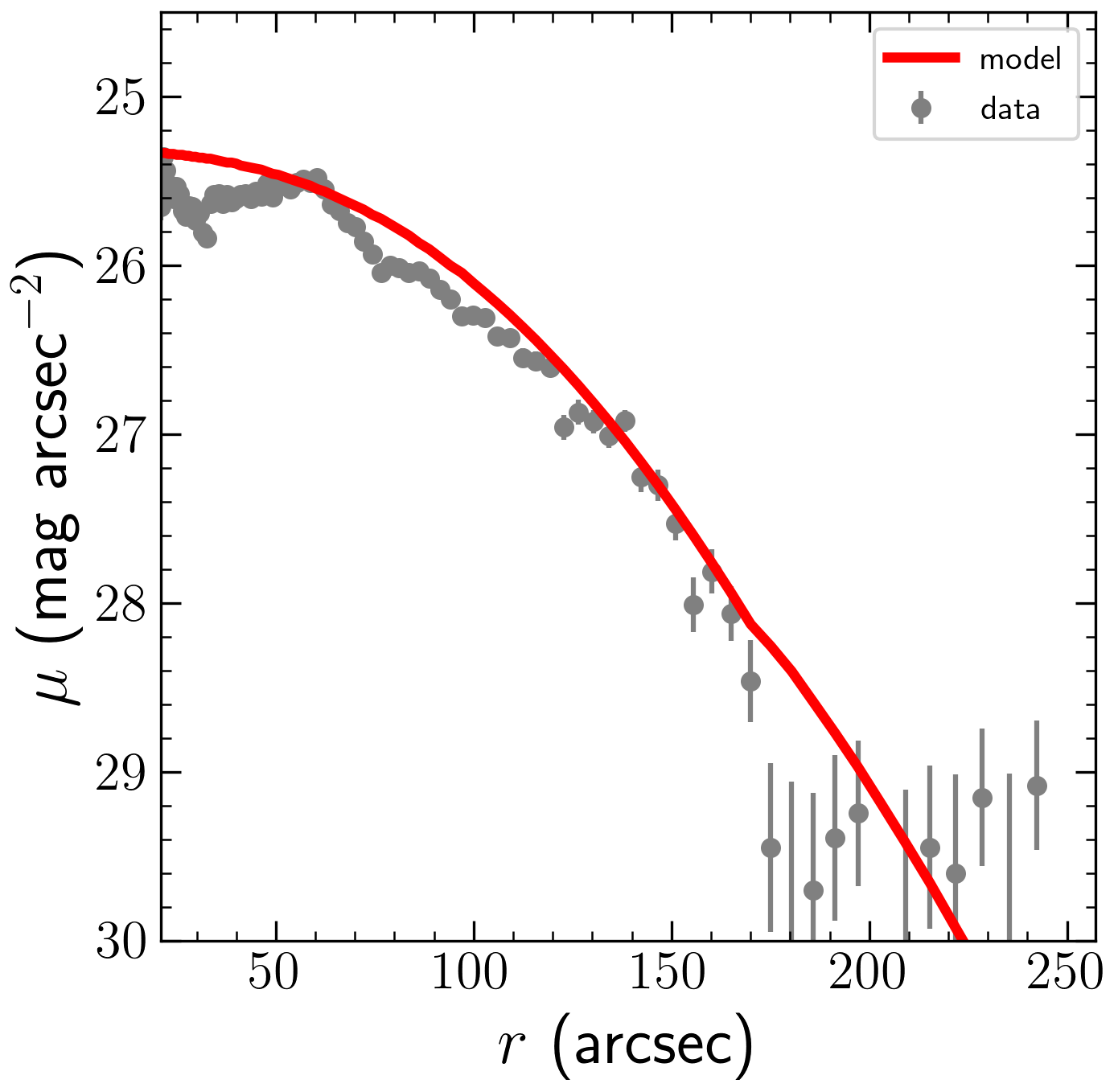}
    \end{center}
    \caption{Azimuthally-averaged brightness profiles of the IGL in HCG\,74. The gray dots correspond to the profile created for the residual image (`image of the group minus model of the galaxies'). The red line corresponds to the model profile of the IGL, convolved with the PSF. The \texttt{IRAF/ELLIPSE} routine \citep{1987MNRAS.226..747J} was used to create these profiles.}
\end{figure}

For convenience of the identification of the model parameters with the objects in the images, we label the group members in the image using the labels in Table~\ref{tab:result_1} (see Fig.~\ref{fig:hcg8_deco}, \ref{fig:hcg17_deco}, \ref{fig:hcg35_deco}, \ref{fig:hcg37_deco}, \ref{fig:hcg74_deco}). In Fig.~\ref{fig:IGL_HCG74}, we provide an example of azimuthally-averaged surface brightness profiles of the IGL for HCG\,74.
As can be seen, the depicted profiles are close to Gaussian and, as expected, are well-consistent.

\citet{2005MNRAS.364.1069D,2008MNRAS.388.1433D} studied the IGL in six compact groups (HCG\,15, 35, 51, 79, 88 and 95) using a wavelet analysis. For the common group HCG\,35, they estimated the $f_{\mathrm{IGL}}=11\pm2\%$, which is perfectly consistent with our result ($12.8\%$). The authors attempted to fit their reconstructed IGL using the \texttt{IRAF/ELLIPSE} routine but in their residual image they noted large discrepancies between the model and the image. On the contrary, our simultaneous fitting of the galaxy profiles and the IGL resulted in the models which provide a quite good relative residual $(I_\mathrm{obs}-I_\mathrm{mod}) / I_\mathrm{obs}$ within the isophote 26~mag\,arcsec$^{-2}$ with less than 30\% of the pixels which show a deviation more than 30\%.
Therefore, we can conclude that our parametrisation of the IGL with a generalised S\'ersic function is robust for the selected groups.

To study the relation between the dynamical properties and possible signs of the diffuse light, we estimated a mean surface brightness value at the geometric centre of the group (hereafter called the mean surface brightness of the IGL) for $33$ out of the $36$ targets listed in Table~\ref{tab:gen_prop_1}. To obtain this estimate, we selected a circular area with a radius equal to half of the median separation in the centre of the group and calculated the mean value and standard deviation within this area. In so doing, all non-IGL sources, including the galaxies of the group, were masked. Since the number of unmasked pixels was usually less than $1000$, we consistently increased the radius of the circle by a factor of $\sqrt[4]{2}$ for a more robust result. To avoid the effect of outlying pixels, we used $\sigma$-clipping ($\sigma = 3$ with the number of iterations $k=5$). If we compare the mean surface brightnesses of the IGL for the five decomposed groups with those of the remaining groups ($25.9\pm0.15$\,mag\,arcsec$^{-2}$ versus $26.9\pm0.7$\,mag\,arcsec$^{-2}$), we can see that the brightness of the IGL in the non-decomposed groups (except HCG\,84, 86, 94, 98) is 3 times fainter, on average, than in the groups with the visual presence of the IGL.

During our visual inspection of the images, apart from the diffuse light in the compact groups we noted other low surface brightness features in and around the galaxies of the groups, such as tidal tails and streams, warped discs and tilted envelopes (for edge-on galaxies), Galactic cirri, shells, bridges, and faint polar structures. 
A detailed analysis of these low surface brightness structures will be carried out in Brosch et al. (in prep), therefore here we do not provide the images of these groups, nor do we discuss these features. However, in this paper we compare the presence of these features with the characteristics of the IGL and galaxies of the groups. We point out that only five groups in our sample (HCG\,71, 76, 86, 88, 89) do not show any signs of tidal features and a diffuse light at a level of 28\,mag\,arcsec$^{-2}$. The median separation in these groups does not differ from the average median separation in the remaining groups. We surmise that the galaxies in these groups are either only start interacting, or are gas-poor and thus do not produce outstanding tidal features, or have gone through the active phase of interaction and, thus, are ready to coalesce. Overall, the ubiquitous presence of fine structures in compact groups makes them ideal targets for studying galaxy interactions in dense environments.

\begin{figure}
	\label{fig:hist_2}
    \begin{center}
        \includegraphics[width = 0.5\textwidth]{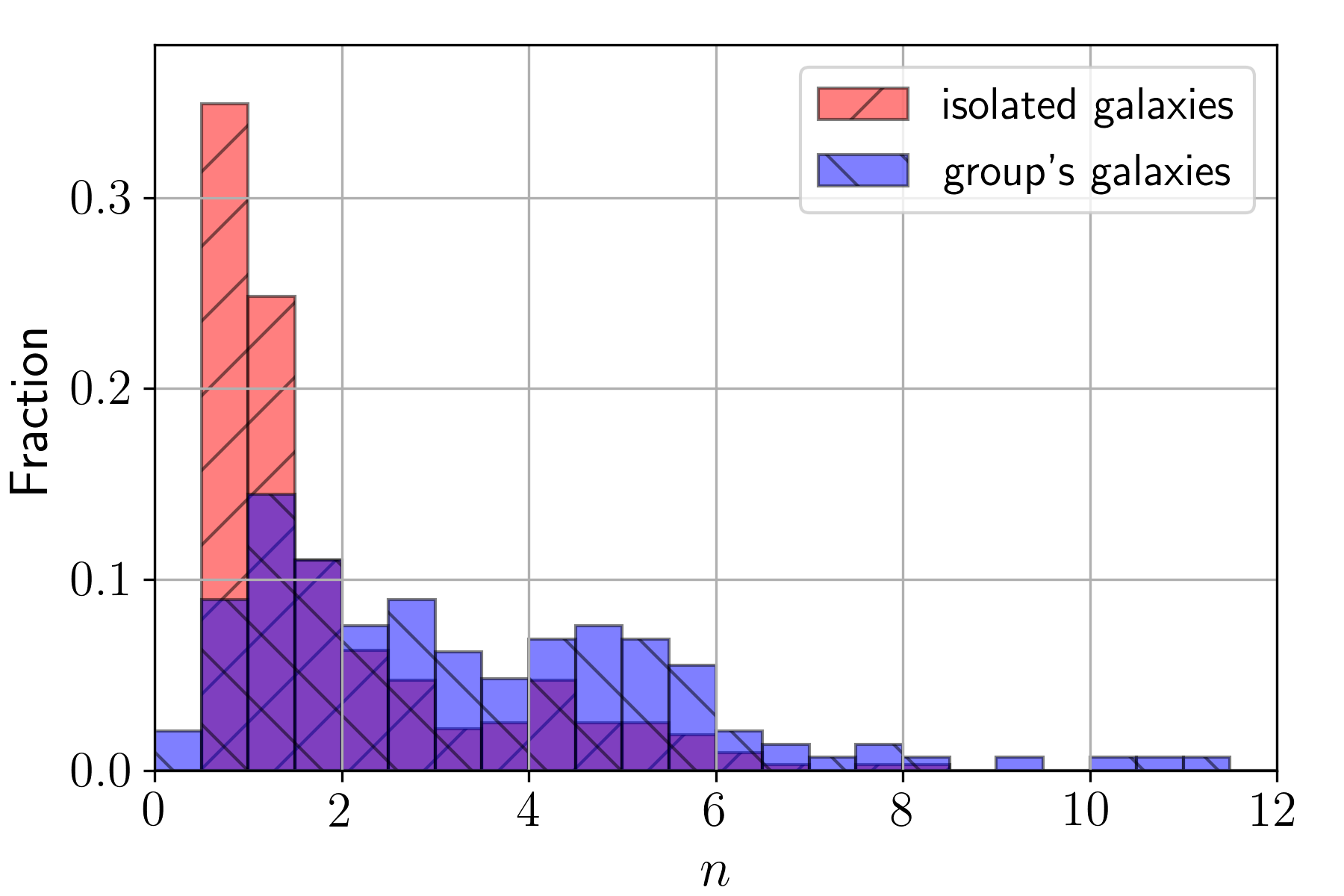}
    \end{center}
    \caption{Distribution by the S\'ersic index for individual 147 galaxies in all 36 compact groups and 318 isolated galaxies from \citet{2011ApJS..196...11S} (see text).}
\end{figure}

We also estimated the S\'ersic parameters of the group members in all 36 groups (see Fig.~\ref{fig:hist_2}, the blue histogram). This was done to quantify the morphology of the galaxies in compact groups and to estimate their general structural properties (size, total magnitude). As to the five groups with an IGL, all galaxy profiles in these groups were fitted simultaneously using a single S\'ersic function for each group member. However, here we did not take into account the diffuse component as we assume it to be very faint in comparison with the individual members of the group and, thus, it should have little effect on the estimated galaxy parameters. We discuss in Sect.~\ref{sec:discussion} that the unaccounted diffuse light does affect the results of the fitting for individual galaxies but its influence on the obtained results can be estimated. The fitting was successful for 147 out of the total 163 galaxies in all 36 groups. For the remaining 16 galaxies, the fitting yielded unreliable results due to a small angular size of the galaxy or due to a close proximity to another source.   

As one can see in Fig.~\ref{fig:hist_2}, although the distribution by the S\'ersic index has a major peak at $n \sim 1$, the fraction of such disc galaxies in compact groups is estimated to be less than $36\%$. Therefore, we qualitatively show that the dominated morphological types of galaxies in compact groups are galaxies with a luminous spheroidal component (64 \% of the galaxies with $n>2$ -- spiral galaxies with a classical bulge and early-type galaxies, see Sect.~\ref{sec:relation}).  

\section{Discussion}
\label{sec:discussion}

\subsection{On the robustness of our fitting method}
\label{sec:rob_method}
Our multicomponent decomposition approach allowed us to simultaneously quantify the IGL and the galaxies in the five compact groups. One of our advantages over the wavelet-based approach \citep{2008MNRAS.388.1433D} is a possibility to roughly compare the IGL surface brightness profiles with the model distributions obtained in cosmological simulations \citep{2015RMxAA..51...13A}. However, this method has some restrictions. It is only applicable for an IGL of a symmetric shape. Objects with significantly asymmetric photometric profiles can also be fitted, but using Fourier modes (for example, it is possible with \texttt{GALFIT} as shown in \citealt{2010AJ....139.2097P} for several galaxies with significantly twisted and bound isophotes). However, as has been above noted, this would greatly increase the chance of parameter degeneracies. Therefore, in this paper we only fitted groups with an IGL of a symmetric shape. Also, our approach faces difficulties in the case of closely spaced galaxies in very tight compact groups and does not guarantee finding the optimal parameters for the individual profiles due to their mutual overlapping. This is also important in the case of tightly bound non-decomposed groups while estimating the mean surface brightness of the IGL between the individual galaxies (HCG\,1, 3, 97).

It is interesting to consider the fitted central surface brightness for the IGL model profiles. For HCG\,17, HCG\,35, HCG\,37, and HCG\,74, these values lie in the relatively narrow interval $24.4\leq \mu_{0,r} \leq 26.7$ mag\,arcsec$^{-2}$. For HCG\,8, this value is significantly brighter $\mu_{0,r} = 21.4$~mag\,arcsec$^{-2}$. If we consider the decomposition parameters for HCG\,8 (see Tables~\ref{tab:result_1}, \ref{tab:result_1_2}), we find that 
the centre of the IGL model is very close to the centre of the galaxy \textit{(d)} and $r_\mathrm{e}$ for its model is significantly larger than for the other galaxies in the group. This might mean that either the profiles of the IGL and the galaxy \textit{(d)} were not clearly separated during the decomposition or that the galaxy \textit{(d)} is the dynamical centre similar to the brightest cluster galaxy (BCG) in rich clusters. According to the cosmological simulations in \citet{2017MNRAS.464.1659Q} and \citet{2018MNRAS.475..648P}, $\sim 70\%$ of the stellar mass in BCGs is accreted. Thus, the diffuse light component in this group may be an extended halo of the BCG or, at least, cannot be well-separated from the light of the BCG. Since HCG\,8 consists of 8 closely spaced galaxies and their brightness profiles overlap, this can negatively affect the quality of the decomposition. Thus, the extraordinary IGL parameters for HCG\,8 should be taken with caution and require an additional verification using alternative methods.

\begin{figure}
	\label{fig:all_isophots}	
	\center{\includegraphics[width=0.85\linewidth]{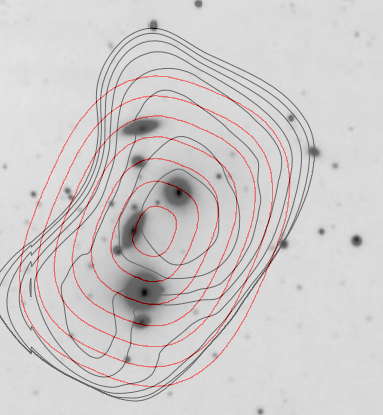}}
	\caption{IGL component of HCG\,35 identified and reconstructed with the OV\_WAV in the $R$ band in \citet{2008MNRAS.388.1433D} (black contours) and our IGL model (red contours). The contours correspond to the levels in the range from $25.75$ to $27.5$~mag\,arcsec$^{-2}$ with a step of $0.25$~mag\,arcsec$^{-2}$.}
\end{figure}

\citet{2008MNRAS.388.1433D} plotted the isophotes for the reconstructed IGL component in HCG\,35 (see Fig.~\ref{fig:all_isophots}). To compare them with our results, we superimposed the isophotes for the model IGL, which we obtained in our study. Although the plausibility of the distribution of the IGL on small scales is questionable, it broadly matches the results of \citet{2008MNRAS.388.1433D}. As stressed by the authors, their IGL fitting using the \texttt{IRAF/ELLIPSE} routine faced difficulties, probably due to the use of the reconstructed image that already contained wavelet decomposition errors. On the contrary, our fitting method works with images which only contain image processing errors. As mentioned earlier, the IGL fraction for HCG\,35 in our model is consistent with the one from \citet{2008MNRAS.388.1433D}. As in this study we consider the groups with rather symmetric 2D profiles, we can conclude that in the case of an approximate symmetry of the IGL, one can robustly quantify its parameters using a multicomponent photometric decomposition.

\subsection{The IGL and dynamical status of the groups}
\label{sec:correlations}

In Sect.~\ref{sec:results}, we analysed the properties of 36 compact groups. It is now interesting to consider the relations between the different quantities which characterise the dynamical status of these objects.

\begin{figure}
    \label{fig:scatter_1}
	\centering
    \includegraphics[width = 0.5\textwidth]{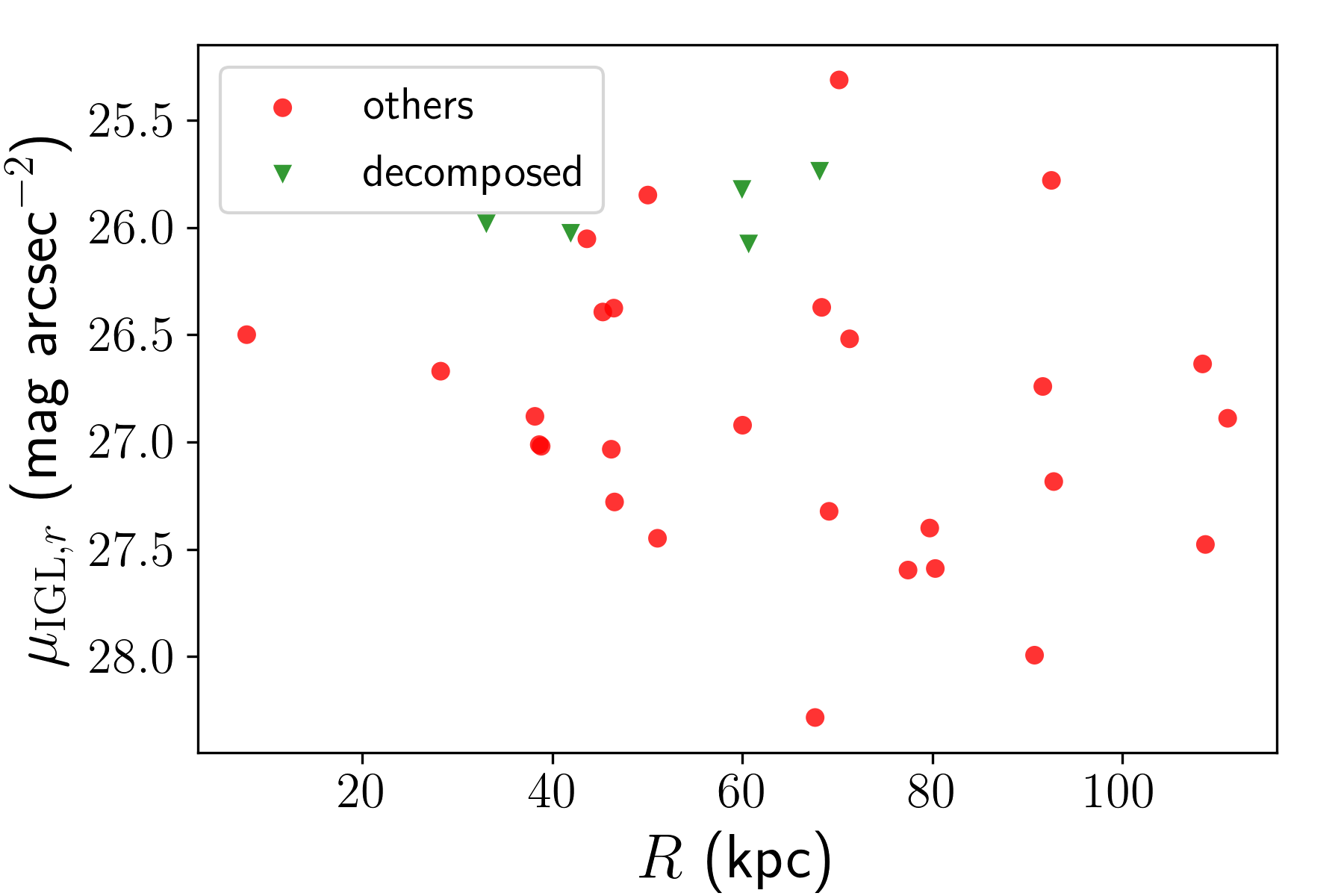}
    \caption{Dependence of the median separation on the mean surface brightness of the IGL in the groups.}
\end{figure}

In Fig.~\ref{fig:scatter_1}, we show the dependence between the median projected separation and the mean surface brightness of the IGL. As one can see, the correlation is barely visible (the Pearson correlation coefficient $\rho=0.3$) and not statistically significant ($p=0.17$). Obviously, the median separation does not only depend on the physical properties of the group, but also on the geometry and orientation of the group in space relative to the observer. For example, HCG\,5 is a very elongated ($b/a\sim0.3$) group, whereas HCG\,17 is almost round. Consequently, the projected median separation should not necessarily correlate with the dynamical properties of the group. We assume that this correlation might be stronger if real distances between the galaxies had been known.

\begin{figure}
	\label{fig:pec_vs_mu}
    \begin{center}
        \includegraphics[width = 0.5\textwidth]{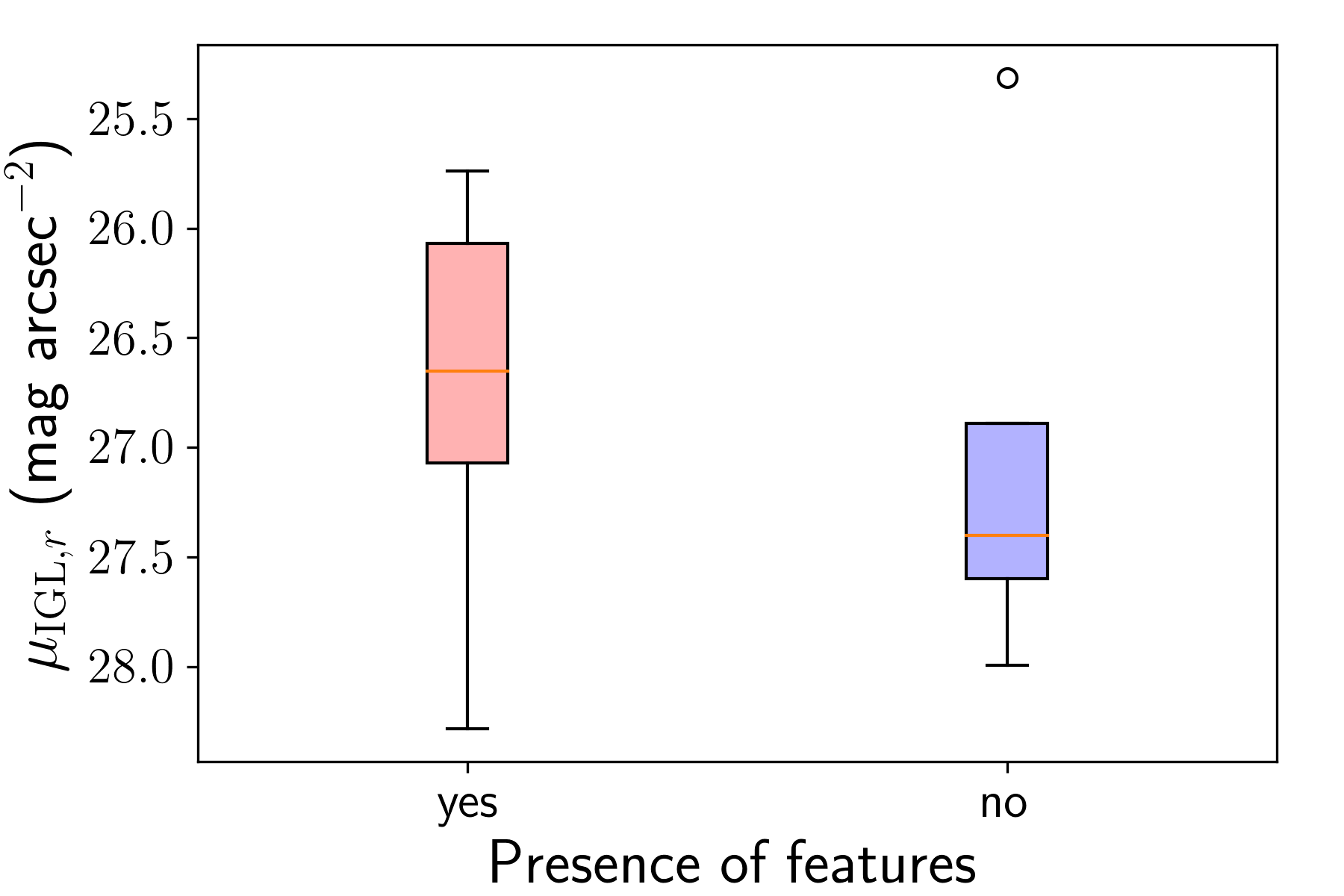}
    \end{center}
    \caption{Difference in the mean surface brightness of the IGL between the groups with visible faint features and without them. The sub-sample with faint features contains 28 entries, whereas the one without low-surface brightness features -- only 5 entries.}
\end{figure}

As the mean surface brightness of the IGL in a compact group characterises the diffuse light in the group, we can now address how it relates to the presence of low surface brightness features in and around the galaxies of our compact groups. We conducted a one-way analysis of variance, where $\mu_{\mathrm{IGL},r}$ is the outcome and the presence of faint features or diffuse light is the factor. This analysis is used to compare two means from two independent (unrelated) groups using the F-distribution. The null hypothesis for the test is that the two means are equal. The estimated p-value $0.985$ is greater than the significance level $\alpha = 0.05$ and, therefore, we cannot reject the null hypothesis (see Fig.~\ref{fig:pec_vs_mu}). Thus, the differences in $\mu_{\mathrm{IGL},r}$ between the groups are likely due to random chance. We note here that the sub-sample without faint features contains only 5 objects.

\begin{figure}
	\label{fig:scatter_2}
    \begin{center}
        \includegraphics[width = 0.5\textwidth]{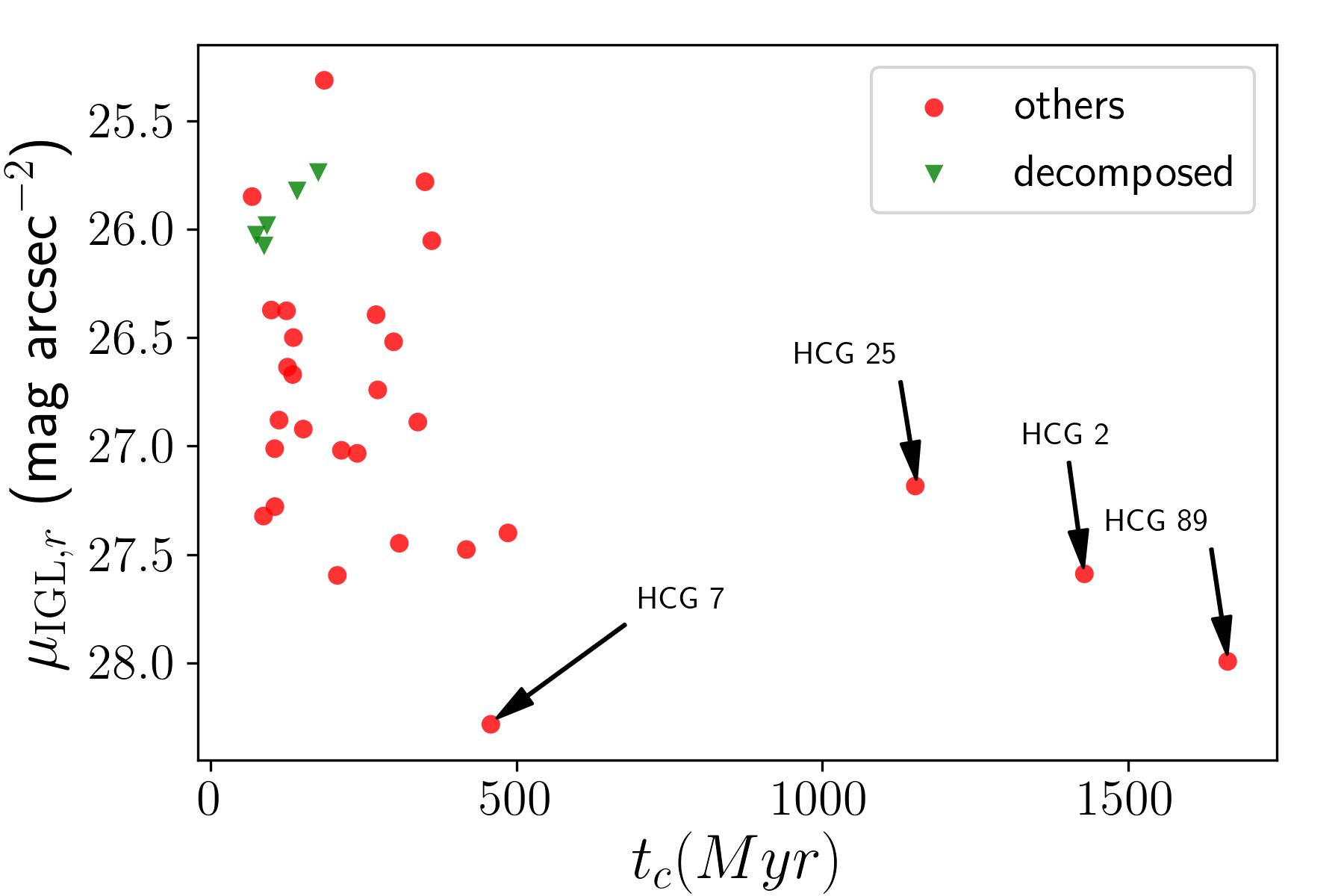}
    \end{center}
    \caption{Dependence between the crossing time and mean surface brightness of the IGL.}
\end{figure}

According to \citet{1992ApJ...399..353H}, the effects of the dynamical evolution should be most pronounced in groups with small crossing times $t_\mathrm{c} $. Therefore, one can expect to see a significant correlation between $t_\mathrm{c} $ and $\mu_{\mathrm{IGL},r}$. However, the scattering diagram for these two quantities in Fig.~\ref{fig:scatter_2} shows a very weak and not statistically significant correlation ($\rho=0.28$ and $p=0.14$). One reason for this fact can be that the intensive dynamical evolution does not always lead to the formation of a bright diffuse component in the centre of the group. For example, in HCG\,78 the diffuse light is essentially comparable to the image depth limit. Also, as has been noted by \citet{2008MNRAS.388.1433D}, compact groups are far from being relaxed and virialized. This is also seen for our much larger sample: the geometry of the groups is mostly irregular, without a certain brightest galaxy in the centre. The absence of fine structures and tidal features is evidence that the members of most groups have just started intensive interactions. Also, the diffuse light is non-symmetrically distributed in the vast majority of the groups in our sample (excluding the ones which we selected for the decomposition of the regular IGL). Finally, as we discuss below, the formation of these groups started 2-3~Gyr ago which is much longer than the crossing time for almost all compact groups in the sample. Therefore, the distribution of the IGL should be mainly governed by the formation age of the group.

\begin{figure}
	\label{fig:scatter_3}
    \begin{center}
        \includegraphics[width = 0.5\textwidth]{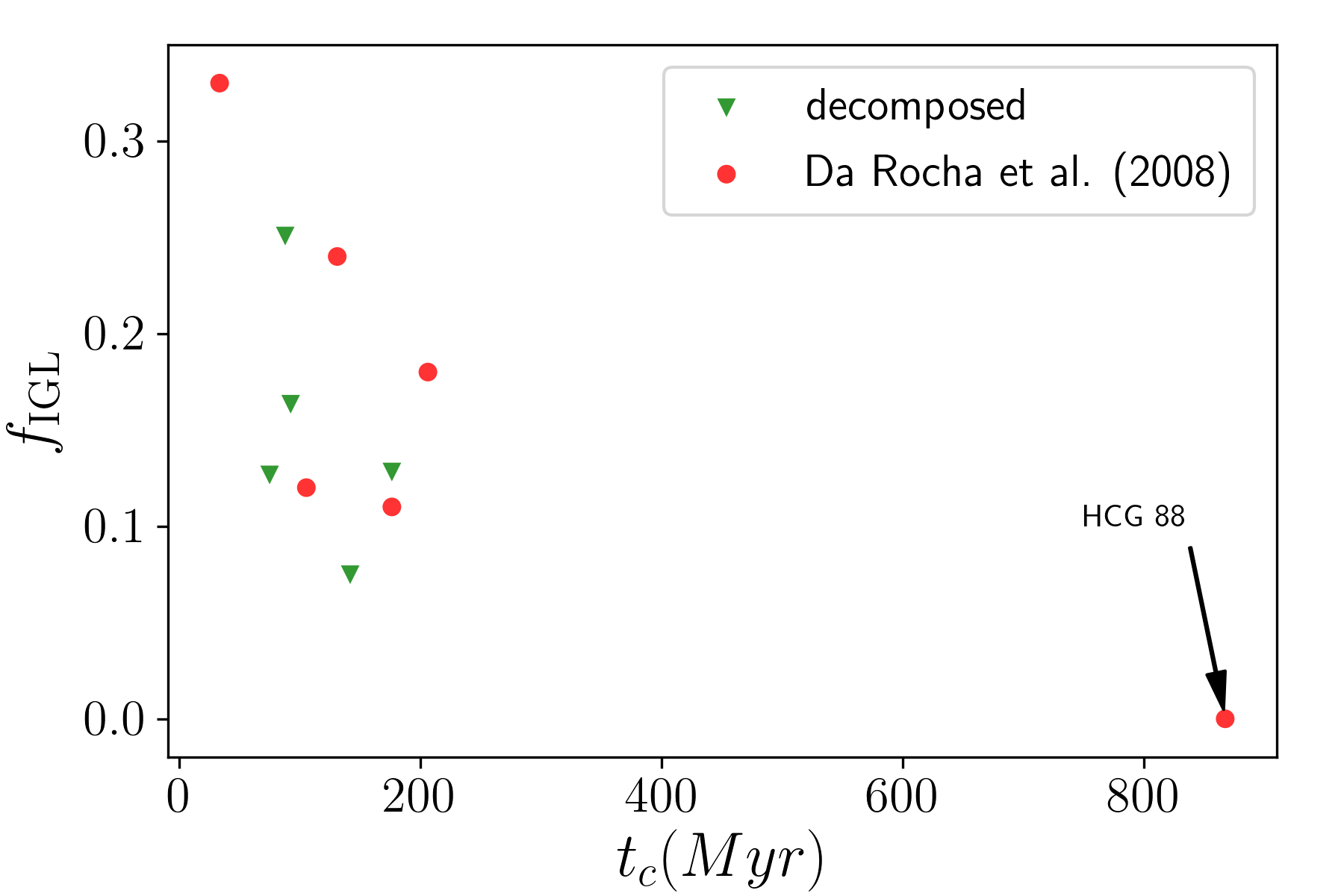}
    \end{center}
    \caption{Dependence between the crossing time and the IGL fraction for the decomposed groups and the six groups from \citet{2008MNRAS.388.1433D}.}
\end{figure}

Similarly, no correlation is seen between the IGL fraction $f_{\mathrm{IGL}}$ and $t_\mathrm{c} $ for the decomposed groups ($p=0.4$, see Fig.~\ref{fig:scatter_3}). Our conclusion contrasts with the result obtained for a sample of 6 compact groups in \citet{2008MNRAS.388.1433D} (their results are also depicted in Fig.~\ref{fig:scatter_3}). Interestingly, they note no signs of an IGL in HCG\,88. This group has a relatively large crossing time, but according to Table~\ref{tab:gen_prop_1} and Fig.~\ref{fig:scatter_2}, many groups, that do not show signs of a diffuse light, have significantly shorter crossing times than HCG\,88. Therefore, for a reliable detection of a correlation between $f_{\mathrm{IGL}}$ and $t_\mathrm{c} $, one needs to increase the sample size. For our subsample with the IGL, no correlation is detected. If this is true, then this may indicate that the active phase of the dynamical evolution in such groups has ended after a few $t_\mathrm{c}$. Therefore, the IGL formation ended far in the past and, thus, $f_{\mathrm{IGL}}$ is no longer dependent on the crossing time. According to a spectral stellar population synthesis modelling \citep{2012A&A...546A..48P}, HCGs were most likely formed $\sim 3$~Gyr in the past. Furthermore, using cosmological hydrodynamical simulations in the framework of the \texttt{EAGLE} project \citep{2015MNRAS.446..521S,2015MNRAS.450.1937C}, \citet{2020MNRAS.491L..66H} found that the typical coalescence time for compact groups that merge between $z = 1$ and $z = 0$ is $2-3$\,Gyr. 

As shown in N-body simulations by \citet{2001ApJ...549L.187G}, dynamical timescales for the infall of the gravitationally bound galaxies in a compact group, and the subsequent vanishing of the compact group are not controlled by the galaxy-galaxy interactions if binary interactions between the group galaxies are unimportant relative to the global field of forces caused by the common massive halo that determines their trajectories. 

\citet{2008MNRAS.388.1433D} found the correlation between the fraction of early-type galaxies and the crossing time. As the fraction of early-type galaxies is a reliable dynamical evolution indicator, in contrast to $t_\mathrm{c}$, we inspected this correlation for our sample and the sample from \citet{2008MNRAS.388.1433D} and did not find any dependence (for all groups from both samples $p=0.13$, see Fig~\ref{fig:f_Egal_vs_tc_our}). Therefore, all the aforementioned facts suggest that the crossing time does not characterise the dynamical status of a compact group and its age may exceed the crossing time by several dozen times (see Fig.~\ref{fig:scatter_2},\,\ref{fig:scatter_3}).

\begin{figure}
	\begin{minipage}{1.0\linewidth}
 	   	\begin{center}
			\includegraphics[width = 1.0\textwidth]{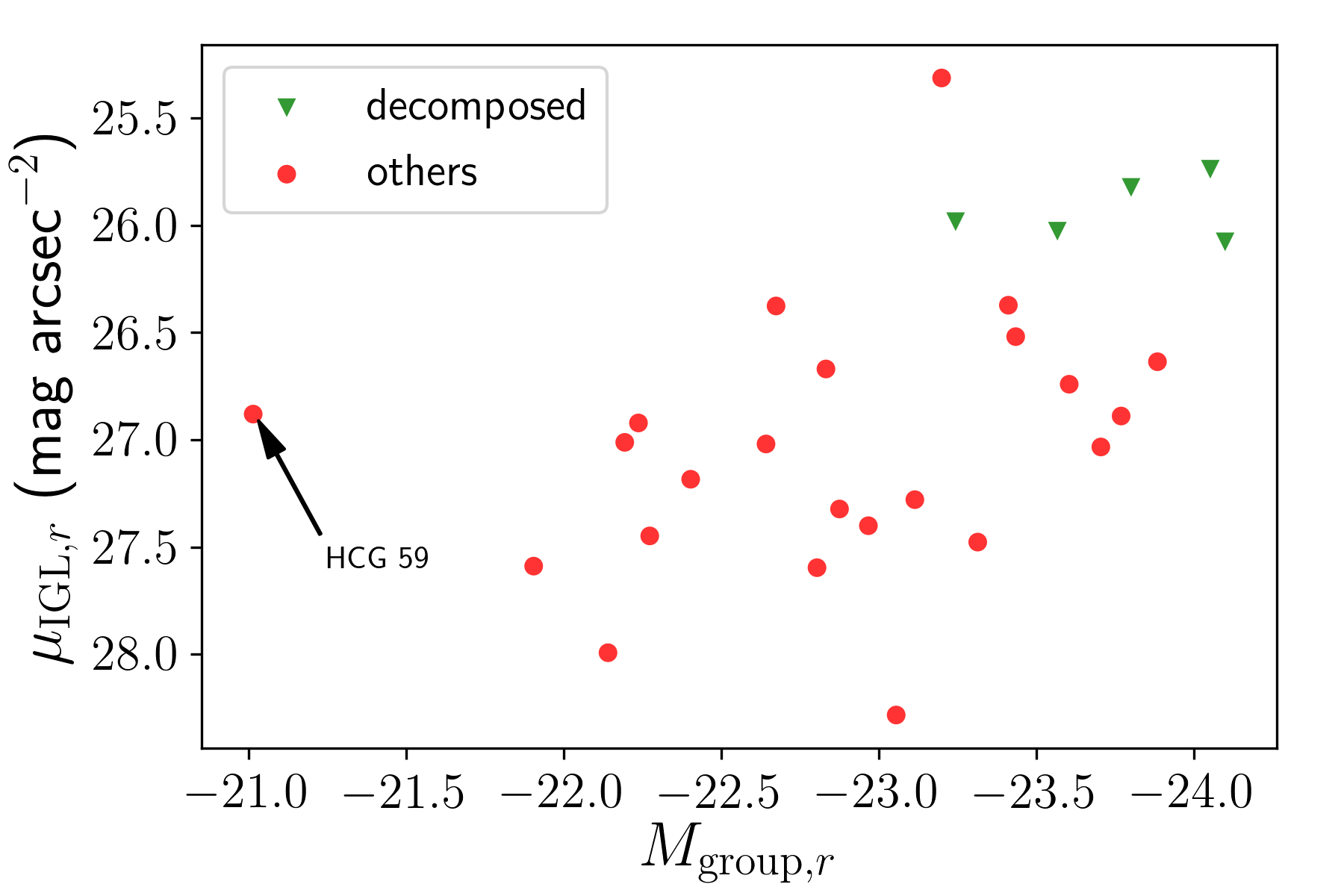}
    		\end{center}
    		\label{fig:M_group_vs_mu}
    \end{minipage}
    \vfill
    \begin{minipage}{1.0\linewidth}
 	   	\begin{center}
        		\includegraphics[width = 1.0\textwidth]{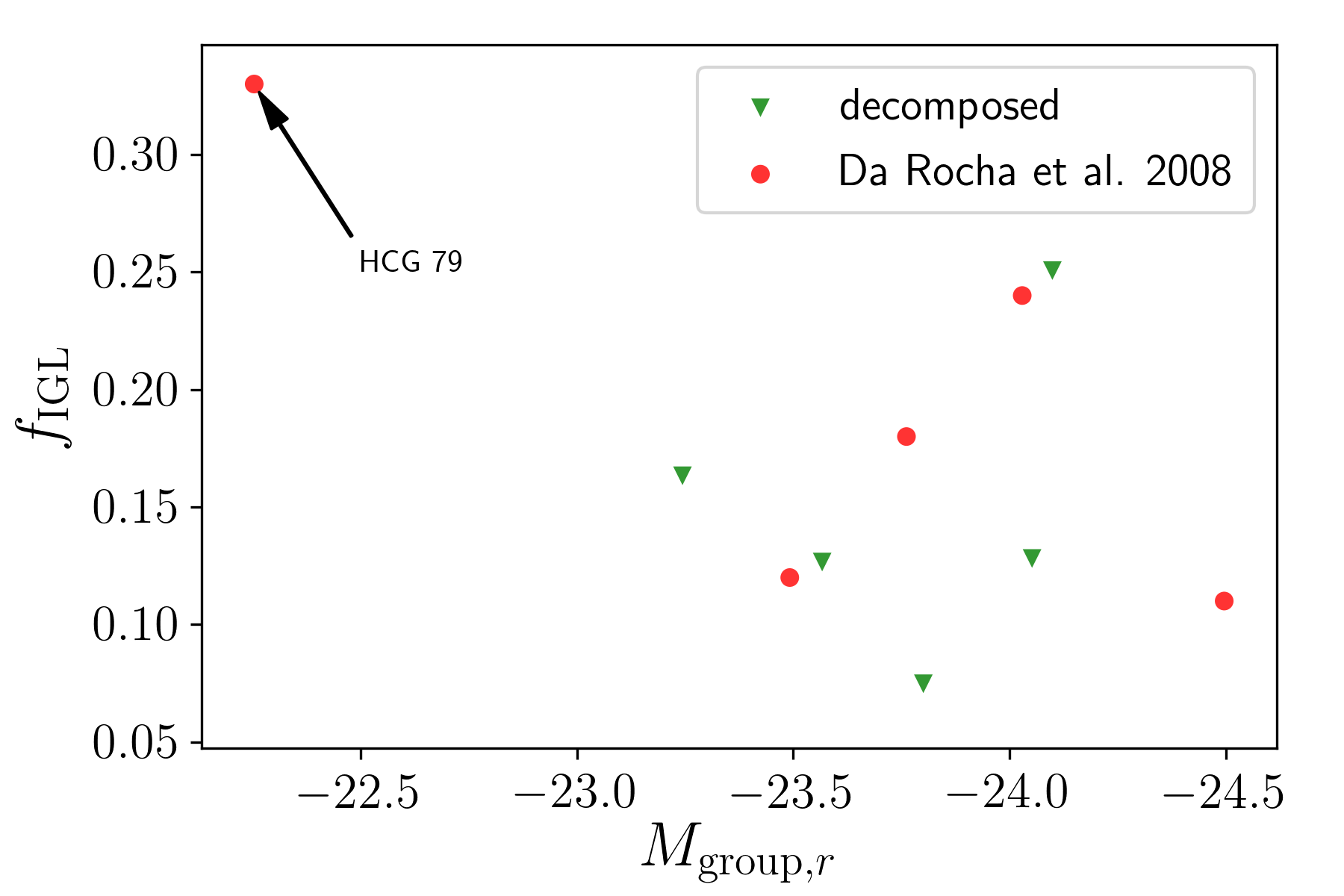}
    		\end{center}
   		\label{fig:M_group_vs_f_IGL}
    \end{minipage}
    \caption{The top panel shows the dependence between the absolute group magnitude and the mean surface brightness of the IGL. The bottom panel shows the dependence between the absolute magnitude of the group and the IGL fraction for our decomposed groups and the five groups from \citet{2008MNRAS.388.1433D}.}
\end{figure}

\begin{figure}
	\label{fig:f_IGL_vs_M_IGL}
    \begin{center}
        \includegraphics[width = 0.5\textwidth]{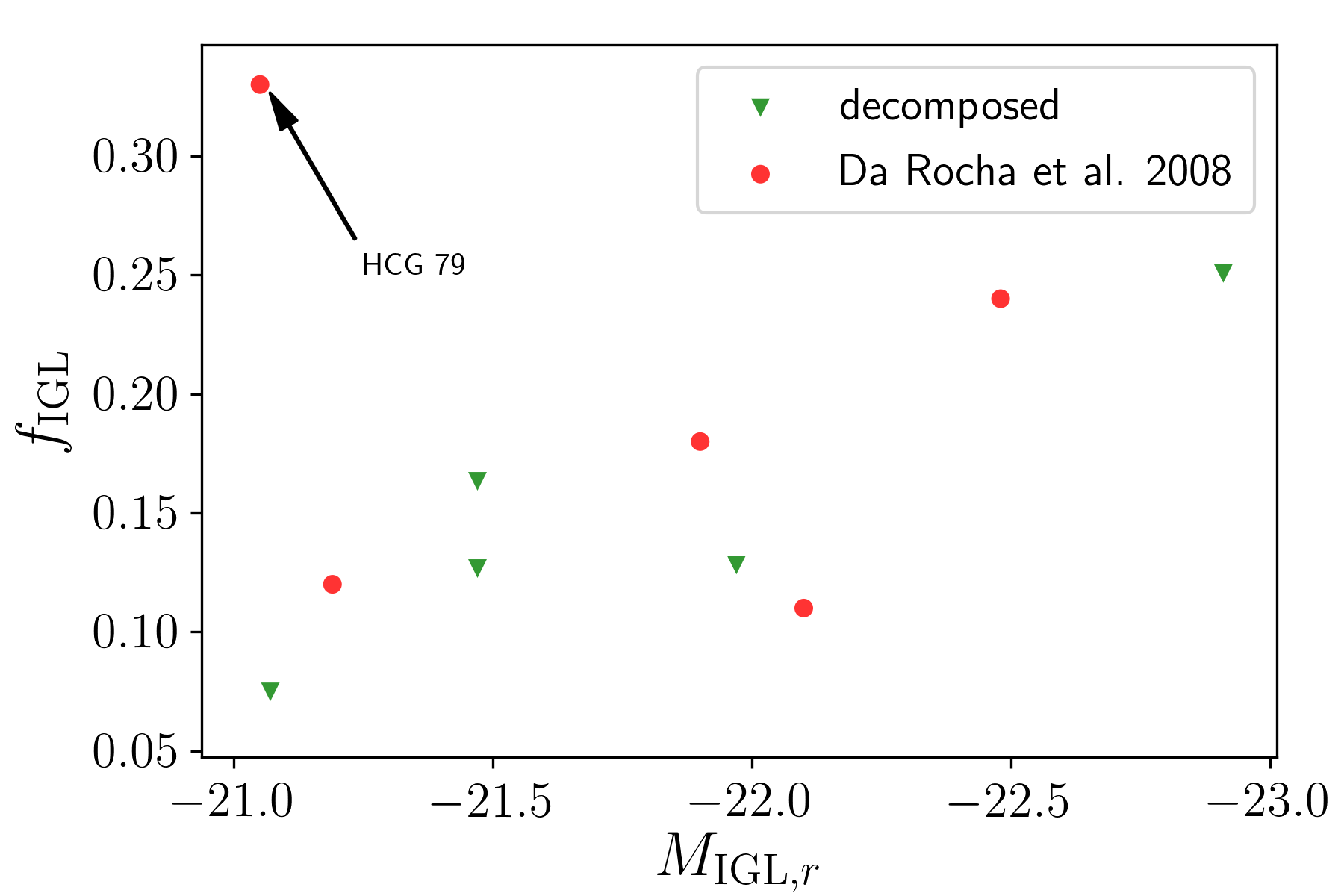}
    \end{center}
    \caption{Dependence between the IGL
fraction and the absolute IGL magnitude for our decomposed groups and the five groups from \citet{2008MNRAS.388.1433D}.}
\end{figure}

Recent cosmological hydrodynamical simulations from \citet{2020MNRAS.494.4314C} for systems with a wide range of masses from $10^{9}$\,$M_{\odot}$ to $10^{13}$\,$M_{\odot}$ focus on the Intra-Halo Stellar Component (IHSC). These authors show that, on average, the IHSC mass fraction increases with the total stellar mass of the system from $10^{10}$\,$M_{\odot}$ up to $10^{12}$\,$M_{\odot}$. This mass range includes mid-size galaxy groups. As the stellar mass of galaxy systems correlates with their luminosity, we consider the correlation between the absolute group magnitude and 1) $\mu_{\mathrm{IGL},r}$ and 2) $f_{\mathrm{IGL}}$ for the five decomposed groups and the five groups from \citet{2008MNRAS.388.1433D} (see Fig.~\ref{fig:M_group_vs_mu}). 
We find a moderate correlation between the absolute magnitude of the groups and $\mu_{\mathrm{IGL},r}$ ($\rho=0.48$ and $p=0.009$), but the correlation with $f_{\mathrm{IGL}}$ is not statistically significant ($p=0.76$ without the outlier).  Therefore, we can conclude that the mean surface brightness of the IGL $\mu_{\mathrm{IGL},r}$ increases with the luminosity of the group. It is generally consistent with the results of cosmological hydrodynamical simulations from \citet{2020MNRAS.494.4314C}. We also find a tight correlation between $f_{\mathrm{IGL}}$ and the IGL absolute magnitude $M_{\mathrm{IGL},r}$ for the decomposed groups and the five groups from \citet{2008MNRAS.388.1433D} with the outlier HCG\,79 ($\rho=-0.82$ and $p=0.007$, see Fig.~\ref{fig:f_IGL_vs_M_IGL}). The inspection of the dependence between $\mu_{0,r}$ and the absolute group magnitude shows no correlation ($\rho=0.18$ and $p=0.78$), but the dependence between $\mu_{0,r}$ and $f_{\mathrm{IGL}}$ exhibits some trend for our five decomposed groups ($\rho=-0.97$ and $p=0.006$, see Fig.~\ref{fig:m_central_vs_M}). 

According to our decomposition results for the five groups, the model profile of the IGL has a S\'ersic index ranging from $0.4$ to $1.5$ and tend to be more boxy than discy (the IGL isophotes are boxy in three out of the five groups, see Table~\ref{tab:result_1_2}). This means that the profiles of the IGL (at least for the selected groups) are close to a Gaussian or exponential profile and do not show extended wings. At first glance, this is vastly different from what we observe in clusters for a subsystem consisting of a BCG and ICL: \citet{2019lssc.confE...9K} derived large S\'ersic indices $n \geq 4$ for their general profiles. They explain such large indices by accretion that is predominantly happening in the outskirts of the clusters, which subsequently increases the upward curvature of their surface brightness profiles. However, the dissection of BCG and ICL using a double S\'ersic decomposition yields $n = 1.16 \pm 1.28$ for the ICL in observations by \citet{2021ApJS..252...27K} and $n = 1.89 \pm 1.01$ in simulations by \citet{2015MNRAS.451.2703C}. These studies also provide an estimate of the ICL fraction $f_{\mathrm{ICL}} \sim 20\%$. Thus, our measurements of $f_{\mathrm{IGL}}$ and $n$ for the IGL (see Table~\ref{tab:result_1_2}) are comparable with the measurements of these parameters for the ICL. Small indices in our case may potentially indicate that the dark haloes in the decomposed groups are far from being relaxed and the merger process is not yet complete \citep{2015RMxAA..51...13A}. 

\begin{figure}
	\label{fig:brightness_vs_density}
    \begin{center}
        \includegraphics[width = 0.5\textwidth]{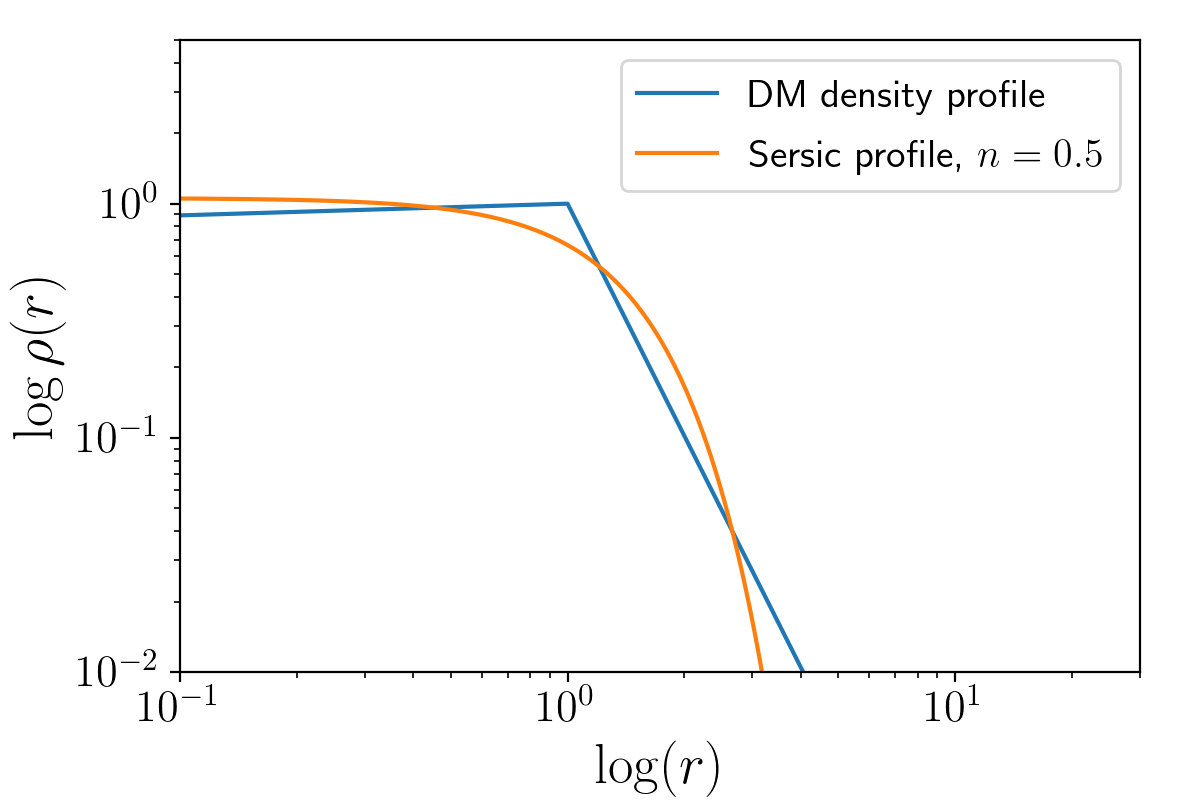}
    \end{center}
    \caption{Model intra-group dark matter density profile of compact groups from \citet{2015RMxAA..51...13A} and its approximation by a S\'ersic law with $n = 0.5$.}
\end{figure}

Using cosmological simulations, \citet{2015RMxAA..51...13A} obtained a density profile for compact associations (groups). Their density profile is given by a flat decline within the association ($\rho \propto r^{0.02}$) and a rapid density decay afterwards ($\rho \propto r^{-3.29}$). We decided to approximate their law with a S\'ersic function (see Fig.~\ref{fig:brightness_vs_density}). As one can see, the S\'ersic index appears to be small ($\sim 0.5$), which is generally consistent with the results obtained by us but for the IGL. Also, the centre of the IGL in our models is relatively close to the geometric centre of the groups (see Table~\ref{tab:result_1_2}). Therefore, we can conclude that there is no inconsistency with the hypothesis that the distribution of the IGL traces the shape of the total mass distribution of the group. Nevertheless, our work is limited by the small number of decomposed compact groups with a rather symmetric IGL and the absence of corresponding gravitational lensing data or X-ray observations for them. That precludes us from making any robust conclusion on the direct relation between the IGL and the total mass distribution in compact groups of galaxies.

\subsection{Morphology-IGL relation}
\label{sec:relation}
According to the morphology-density relation \citep{1980ApJS...42..565D,2008A&A...484..355D}, early-type galaxies are more clustered than later-type ones. As compact groups are very concentrated objects, the fraction of early-type galaxies is generally higher than in the field \citep{1982ApJ...255..382H,1988ApJ...331...64H}. Similar conclusion about the large fraction of early-type galaxies in compact groups was made in \citet{2008MNRAS.388.1433D} for a sample of 6 groups and in \citet{2012A&A...546A..48P} for a sample of 55 groups. Our multicomponent decomposition of the five groups with a symmetric IGL provided S\'ersic indices for 25 galaxies. Also, we estimated S\'ersic indices by fitting the profiles for 122 galaxies in the remaining compact groups. $64\%$ of these galaxies demonstrate a S\'ersic index greater than $2$ (the mean value is $3.1$ and standard deviation is $1.86$). Such indices are typical for early-type galaxies or early-type spirals \citep[see e.g. figure~9 in][]{2019A&A...622A.132M}, but the relation between the morphology and the S\'ersic index is not straightforward \citep{2008ApJ...675L..13V}. 

To explore the impact of the high galaxy density in compact groups on the formation of early-type galaxies, we investigate the distributions by the S\'ersic index for the total number of $147$ galaxies in our compact groups and for a sample of isolated galaxies, which we define as follows. An isolated galaxy must have no companions with measured redshifts within a projected distance of 1 Mpc and with a redshift difference lower than 0.001. To select such galaxies, we exploited the catalogue of pure S\'ersic decompositions in the SDSS $r$ band for a sample of 1.12 million galaxies provided in \citet{2011ApJS..196...11S}. For a consistent comparison with the objects from our sample, we only selected galaxies with redshifts $0.01\leq z \leq 0.06$. For the selection, we use the SDSS and NED databases which provide information on the redshifts.
Finally, we created a random sample of 300 galaxies which yield the above described criteria.
In Fig.~\ref{fig:hist_2}, we present the distribution of the S\'ersic index for both our sample and the sample of isolated galaxies. 
The p-value ($\sim 5.7 \times 10^{-14}$) of the Smirnov homogeneity test for these samples allows us to reject the null hypothesis. Therefore, using the distributions in the light galaxy profiles instead of a subjective morphological classification of galaxies, we quantitatively confirm the previous results that galaxies in compact groups tend to have larger S\'ersic indices and, thus, are dominated by early-type galaxies. This is especially well-seen in comparison with isolated galaxies, which are mostly inhabited by pure disc galaxies, although early-type galaxies can also be found among them \citep[see also][and references therein]{2020A&A...640A..38R}. 

Since we obtained the S\'ersic parameters for the isolated galaxies from the catalogue \citep{2011ApJS..196...11S}, we decided to compare them with the parameters for the compact group galaxies (see Fig.~\ref{fig:early-type_M_abs}). We plotted histograms by the effective radius, the absolute magnitude and the central surface brightness, separately for galaxies with the S\'ersic index $n > 2$ and $n \leq 2$. According to the histograms, the galaxies in our compact groups are systematically bigger and larger than the isolated galaxies, for both $n > 2$ and $n \leq 2$. This result is partially inconsistent with the results by \citep{2012A&A...543A.119C} who compared the properties of galaxies in compact groups and in the field. These authors found that, on average, galaxies in compact groups, being brighter and more massive, are systematically smaller than galaxies in the field. \citet{2008A&A...484..355D} did not find a strong dependence of galaxy size on environment. The inconsistency of these conclusions is likely caused by the selection effect (our samples are quite modest) or the differences in the parameters which describe the size of a galaxy. Note that \citet{2008A&A...484..355D} and \citet{2012A&A...543A.119C} use the $R_{50}$ radius which encloses 50\% of the Petrosian flux whereas in our study we use parametric S\'ersic modelling. 

\begin{figure}
	\begin{minipage}{1.0\linewidth}
 	   	\begin{center}
			\includegraphics[width = 1.0\textwidth]{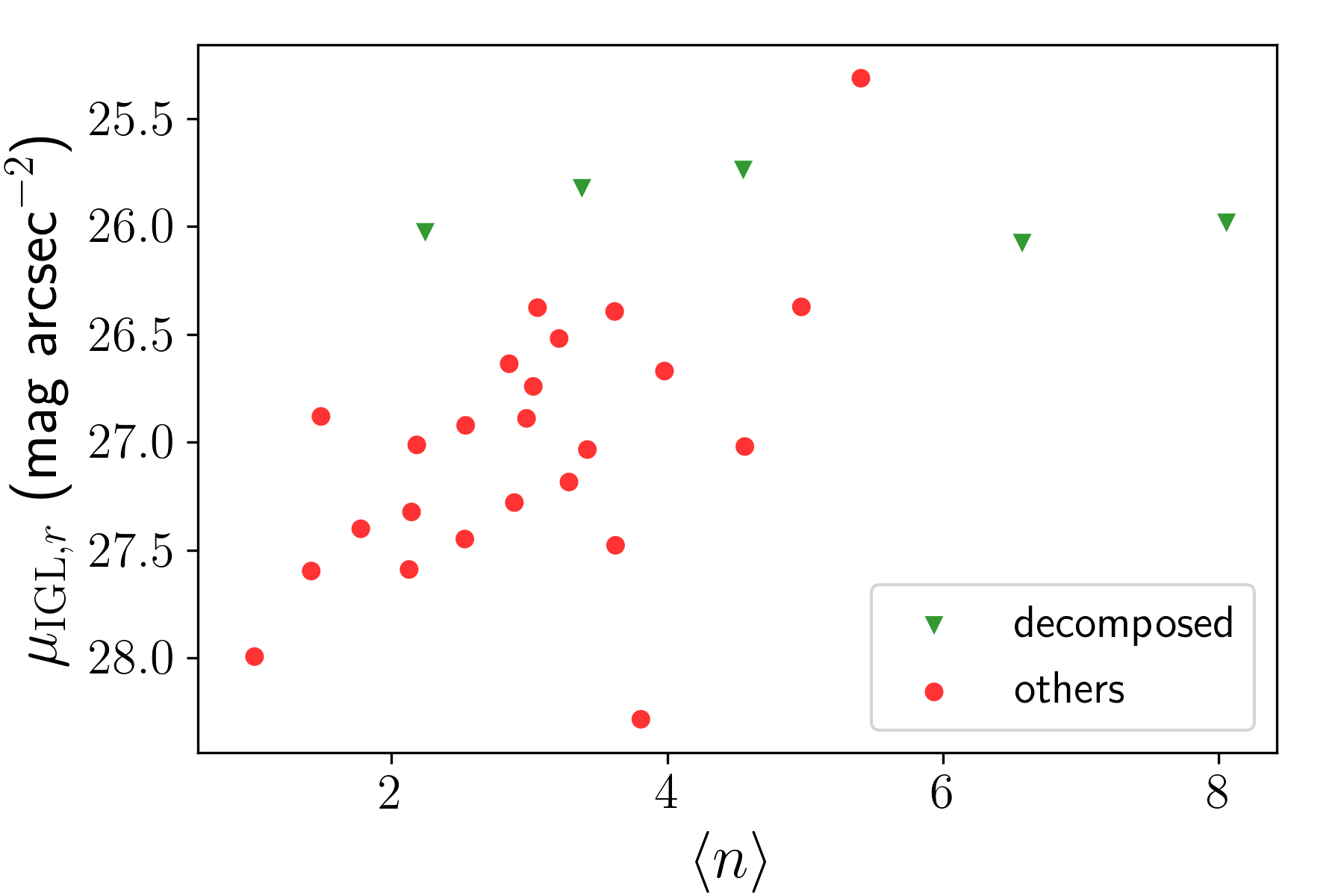}
    		\end{center}
    		\label{fig:mean_ser_vs_m}
    \end{minipage}
    \vfill
    \begin{minipage}{1.0\linewidth}
 	   	\begin{center}
        		\includegraphics[width = 1.0\textwidth]{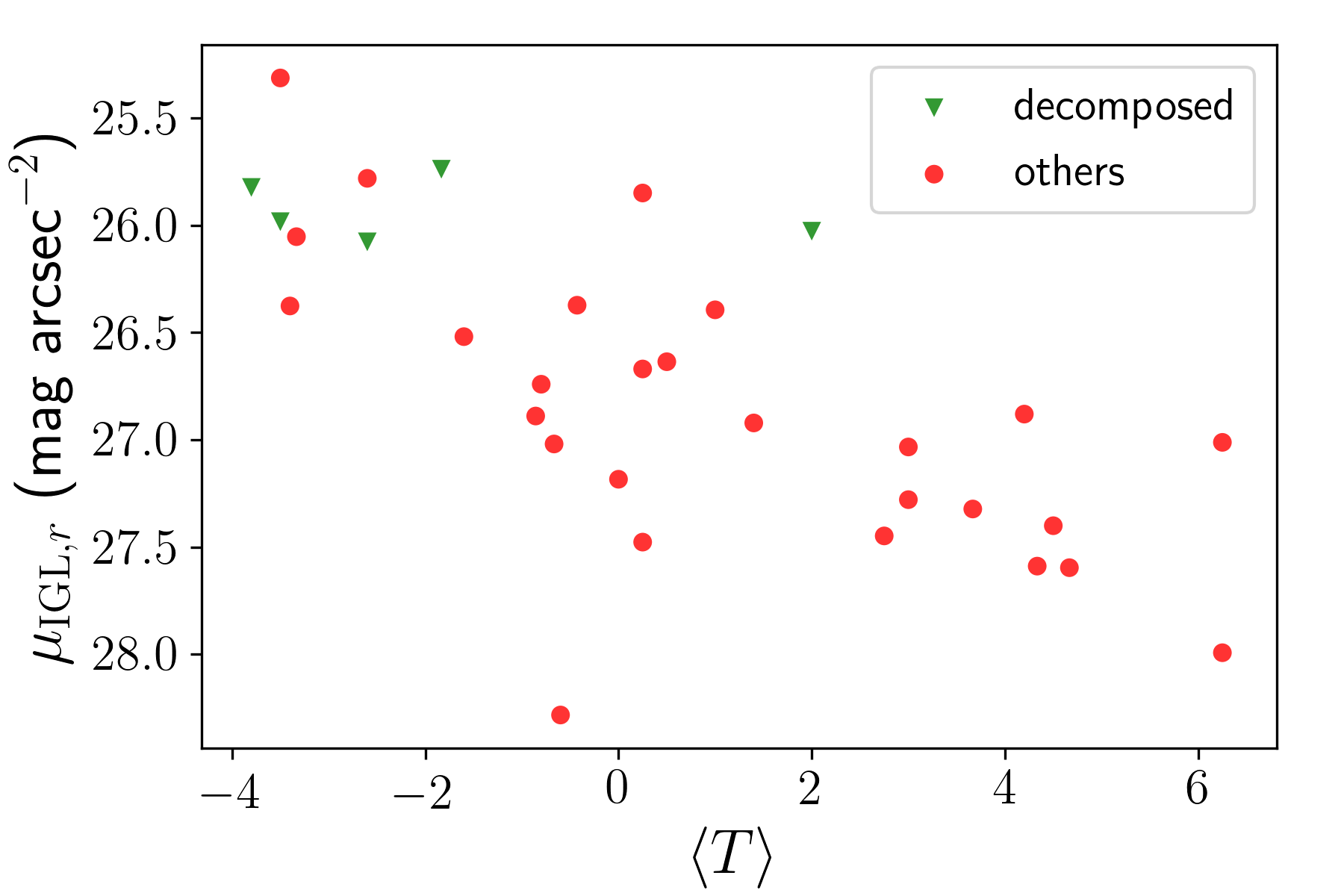}
    		\end{center}
   		\label{fig:mu_vs_num_type}
    \end{minipage}
    \caption{Dependence between the mean S\'ersic index and the mean surface brightness of the IGL shows the moderate correlation with the Pearson correlation coefficient $\rho=0.58$ and $p=0.001$ (top panel). Dependence between the mean numerical galaxy type and the mean surface brightness of the IGL shows a strong correlation with $\rho=0.68$ and $p=1.4 \times 10^{-5}$ (bottom panel).}
\end{figure}

It is reasonable to suggest that the fraction of disc galaxies should be related to the dynamical status of the groups. Therefore, the mean S\'ersic index and the mean numerical morphological type, computed for all members of the group, can serve indicators of the dynamical status of the groups. We examine the relation between the IGL and the dynamical status of the groups in two ways: we consider the correlation between the mean surface brightness of the IGL and the mean S\'ersic index of each group (see the top panel of Fig.~\ref{fig:mu_vs_num_type}), and we find a strong correlation between the mean surface brightness of the IGL and the mean numerical morphological type suggested in \citet{1959HDP....53..275D} (see the bottom panel of Fig.~\ref{fig:mu_vs_num_type}).
The numerical galaxy type varies from -8 to 10, where negative values correspond to early-type galaxies (ellipticals and lenticulars) and positive values -- to late types (spirals and irregulars). We use the NED database to extract the numerical morphological types for all members of our compact groups. The apparent trends in both panels in Fig.~\ref{fig:mean_ser_vs_m} demonstrate the existence of a relationship between the surface brightness of the IGL and the mean morphological type of the group: the brighter the diffuse light, the larger the average S\'ersic index of the group and the smaller the mean numerical galaxy type.
The simple interpretation of this result suggests that a high galaxy density in a compact group affects the morphology of its members and leads to galaxy merging and tidal stripping \citep{2007AJ....133.2630C,2009ApJ...699.1518R}, which naturally increases the brightness of the IGL. This conclusion is concordant with results of \citet{2006A&A...457..771A,2008MNRAS.388.1433D}, which found the correlation between $f_{\mathrm{IGL}}$ and other indicators of dynamical evolution, such as the fraction of early-type galaxies.
 
However, we should point out several reasons why this correlation (and its direct interpretation) should be taken with caution. 
First, the computed S\'ersic indices in our simple decomposition might be slightly overestimated due to the presence of a diffuse light component, which does not belong to any galaxy of the group (note that it was only accounted for in our decomposition of the five HCGs with a bright, regular IGL). To demonstrate that the correlation in Fig.~\ref{fig:mean_ser_vs_m} does not suffer from an unaccounted diffuse light, we subtracted the mean surface brightness of the IGL $\mu_{\mathrm{IGL},r}$ and then carried out decomposition for all groups except the five groups with the estimated diffuse light. Taking into account the possible diffuse light contamination does not significantly affect the S\'ersic indices of the galaxies, although the influence of the background subtraction becomes more pronounced for galaxies with larger S\'ersic indices, as expected.  Nonetheless, the systematic bias in our simple decomposition of the group galaxies does not severely change the correlation between the mean surface brightness of the IGL and the mean S\'ersic index  (see  Fig.~\ref{fig:mean_ser_vs_m_comparison_n_var}).

Second, galaxies with a large S\'ersic index show extended wings in their surface brightness profiles. This light can potentially ``pollute'' the diffuse light in the groups and, thus, increase $\mu_{\mathrm{IGL},r}$, especially in the case of a more compact (denser) group. Such groups should have a smaller ratio of the median separation $R$ to the mean effective radius of the galaxies $\langle r_\mathrm{e}\rangle$ ($R_\mathrm{n} = R / \langle r_\mathrm{e}\rangle$). To examine this relation, we plot a scattering diagram between $R_\mathrm{n}$ and $\mu_{\mathrm{IGL},r}$ (see Fig.~\ref{fig:mu_vs_R_norm}). The estimated p-value in Student's $t$-test $0.16$ is greater than the significance level $\alpha = 0.05$ and, therefore, we cannot reject the null hypothesis. Thus, the correlation between $\mu_{\mathrm{IGL},r}$ and $R_\mathrm{n}$ is likely due to random chance. Consequently, we suppose that the correlation between the mean surface brightness of the IGL and the mean S\'ersic index of the group may imply the real physical basis. However, to ensure that this correlation is not affected by the above mentioned systematic biases, deeper observations of HCGs are highly needed, as well as a more robust technique for extracting the IGL from the compact galaxy group profiles. 

\section{Conclusions}
\label{sec:conclusion}
We summarise the results of our study as follows.
\begin{enumerate}
	\item We obtained and prepared 41 deep images for 39 compact groups of galaxies from the Hickson catalogue. The median depth of our images is $28.07$~mag\,arcsec$^{-2}$ ($3\sigma$, $10\times10$\,arcsec$^2$, the $r$ band). 
	\item For our sample of compact groups, we estimated the mean surface brightness of the IGL (see Table~\ref{tab:gen_prop_1}) and carried out photometric fitting using a pure S\'ersic model for each member of 36 compact groups. We present the fit parameters of the S\'ersic model for all 147 galaxies in Table~\ref{tab:sersic_1}.
	\item The large S\'ersic indices ($64\%$ of galaxies in our groups demonstrate a S\'ersic index greater than $2$) indicate that compact groups mostly consist of early-type galaxies or early spirals. We also found that the S\'ersic indices of galaxies in compact groups are systematically larger than in isolated galaxies. Moreover, the galaxies in compact groups from our sample are brighter and larger than isolated galaxies -- the latter result disagrees with \citet{2008A&A...484..355D} (no strong dependence of galaxy size on environment) and \citet{2012A&A...543A.119C} (galaxies in compact groups are smaller than galaxies in the field). Based on our results, we conclude that tidal interactions in compact groups lead to intensive galaxy mergers and finally to the formation of bright and large early-type galaxies.	
	\item We found that the mean surface brightness of the IGL demonstrates significant correlations with both the mean S\'ersic index of the galaxies in our groups and the mean numerical galaxy type. These indicate a relationship between the presence of the diffuse light and the fraction of early-type galaxies in these systems. A more intensive conversion of late-type and dwarf galaxies in compact groups to early-type galaxies through merging is accompanied by an increasing number of stripped stars which naturally form a brighter IGL component. Using our quantitative analysis of the IGL, we proved that the IGL is an indicator of the dynamical status of compact groups.
	\item We do not see a dependence between the presence of faint tidal features in compact groups and the brightness of their IGL.
	\item We found a weak trend between the mean surface brightness of the IGL and the median projected separation in the groups that is the tighter the galaxies in a compact group, the brighter its IGL. We assume that this correlation would be stronger if a real (not projected) median separation between the galaxies had been known.
	\item We employed photometric decomposition for five, relatively symmetric compact groups with a bright IGL (HCG\,8, 17, 35, 37, 74)  to separate the IGL from the light of the group galaxies and to quantify the profile of the IGL (see Tables~\ref{tab:result_1}~and~\ref{tab:result_1_2}). We found that for HCG\,35, our results are consistent with the results of \citet{2008MNRAS.388.1433D} based on a wavelet analysis. The S\'ersic index of the IGL in the decomposed groups appeared to be $n\sim0.5-1$ which is generally consistent with the mass density profile of dark matter haloes in compact groups obtained from cosmological simulations.
	\item The total luminosity of a compact group correlates with the brightness of its IGL: the brighter the group, the brighter its IGL and the larger its contribution to the total luminosity of the group. Therefore, the characteristics of the IGL do not only depend on the dynamical status of the group, but also on the stellar (and possibly total) mass of the group.
\end{enumerate}

In our future work, we are about to exploit deep observations for a larger sample of HCGs with a detectable IGL, which will be quantified to expand the statistics obtained in this study.

\section*{Acknowledgements}
We thank the anonymous reviewer for their helpful comments on the paper.

Aleksandr Mosenkov acknowledges financial support from the Russian Science Foundation (grant no. 20-72-10052).

This research made use of the NASA/IPAC Extragalactic Database (NED; https://ned.ipac.caltech.edu/), which is operated by the Jet Propulsion Laboratory, California Institute of Technology, under contract with the National Aeronautics and Space Administration and montage (http://montage.ipac.caltech.edu/), which is funded by the National Science Foundation under Grant Number ACI-1440620, and was previously funded by the National Aeronautics and Space Administration's Earth Science Technology Office, Computation Technologies Project, under Cooperative Agreement Number NCC5-626 between NASA and the California Institute of Technology. This work has made use of data from the European Space Agency (ESA) mission {\it Gaia} (\url{https://www.cosmos.esa.intgaia}), processed by the {\it Gaia} Data Processing and Analysis Consortium (DPAC, \url{https://www.cosmos.esa.int/web/gaia/dpac/consortium}). Funding for the DPAC has been provided by national institutions, in particular the institutions participating in the {\it Gaia} Multilateral Agreement. 

\section*{Data availability}
The data underlying this article will be shared on reasonable request to the corresponding author.

\bibliographystyle{mnras}
\interlinepenalty=10000
\bibliography{article} % if your bibtex file is called example.bib

\appendix
\section{Fitting results visualization, tables and plots}
\label{sec:appendix}

\begin{table*}
 \begin{minipage}{180mm}
	\label{tab:sersic_1}
    \caption{S\'ersic parameters for 147 decomposed galaxies. It lists the names of the groups containing the galaxy, the principal galaxy
names from NED, coordinates (from NED), redshifts (from NED), and the S\'ersic fitting parameters: effective radius $r_\mathrm{e}$, the central surface brightness $\mu_{0,r}$, the absolute magnitude $M_{r}$, and the S\'ersic index $n$. $*$ denotes galaxies with either a large $n>8$ or a tiny $r_\mathrm{e}\mathrm{[arcsec]}<\mathrm{FWHM}/2$. For these galaxies our decomposition is likely to be unreliable due to either a poor angular resolution or a strong overlapping with a neighbouring object.}
	\centering
	\begin{tabular}{r c c c c c c c}
        \hline \hline
        Group & Name           & RA (J2000) & Dec. (J2000) & $r_\mathrm{e}$ & $n$ & $\mu_{0,r}$                       & $M_{r}$     \\
              &                & deg        & deg          & kpc            &     & mag\,arcsec$^{-2}$             & mag         \\
\hline
HCG\,1  &UGC\,00248 NED01&$6.52971$   &$25.72519$  &$18.38$ &$2.86$  &$18.23$ &$-22.19$ \\
HCG\,1  &UGC\,00248 NED02&$6.52479$   &$25.71933$  &$4.03$  &$3.34$  &$14.84$ &$-21.34$ \\
HCG\,1  &WISEA\,J002558.82+254331.0&$6.49504$   &$25.72519$  &$2.13$  &$2.85$  &$15.55$ &$-20.23$ \\
$*$HCG\,1  &WISEA\,J002554.42+254325.2&$6.4767$&$25.72366$  &$17.0$  &$11.22$ &$0.63$  &$-22.2$  \\
HCG\,2  &UGC\,00314   &$7.87237$   &$8.40061$   &$5.76$  &$1.9$   &$19.47$ &$-20.22$ \\
HCG\,2  &UGC\,00312   &$7.84962$   &$8.46683$   &$9.12$  &$1.75$  &$19.83$ &$-21.14$ \\
HCG\,2  &UGC\,00312 NOTES01        &$7.82855$   &$8.47482$   &$1.91$  &$2.74$  &$15.08$ &$-20.58$ \\
HCG\,3  &MCG\,-01-02-032 &$8.55492$   &$-7.56483$  &$7.06$  &$0.75$  &$21.27$ &$-20.97$ \\
HCG\,3  &WISEA\,J003409.83-073608.5&$8.54103$   &$-7.60224$  &$1.74$  &$2.7$   &$14.76$ &$-20.82$ \\
HCG\,3  &WISEA\,J003425.11-073558.2&$8.60468$   &$-7.5996$   &$2.35$  &$2.81$  &$14.95$ &$-21.08$ \\
HCG\,5  &NGC\,0190 NED01 &$9.72783$   &$7.06269$   &$15.14$ &$4.02$  &$14.57$ &$-22.14$ \\
HCG\,5  &NGC\,0190 NED02 &$9.72804$   &$7.05672$   &$4.16$  &$4.06$  &$12.75$ &$-21.06$ \\
HCG\,5  &UGC\,00397 NOTES03        &$9.71992$   &$7.07297$   &$4.88$  &$5.61$  &$10.47$ &$-20.51$ \\
HCG\,7  &NGC\,0201&$9.89508$   &$0.85989$   &$11.62$ &$1.82$  &$19.71$ &$-21.64$ \\
HCG\,7  &NGC\,0192&$9.80598$   &$0.86434$   &$11.89$ &$4.65$  &$13.61$ &$-22.14$ \\
HCG\,7  &NGC \,0197&$9.8283$&$0.89192$   &$4.57$  &$2.49$  &$17.92$ &$-20.13$ \\
HCG\,7  &NGC\,0196&$9.82433$   &$0.91276$   &$4.29$  &$5.6$   &$10.13$ &$-21.44$ \\
HCG\,7  &WISEA\,J003915.46+005633.2&$9.81443$   &$0.94258$   &$0.73$  &$4.48$  &$13.12$ &$-16.94$ \\
$*$HCG\,8  &VV\,521 NED01&$12.39229$  &$23.57825$  &$14.05$ &$10.7$  &$0.9$   &$-22.73$ \\
HCG\,8  &VV\,521 NED02&$12.39683$  &$23.59156$  &$4.07$  &$4.86$  &$11.13$ &$-22.07$ \\
HCG\,8  &VV\,521 NED03&$12.39893$  &$23.58415$  &$4.1$   &$4.32$  &$12.22$ &$-22.11$ \\
$*$HCG\,8  &VV\,521 NED04&$12.40263$  &$23.57339$  &$35.39$ &$12.32$ &$-0.75$ &$-22.94$ \\
HCG\,8  &MCG\,+04-03-009 &$12.38667$  &$23.57361$  &$3.63$  &$0.67$  &$21.0$  &$-20.05$ \\
HCG\,12 &MCG\,-01-04-052 &$21.88915$  &$-4.68301$  &$9.95$  &$4.56$  &$12.5$  &$-23.23$ \\
HCG\,12 &WISEA\,J012727.83-044038.6&$21.86604$  &$-4.67742$  &$1.4$   &$3.44$  &$13.76$ &$-20.01$ \\
HCG\,12 &WISEA\,J012737.28-044006.2&$21.90483$  &$-4.66875$  &$5.35$  &$1.27$  &$20.68$ &$-20.2$  \\
HCG\,12 &WISEA\,J012737.39-043940.3&$21.90588$  &$-4.66119$  &$2.02$  &$1.7$   &$17.55$ &$-20.42$ \\
HCG\,12 &WISEA\,J012734.33-043906.6&$21.89337$  &$-4.65211$  &$4.39$  &$4.18$  &$13.13$ &$-21.6$  \\
HCG\,13 &WISEA\,J013219.13-075345.8&$23.07975$  &$-7.89606$  &$1.2$   &$3.24$  &$13.61$ &$-20.18$ \\
HCG\,13 &MCG\,-01-05-004 &$23.09662$  &$-7.87958$  &$5.08$  &$3.71$  &$15.02$ &$-20.96$ \\
HCG\,13 &MCG\,-01-05-003 &$23.09296$  &$-7.87319$  &$13.72$ &$4.89$  &$13.2$  &$-22.52$ \\
HCG\,13 &WISEA\,J013225.97-075208.9&$23.10829$  &$-7.86953$  &$3.13$  &$2.56$  &$17.12$ &$-20.11$ \\
HCG\,13 &MCG\,-01-05-002 &$23.08537$  &$-7.85956$  &$13.19$ &$1.69$  &$19.67$ &$-22.35$ \\
HCG\,17 &HCG\,017A&$33.52135$  &$13.31104$  &$6.71$  &$6.57$  &$8.61$  &$-22.16$ \\
$*$HCG\,17 &WISEA\,J021403.85+131847.2&$33.51602$  &$13.31313$  &$7.54$  &$10.23$ &$1.27$  &$-22.04$ \\
$*$HCG\,17 &WISEA\,J021405.04+131902.2&$33.5209$   &$13.3173$   &$5.07$  &$8.02$  &$5.62$  &$-21.5$  \\
HCG\,17 &WISEA\,J021407.58+131823.6&$33.53163$  &$13.30656$  &$1.77$  &$7.4$   &$5.67$  &$-20.47$ \\
HCG\,20 &WISEA\,J024417.16+260639.2&$41.07158$  &$26.11133$  &$4.57$  &$1.69$  &$20.13$ &$-19.57$ \\
HCG\,20 &2MFGC\,02173 &$41.05229$  &$26.10953$  &$2.86$  &$2.53$  &$16.02$ &$-21.05$ \\
HCG\,20 &WISEA\,J024413.62+260620.4&$41.05675$  &$26.10561$  &$8.36$  &$4.62$  &$13.55$ &$-21.63$ \\
HCG\,20 &WISEA\,J024416.74+260610.5&$41.06975$  &$26.10286$  &$2.49$  &$3.54$  &$14.57$ &$-20.19$ \\
HCG\,20 &WISEA\,J024412.04+260556.2&$41.05029$  &$26.09939$  &$5.21$  &$2.92$  &$16.5$  &$-21.1$  \\
HCG\,25 &CGCG\,390-067&$50.16065$  &$-1.03502$  &$1.48$  &$2.2$   &$16.29$ &$-19.9$  \\
HCG\,25 &UGC\,02691 NED01&$50.18921$  &$-1.04469$  &$5.97$  &$3.5$   &$15.11$ &$-21.53$ \\
HCG\,25 &UGC\,02691 NED02&$50.18888$  &$-1.05397$  &$2.88$  &$6.58$  &$8.79$  &$-19.92$ \\
HCG\,25 &WISEA\,J032034.84-010545.0&$50.14511$  &$-1.09587$  &$2.7$   &$2.79$  &$18.16$ &$-18.19$ \\
HCG\,25 &UGC\,02690   &$50.17892$  &$-1.10858$  &$8.07$  &$1.37$  &$20.24$ &$-21.2$  \\
HCG\,35 &WISEA\,J084521.24+443114.0&$131.33855$ &$44.52057$  &$5.66$  &$3.54$  &$14.41$ &$-22.21$ \\
HCG\,35 &WISEA\,J084520.59+443032.1&$131.3358$  &$44.50895$  &$11.09$ &$5.4$   &$11.21$ &$-23.05$ \\
$*$HCG\,35 &WISEA\,J084518.42+443139.5&$131.32675$ &$44.52768$  &$8.6$   &$9.43$  &$2.61$  &$-22.64$ \\
HCG\,35 &2MASX\,J08452066+4432231  &$131.33629$ &$44.53975$  &$5.98$  &$1.65$  &$19.07$ &$-21.38$ \\
$*$HCG\,35 &WISEA\,J084520.75+443012.2&$131.33648$ &$44.50333$  &$0.75$  &$4.49$  &$9.77$  &$-20.52$ \\
HCG\,35 &WISEA\,J084520.87+443159.5&$131.33697$ &$44.53324$  &$1.65$  &$2.8$   &$15.56$ &$-19.86$ \\
HCG\,37 &NGC\,2783&$138.41444$ &$29.99297$  &$18.56$ &$5.37$  &$11.96$ &$-23.31$ \\
HCG\,37 &NGC\,2783B   &$138.38813$ &$30.00014$  &$12.74$ &$1.01$  &$21.73$ &$-21.34$ \\
HCG\,37 &MCG\,+05-22-020 &$138.40563$ &$29.99956$  &$2.29$  &$1.61$  &$18.36$ &$-19.91$ \\
HCG\,37 &MCG\,+05-22-016 &$138.39082$ &$30.01578$  &$2.73$  &$1.22$  &$19.92$ &$-19.44$ \\
HCG\,37 &MCG\,+05-22-018 &$138.39174$ &$30.03983$  &$1.79$  &$2.02$  &$17.49$ &$-19.47$ \\
HCG\,44 &NGC\,3185&$154.41069$ &$21.68825$  &$3.33$  &$1.84$  &$18.62$ &$-19.94$ \\
\hline
\end{tabular}
\end{minipage}
\end{table*}

\begin{table*}
 \begin{minipage}{180mm}
\contcaption{}
\label{tab:sersic_1:continued_1}
	\centering
	\begin{tabular}{r c c c c c c c}
        \hline \hline
         Group & Name           & RA (J2000) & Dec. (J2000) & $r_\mathrm{e}$ & $n$ & $\mu_{0,r}$                       & $M_{r}$     \\
              &                & deg        & deg          & kpc            &     & mag\,arcsec$^{-2}$             & mag         \\
\hline
HCG\,44 &SDSS\,J101723.29+214757.9 &$154.34706$ &$21.79944$  &$1.37$  &$1.17$  &$21.71$ &$-16.16$ \\
HCG\,44 &NGC\,3190:[WLQ2016] X0001 &$154.52347$ &$21.83229$  &$4.16$  &$3.86$  &$13.81$ &$-21.22$ \\
HCG\,44 &NGC\,3187&$154.44944$ &$21.87333$  &$2.86$  &$0.59$  &$21.91$ &$-18.48$ \\
HCG\,44 &NGC\,3193&$154.60375$ &$21.89397$  &$3.74$  &$5.24$  &$10.61$ &$-21.37$ \\
HCG\,59 &KUG\,1145+130&$177.12775$ &$12.72981$  &$3.24$  &$0.63$  &$22.05$ &$-18.61$ \\
HCG\,59 &IC\,0737 &$177.11469$ &$12.72739$  &$1.79$  &$2.18$  &$16.39$ &$-20.22$ \\
HCG\,59 &SDSS\,J114817.89+124333.1 &$177.07455$ &$12.72588$  &$1.52$  &$1.04$  &$21.61$ &$-16.74$ \\
HCG\,59 &IC\,0736 &$177.08382$ &$12.71657$  &$1.81$  &$2.61$  &$16.18$ &$-19.59$ \\
HCG\,59 &KUG\,1145+129&$177.1352$  &$12.70529$  &$3.33$  &$0.99$  &$21.45$ &$-18.7$  \\
HCG\,69 &UGC\,08842 NOTES01        &$208.88571$ &$25.07436$  &$4.66$  &$7.71$  &$6.23$  &$-21.21$ \\
HCG\,69 &UGC\,08842 NED02&$208.874$   &$25.07367$  &$13.37$ &$1.45$  &$20.93$ &$-21.51$ \\
HCG\,69 &CGCG\,132-048&$208.8933$  &$25.04978$  &$3.26$  &$1.7$   &$17.98$ &$-20.93$ \\
HCG\,71 &IC\,4382 &$212.76058$ &$25.51935$  &$3.59$  &$1.68$  &$17.89$ &$-21.27$ \\
HCG\,71 &NGC\,5008&$212.73849$ &$25.49722$  &$9.53$  &$1.66$  &$18.95$ &$-22.36$ \\
HCG\,71 &KUG\,1408+257&$212.77141$ &$25.48276$  &$4.35$  &$0.92$  &$20.71$ &$-20.23$ \\
HCG\,72 &UGC\,09532 NED01&$221.97245$ &$19.07709$  &$3.62$  &$4.52$  &$11.88$ &$-21.76$ \\
HCG\,72 &WISEA\,J144748.21+190352.2&$221.95096$ &$19.0645$   &$3.15$  &$0.85$  &$21.62$ &$-18.83$ \\
HCG\,72 &UGC\,09532 NED02&$221.97883$ &$19.06003$  &$3.11$  &$5.72$  &$9.35$  &$-21.48$ \\
HCG\,72 &UGC\,09532 NED04&$221.98183$ &$19.05728$  &$4.96$  &$4.14$  &$13.3$  &$-21.8$  \\
$*$HCG\,72 &UGC\,09532 NED07&$221.97962$ &$19.04742$  &$4.46$  &$12.08$ &$-1.55$ &$-19.76$ \\
HCG\,72 &UGC\,09532 NED05&$221.98671$ &$19.04508$  &$10.8$  &$6.04$  &$10.68$ &$-22.17$ \\
HCG\,72 &UGC\,09532 NED06&$221.99442$ &$19.04136$  &$4.79$  &$1.47$  &$21.22$ &$-19.07$ \\
HCG\,74 &NGC\,5910 NED02 &$229.8531$  &$20.89635$  &$17.23$ &$4.41$  &$13.79$ &$-23.4$  \\
HCG\,74 &NGC\,5910 NED01 &$229.85109$ &$20.8908$   &$5.2$   &$5.79$  &$9.86$  &$-21.87$ \\
HCG\,74 &NGC\,5910 NED03 &$229.85765$ &$20.89954$  &$2.43$  &$3.3$   &$14.64$ &$-20.55$ \\
HCG\,74 &WISEA\,J151931.81+205301.0&$229.88244$ &$20.88356$  &$3.13$  &$2.41$  &$16.52$ &$-20.99$ \\
HCG\,74 &WISEA\,J151927.78+205431.8&$229.86579$ &$20.90888$  &$1.79$  &$1.0$   &$19.63$ &$-19.29$ \\
HCG\,76 &2MFGC\,12530 &$232.90023$ &$7.34976$   &$7.19$  &$0.53$  &$22.91$ &$-19.72$ \\
HCG\,76 &NGC\,5941&$232.91766$ &$7.339$ &$12.54$ &$4.57$  &$13.51$ &$-22.63$ \\
$*$HCG\,76 &CGCG\,050-011 NED01       &$232.91457$ &$7.33594$   &$0.01$  &$1.91$  &$5.71$  &$-19.1$  \\
HCG\,76 &NGC\,5942&$232.90342$ &$7.31242$   &$8.15$  &$5.59$  &$10.69$ &$-22.4$  \\
HCG\,76 &WISEA\,J153150.20+071841.8&$232.95916$ &$7.31161$   &$2.8$   &$2.4$   &$17.06$ &$-20.2$  \\
HCG\,76 &NGC\,5944&$232.94835$ &$7.30815$   &$5.21$  &$1.71$  &$18.29$ &$-21.63$ \\
HCG\,76 &CGCG\,050-010&$232.92591$ &$7.28739$   &$4.24$  &$4.14$  &$12.93$ &$-21.73$ \\
HCG\,78 &WISEA\,J154833.18+681412.0&$237.13792$ &$68.23583$  &$4.7$   &$0.88$  &$21.48$ &$-19.65$ \\
HCG\,78 &UGC\,10057   &$237.07233$ &$68.22075$  &$13.8$  &$3.18$  &$16.93$ &$-22.2$  \\
HCG\,78 &CGCG\,319-024&$237.03708$ &$68.20686$  &$3.42$  &$2.38$  &$15.77$ &$-21.91$ \\
HCG\,80 &WISEA\,J155907.33+651401.4&$239.77936$ &$65.23375$  &$3.22$  &$0.62$  &$20.45$ &$-20.27$ \\
HCG\,80 &2MFGC\,12823 &$239.82935$ &$65.23262$  &$6.77$  &$0.43$  &$21.34$ &$-21.25$ \\
HCG\,80 &WISEA\,J155921.65+651323.2&$239.8401$  &$65.223$&$4.38$  &$6.35$  &$9.13$  &$-20.98$ \\
HCG\,80 &WISEA\,J155912.02+651319.5&$239.80017$ &$65.22198$  &$3.27$  &$1.34$  &$20.03$ &$-19.52$ \\
HCG\,81 &UGC\,10319 NED01&$244.55682$ &$12.80287$  &$4.05$  &$1.17$  &$19.42$ &$-21.03$ \\
HCG\,81 &UGC\,10319 NED03&$244.56087$ &$12.79559$  &$6.83$  &$6.2$   &$10.22$ &$-21.31$ \\
HCG\,81 &UGC\,10319 NED04&$244.56175$ &$12.79328$  &$7.79$  &$4.52$  &$14.11$ &$-21.18$ \\
HCG\,81 &UGC\,10319 NED02&$244.55888$ &$12.78867$  &$5.81$  &$4.04$  &$13.96$ &$-21.68$ \\
HCG\,82 &NGC\,6162&$247.09323$ &$32.84933$  &$23.61$ &$5.85$  &$11.53$ &$-23.34$ \\
HCG\,82 &NGC\,6163&$247.11628$ &$32.84638$  &$7.09$  &$2.29$  &$17.27$ &$-22.23$ \\
HCG\,82 &NGC\,6161&$247.08595$ &$32.81064$  &$6.73$  &$0.43$  &$21.12$ &$-21.51$ \\
HCG\,84 &CGCG\,355-020 NED01       &$251.0951$  &$77.83875$  &$22.45$ &$5.08$  &$13.08$ &$-23.39$ \\
HCG\,86 &MCG\,-05-47-001 &$297.96642$ &$-30.80844$ &$2.43$  &$5.62$  &$9.63$  &$-20.68$ \\
HCG\,86 &MCG\,-05-47-003 &$297.9961$  &$-30.81622$ &$7.97$  &$7.68$  &$6.67$  &$-21.91$ \\
HCG\,86 &ESO\,461- G 007 &$298.03652$ &$-30.82575$ &$8.73$  &$4.87$  &$12.26$ &$-22.38$ \\
HCG\,86 &MCG\,-05-47-002 &$297.98943$ &$-30.85695$ &$3.84$  &$3.45$  &$14.8$  &$-20.96$ \\
HCG\,88 &NGC\,6978&$313.14765$ &$-5.71113$  &$8.75$  &$1.86$  &$18.76$ &$-21.94$ \\
HCG\,88 &WISEA\,J205223.16-054326.2&$313.0966$  &$-5.72383$  &$1.65$  &$2.9$   &$17.42$ &$-17.63$ \\
HCG\,88 &NGC\,6977&$313.12377$ &$-5.7461$   &$6.63$  &$2.13$  &$17.69$ &$-21.89$ \\
HCG\,88 &NGC\,6975&$313.10847$ &$-5.7723$   &$6.31$  &$1.56$  &$19.54$ &$-21.02$ \\
HCG\,88 &SDSS\,J205224.43-054714.2 &$313.1018$  &$-5.78729$  &$1.91$  &$1.2$   &$21.27$ &$-17.33$ \\
HCG\,88 &MCG\,-01-53-014 &$313.05321$ &$-5.79834$  &$6.29$  &$1.02$  &$21.56$ &$-19.94$ \\
HCG\,89 &WISEA\,J212019.18-035345.6&$320.07979$ &$-3.89603$  &$6.9$   &$1.26$  &$20.69$ &$-20.65$ \\
HCG\,89 &WISEA\,J212007.99-035429.1&$320.03342$ &$-3.90837$  &$3.35$  &$0.44$  &$21.8$  &$-19.27$ \\
HCG\,89 &2MASX\,J21200830-0355036  &$320.03479$ &$-3.91762$  &$5.11$  &$1.14$  &$20.67$ &$-20.24$ \\
HCG\,89 &MCG\,-01-54-012 &$320.00429$ &$-3.92214$  &$10.86$ &$1.19$  &$21.05$ &$-21.39$ \\
\hline
\end{tabular}
\end{minipage}
\end{table*}

\begin{table*}
 \begin{minipage}{180mm}
\contcaption{}
\label{tab:sersic_1:continued_2}
	\centering
	\begin{tabular}{r c c c c c c c}
        \hline \hline
         Group & Name           & RA (J2000) & Dec. (J2000) & $r_\mathrm{e}$ & $n$ & $\mu_{0,r}$                       & $M_{r}$     \\
              &                & deg        & deg          & kpc            &     & mag\,arcsec$^{-2}$             & mag         \\
\hline
HCG\,93 &NGC\,7553&$348.88789$ &$19.04816$  &$2.67$  &$5.42$  &$10.46$ &$-20.47$ \\
HCG\,93 &NGC\,7549&$348.82179$ &$19.04169$  &$6.45$  &$1.41$  &$19.31$ &$-21.56$ \\
HCG\,93 &NGC\,7547&$348.76418$ &$18.97344$  &$6.78$  &$3.74$  &$14.96$ &$-21.45$ \\
HCG\,93 &NGC\,7550&$348.8167$  &$18.9618$   &$8.79$  &$3.94$  &$13.9$  &$-22.66$ \\
HCG\,94 &2MASXi\,J2317202+184405   &$349.33432$ &$18.73487$  &$10.52$ &$5.81$  &$11.07$ &$-22.17$ \\
HCG\,94 &2MASX\,J23171540+1843385  &$349.31429$ &$18.72742$  &$6.02$  &$1.0$   &$21.47$ &$-20.1$  \\
HCG\,95 &NGC\,7609 NED01 &$349.87521$ &$9.50822$   &$8.9$   &$5.19$  &$11.43$ &$-22.72$ \\
HCG\,95 &NGC\,7609 NED02 &$349.87954$ &$9.50297$   &$8.2$   &$2.07$  &$18.83$ &$-21.41$ \\
HCG\,95 &MCG\,+01-59-046 &$349.866$   &$9.49436$   &$7.24$  &$1.41$  &$20.63$ &$-20.58$ \\
HCG\,96 &NGC\,7674A   &$351.99492$ &$8.78281$   &$1.64$  &$3.06$  &$13.92$ &$-20.83$ \\
HCG\,96 &NGC\,7674&$351.98635$ &$8.77904$   &$16.88$ &$4.65$  &$13.38$ &$-23.21$ \\
HCG\,96 &NGC\,7675&$352.02467$ &$8.7686$&$5.69$  &$5.08$  &$11.0$  &$-22.33$ \\
HCG\,96 &WISEA\,J232800.09+084602.8&$352.00083$ &$8.76732$   &$1.56$  &$0.89$  &$19.48$ &$-19.25$ \\
HCG\,97 &IC\,5352 &$356.83292$ &$-2.28067$  &$1.6$   &$5.33$  &$10.23$ &$-19.81$ \\
HCG\,97 &IC\,5357 &$356.84579$ &$-2.30067$  &$21.26$ &$5.33$  &$12.84$ &$-22.8$  \\
HCG\,97 &IC\,5351 &$356.82887$ &$-2.3135$   &$4.8$   &$3.95$  &$13.45$ &$-21.81$ \\
HCG\,97 &IC\,5359 &$356.90775$ &$-2.31667$  &$8.67$  &$1.48$  &$20.83$ &$-20.56$ \\
HCG\,97 &IC\,5356 &$356.84913$ &$-2.35125$  &$4.01$  &$2.76$  &$15.9$  &$-21.37$ \\
HCG\,97 &WISEA\,J234720.25-022225.9&$356.83458$ &$-2.37411$  &$3.28$  &$1.47$  &$21.36$ &$-17.95$ \\
HCG\,97 &WISEA\,J234723.26-022147.1&$356.84708$ &$-2.36306$  &$2.84$  &$2.47$  &$17.95$ &$-19.12$ \\
HCG\,98 &NGC\,7783 NED01 &$358.542$   &$0.38286$   &$8.93$  &$5.19$  &$11.22$ &$-22.86$ \\
HCG\,100&KUG\,2358+128A  &$0.30575$   &$13.14406$  &$5.89$  &$3.13$  &$17.45$ &$-19.89$ \\
HCG\,100&MRK\,0934&$0.3585$&$13.113$&$3.28$  &$1.31$  &$19.5$  &$-20.09$ \\
HCG\,100&MCG\,+02-01-010 &$0.31238$   &$13.11256$  &$3.39$  &$0.97$  &$21.6$  &$-18.65$ \\
HCG\,100&NGC\,7803&$0.33321$   &$13.11125$  &$5.39$  &$4.72$  &$12.01$ &$-21.92$ \\
\hline
\end{tabular}
\end{minipage}
\end{table*}

\begin{figure}
	\label{fig:mean_ser_vs_m_comparison_n_var}
    \begin{center}
        \includegraphics[width = 0.5\textwidth]{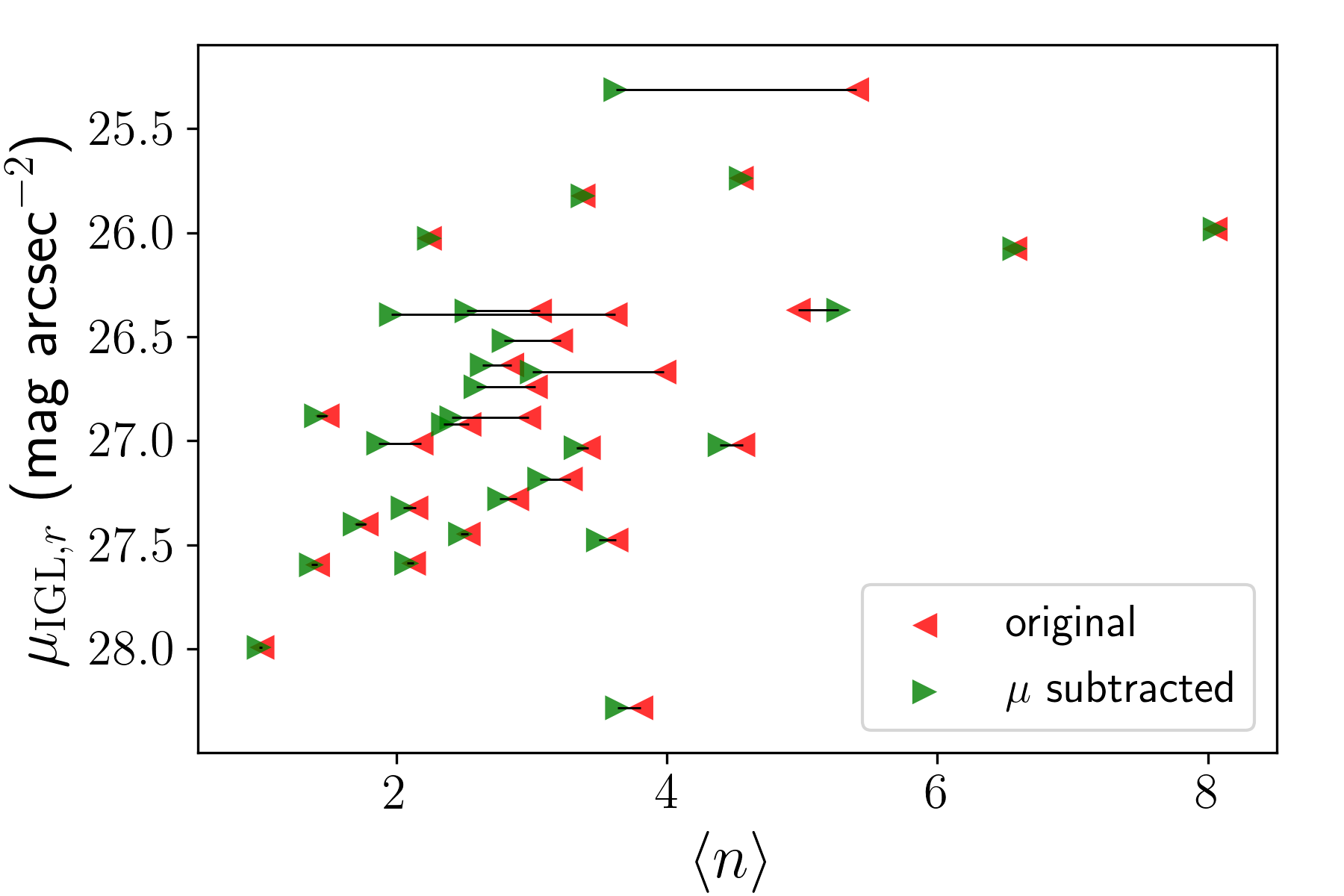}
    \end{center}
    \caption{Possible influence of the diffuse light on the mean S\'ersic indices of our compact groups. The Pearson correlation coefficient $\rho$ changed from $-0.58$ to $-0.48$ and the estimated p-value in Student's $t$-test changed from $0.001$ to $0.009$.}
\end{figure}

\begin{figure}
	\label{fig:mu_vs_R_norm}
    \begin{center}
        \includegraphics[width = 0.5\textwidth]{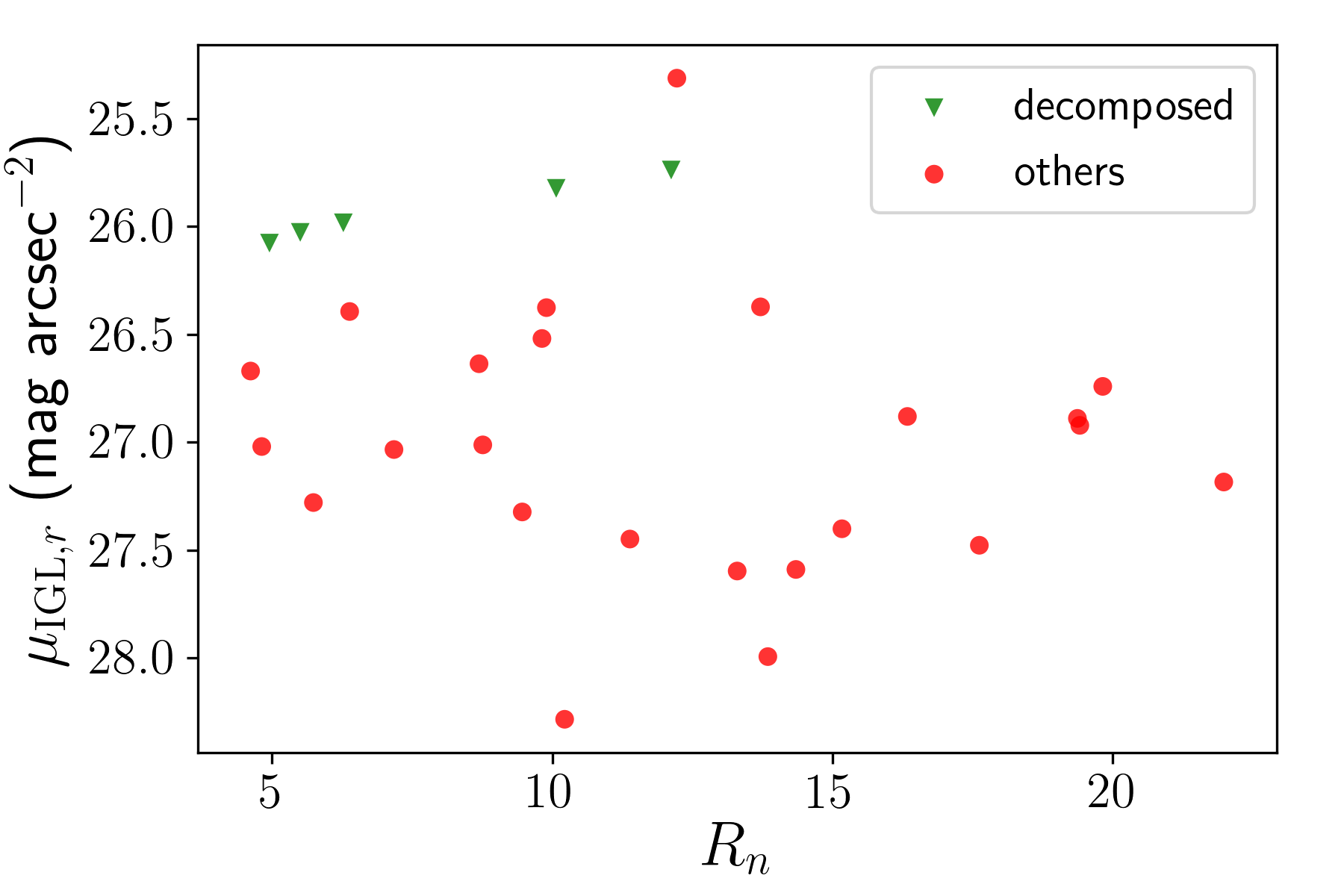}
    \end{center}
    \caption{Dependence between $R_\mathrm{n}$ (see text) and the mean surface brightness of the IGL $\mu_\mathrm{IGL}$.}
\end{figure}

\begin{figure*}
	\begin{minipage}{0.49\linewidth}
 	   	\begin{center}
			\includegraphics[width = 1.0\textwidth]{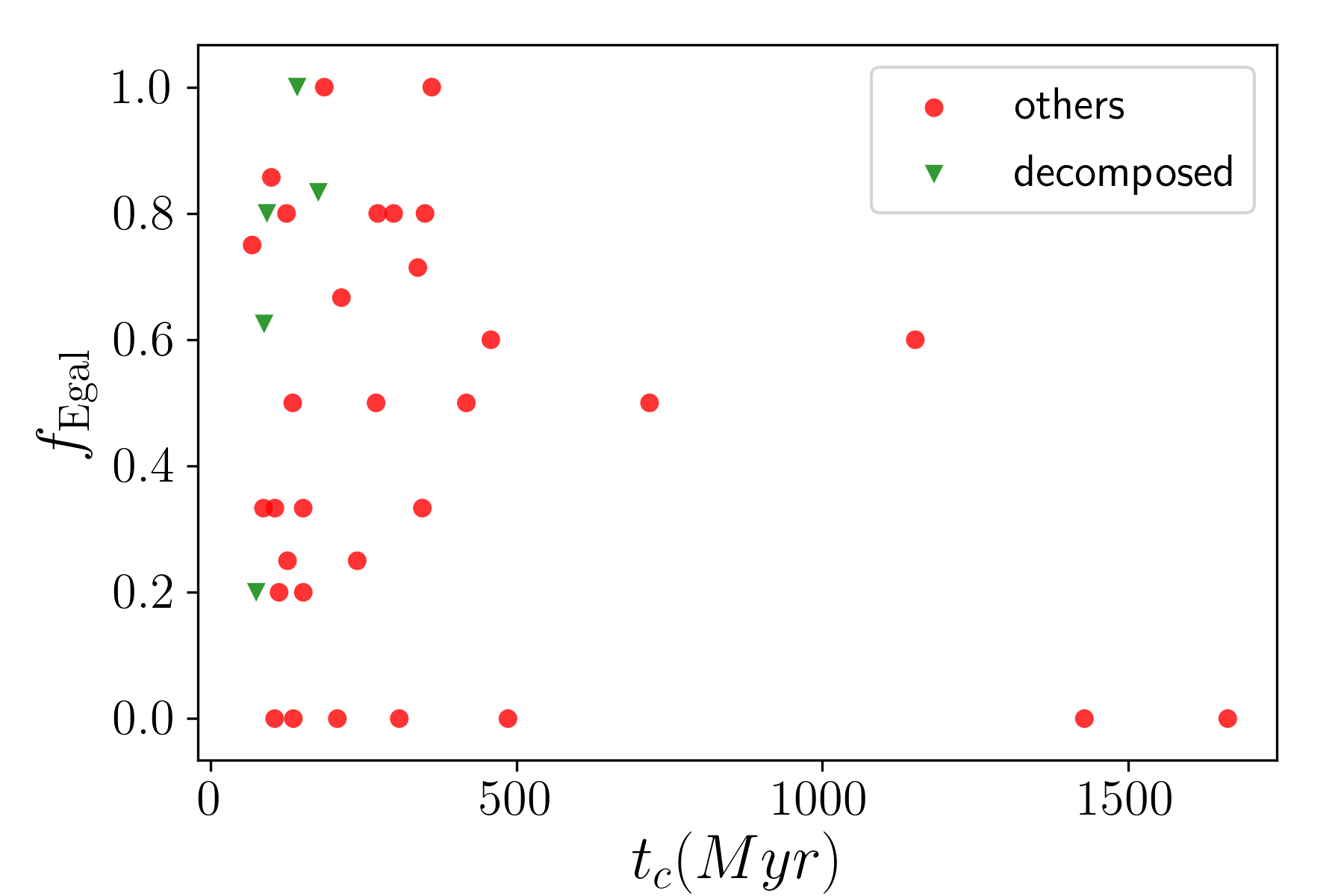}
    		\end{center}
    		\label{fig:f_Egal_vs_tc_our}
    \end{minipage}
    \hfill
    \begin{minipage}{0.49\linewidth}
 	   	\begin{center}
			\includegraphics[width = 1.0\textwidth]{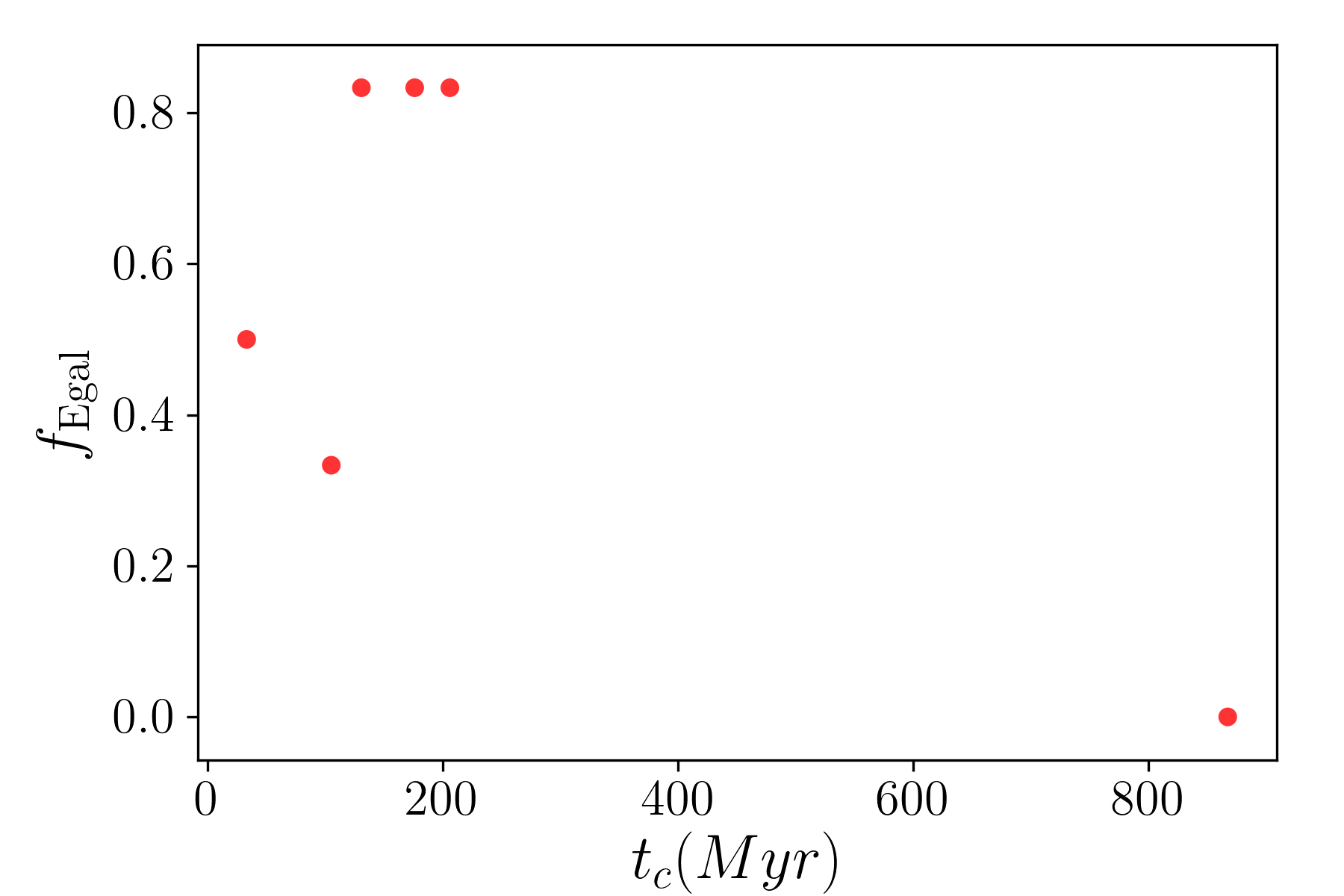}
    		\end{center}
    		\label{fig:f_Egal_vs_tc_DaRocha}
    \end{minipage}
    \caption{The left panel shows the dependence between the fraction of early-type galaxies and the crossing time $t_\mathrm{c} $ for all 36 compact groups in our sample. The right panel shows the same dependence for the groups from \citet{2008MNRAS.388.1433D}.}
\end{figure*}

\begin{figure*}
    \begin{minipage}{0.49\linewidth}
 	   	\begin{center}
			\includegraphics[width = 1.0\textwidth]{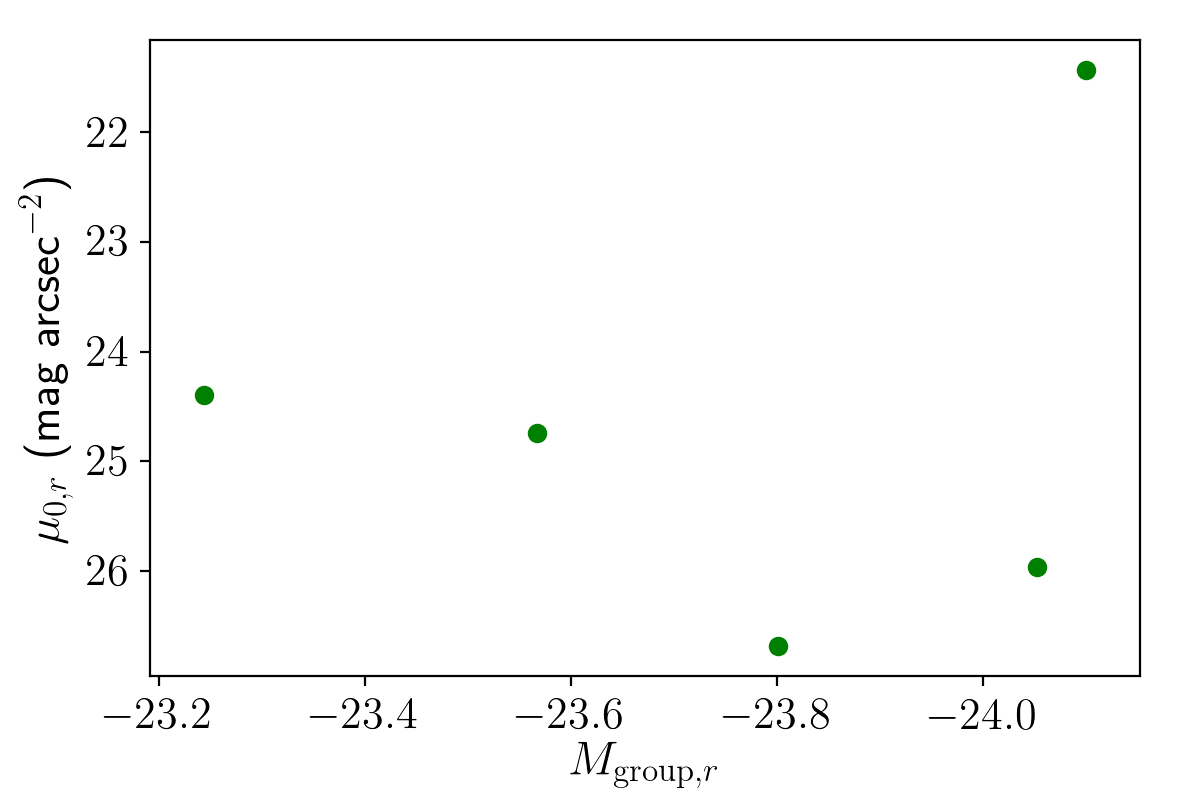}
    		\end{center}
    		\label{fig:m_central_vs_M}
    \end{minipage}
    \hfill
    \begin{minipage}{0.49\linewidth}
 	   	\begin{center}
			\includegraphics[width = 1.0\textwidth]{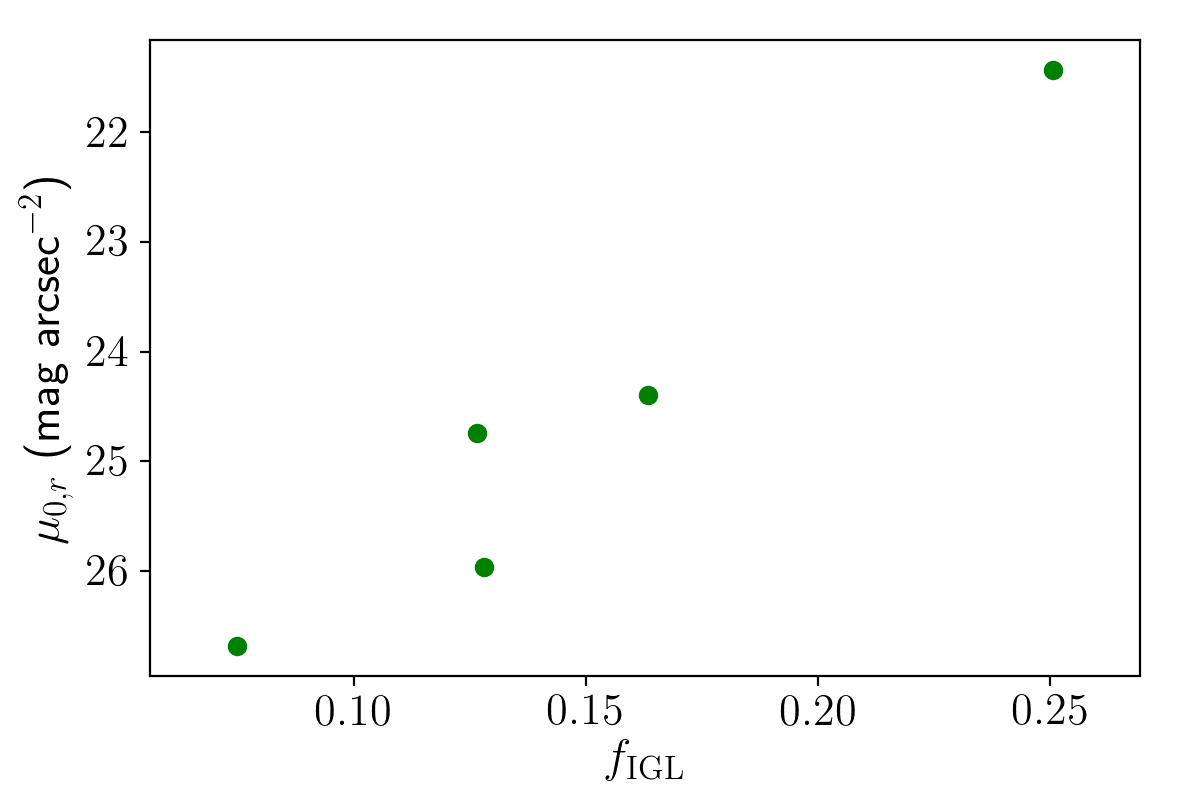}
    		\end{center}
    		\label{fig:m_central_vs_f_IGL}
    \end{minipage}
    \caption{Dependencies between the central surface brightness of the IGL $\mu_{0,r}$ and the absolute magnitude of the group $M_{\mathrm{group},r}$ (the left-hand plot) and the IGL fraction $f_{\mathrm{IGL}}$ (the right-hand plot).}
\end{figure*}

\begin{figure*}
	\begin{minipage}{0.32\linewidth}
 	   	\begin{center}
			\includegraphics[width = 1.0\textwidth]{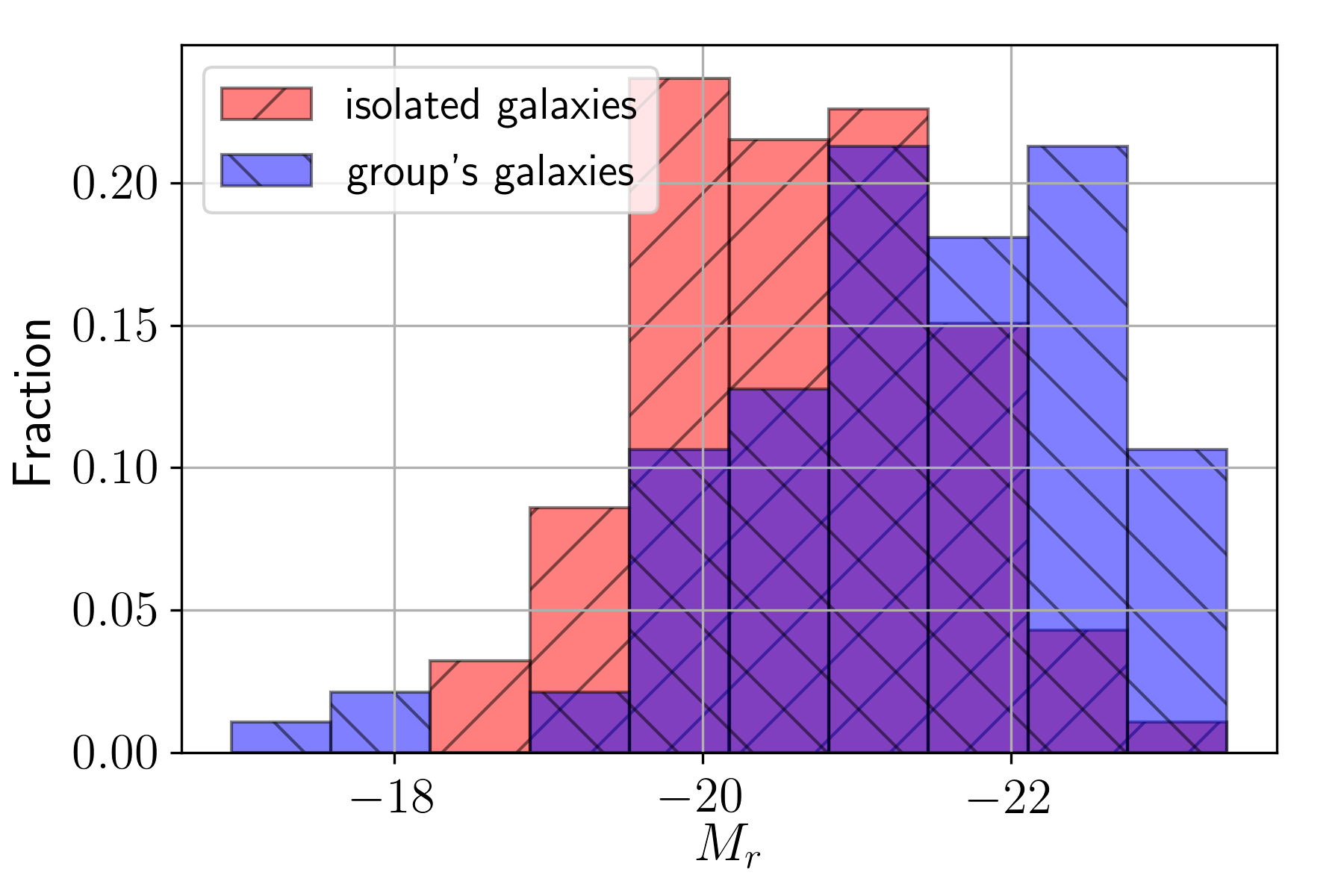}
    		\end{center}
    		\label{fig:early-type_M_abs}
    \end{minipage}
    \hfill
    \begin{minipage}{0.32\linewidth}
 	   	\begin{center}
        		\includegraphics[width = 1.0\textwidth]{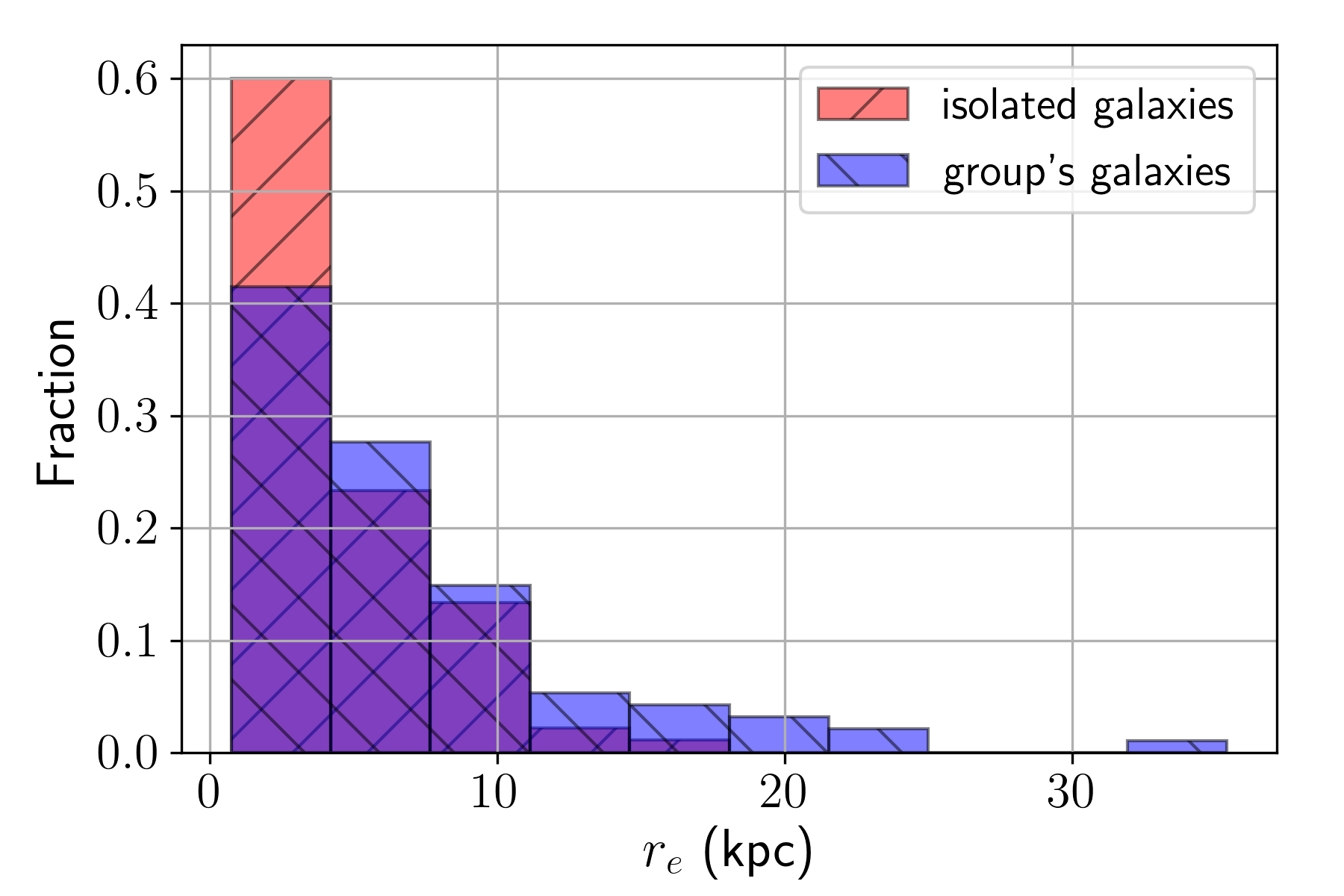}
    		\end{center}
   		\label{fig:early-type_re}
    \end{minipage}
    \hfill
    \begin{minipage}{0.32\linewidth}
 	   	\begin{center}
        		\includegraphics[width = 1.0\textwidth]{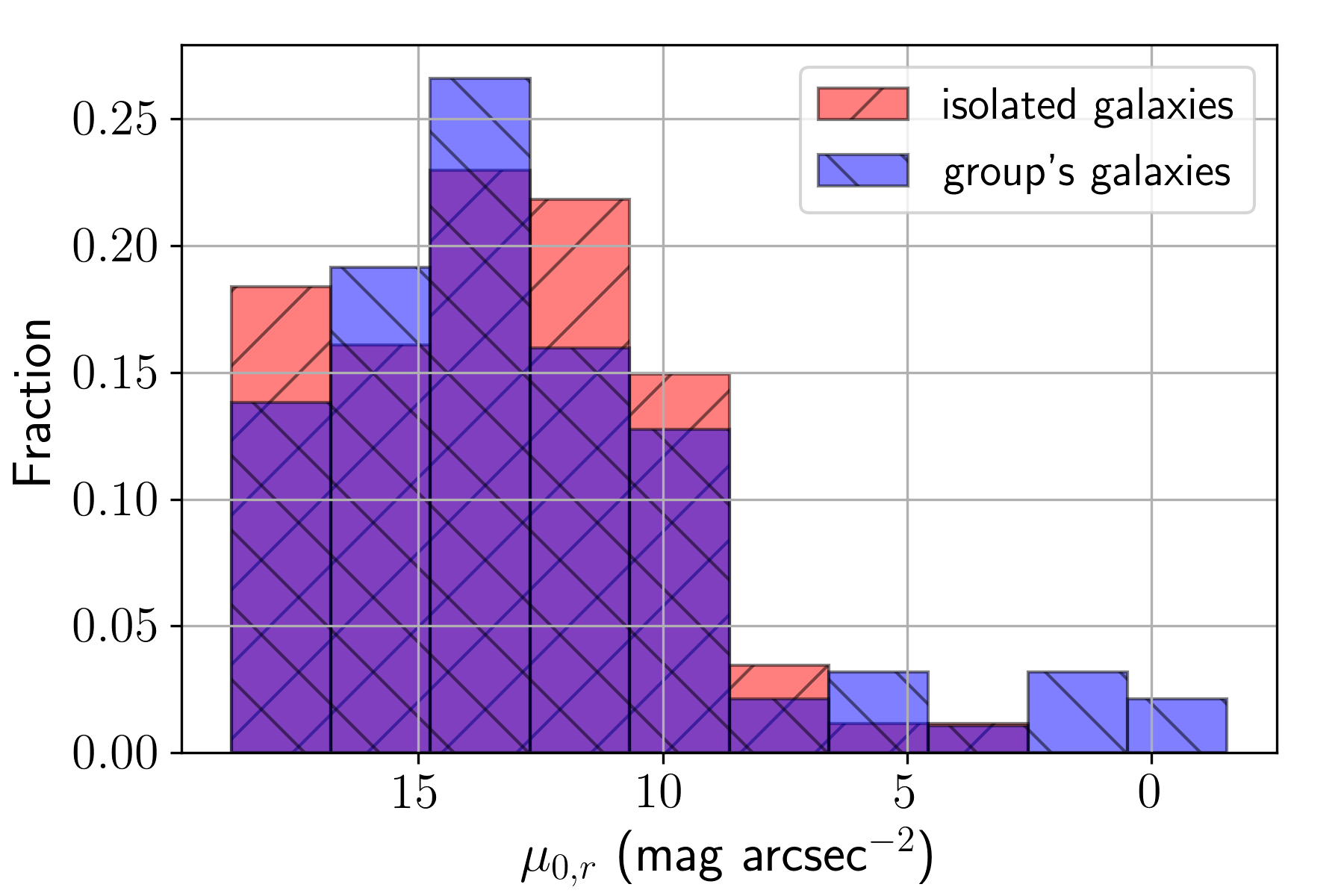}
    		\end{center}
   		\label{fig:early-type_m_central}
    \end{minipage}
    \vfill
    \begin{minipage}{0.32\linewidth}
 	   	\begin{center}
			\includegraphics[width = 1.0\textwidth]{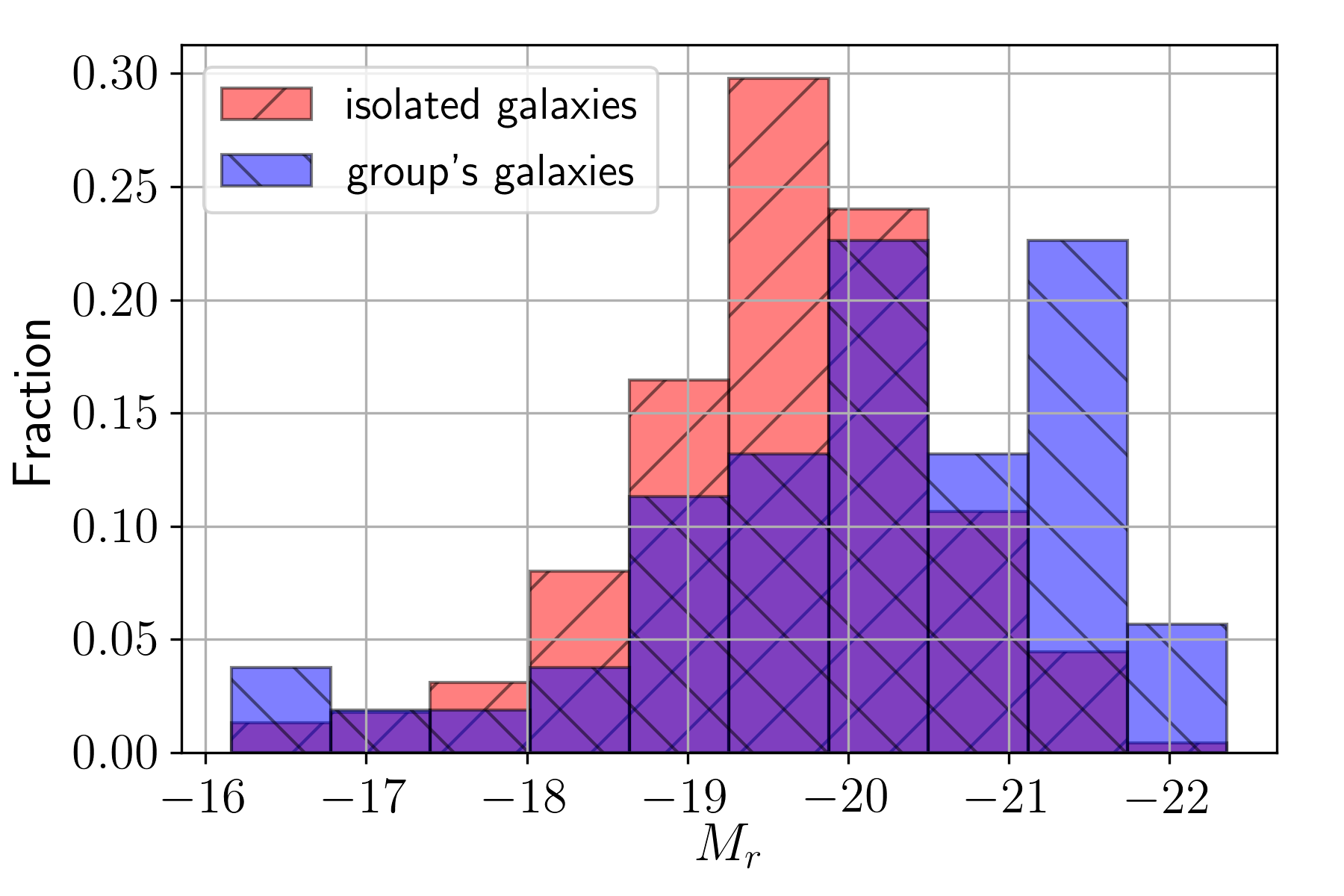}
    		\end{center}
    		\label{fig:late-type_M_abs}
    \end{minipage}
    \hfill
    \begin{minipage}{0.32\linewidth}
 	   	\begin{center}
        		\includegraphics[width = 1.0\textwidth]{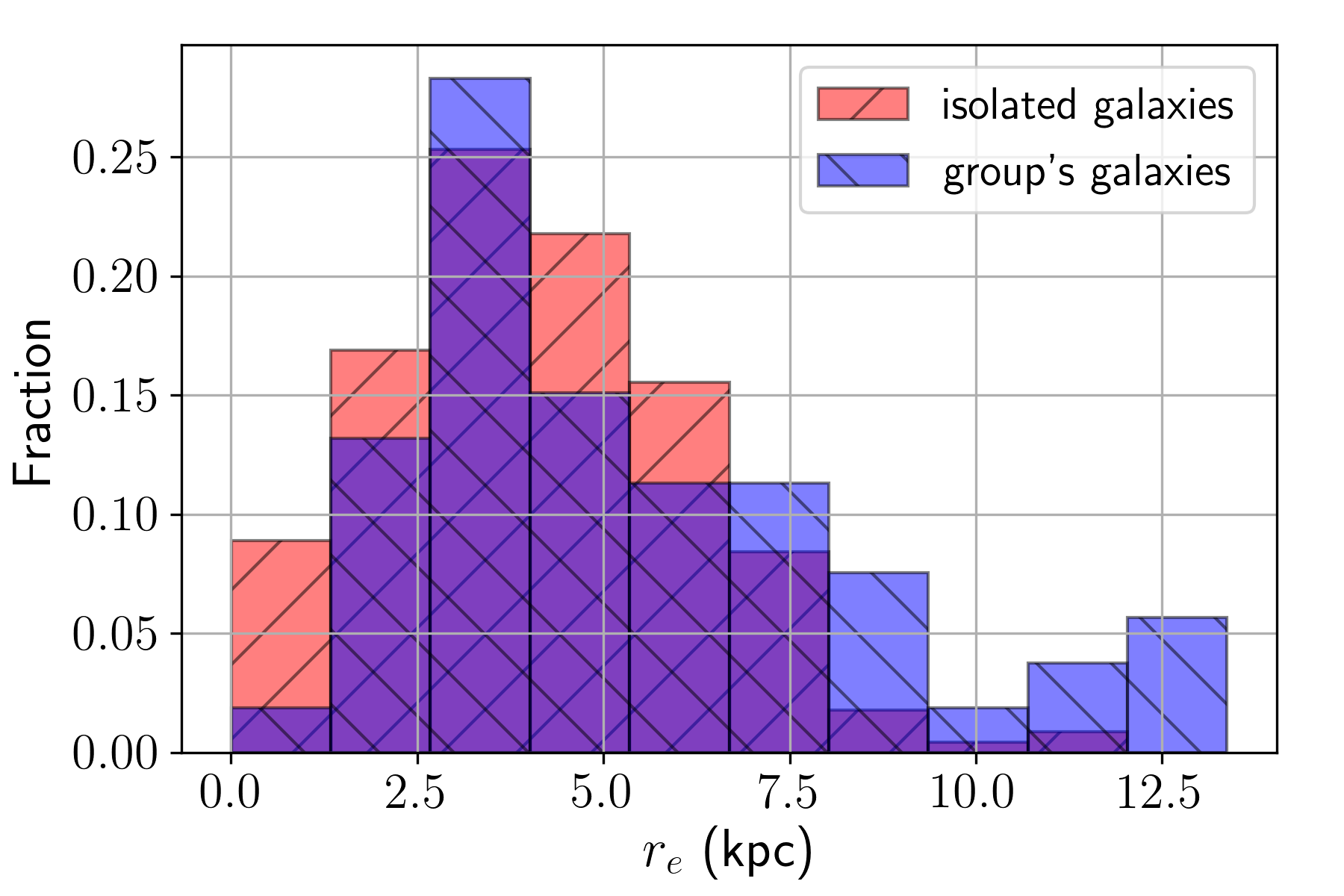}
    		\end{center}
   		\label{fig:late-type_re}
    \end{minipage}
    \hfill
    \begin{minipage}{0.32\linewidth}
 	   	\begin{center}
        		\includegraphics[width = 1.0\textwidth]{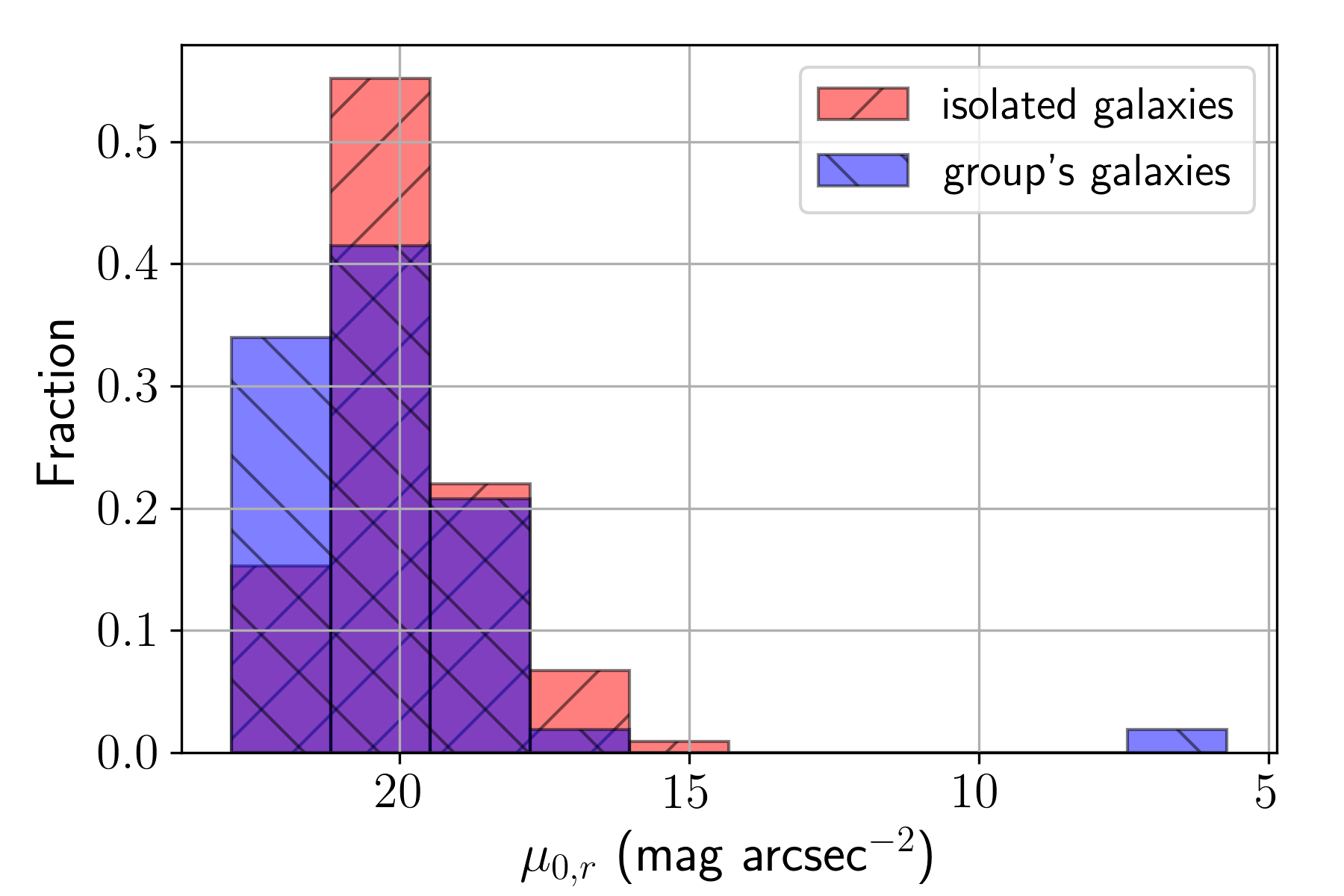}
    		\end{center}
   		\label{fig:late-type_m_central}
    \end{minipage}
    \caption{Distribution by the S\'ersic parameters for individual galaxies in all 36 compact groups and isolated galaxies from \citet{2011ApJS..196...11S}. The top row corresponds to galaxies with the S\'ersic index $n > 2$ and the bottom row -- to galaxies with $n \leq 2$.}
\end{figure*}

\begin{figure*}
	\label{fig:hcg8_deco}
    \begin{center}
        \includegraphics[width = 0.9\textwidth]{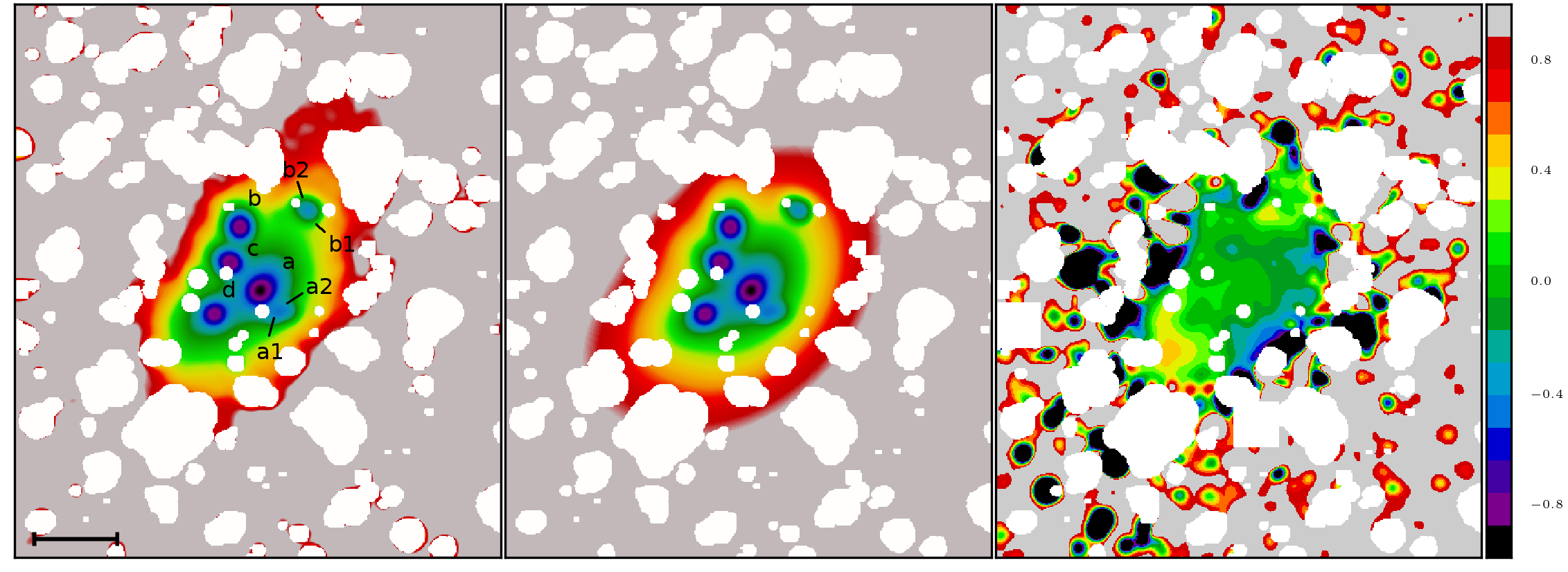}
    \end{center}
    \caption{Original image (left panel), model (middle panel) and relative residue (right panel) for HCG\,8. Dark red boundaries of the coloured areas in the left and middle panels correspond to the minimum surface brightness level $27.3$~mag\,arcsec$^{-2}$. White areas represent masked pixels. The black scale bar is 1$\arcmin$.}
\end{figure*} 

\begin{figure*}
	\label{fig:hcg17_deco}
    \begin{center}
        \includegraphics[width = 0.9\textwidth]{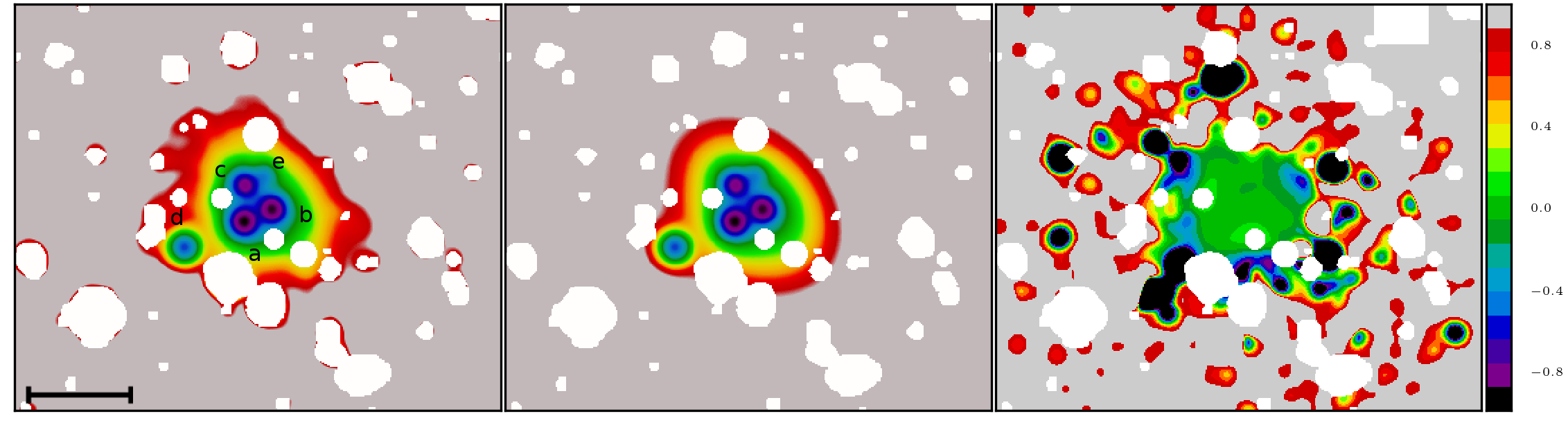}
    \end{center}
    \caption{The same as in Fig.~\ref{fig:hcg8_deco} but for HCG\,17.}
\end{figure*}

\begin{figure*}
	\label{fig:hcg35_deco}
    \begin{center}
        \includegraphics[width = 0.9\textwidth]{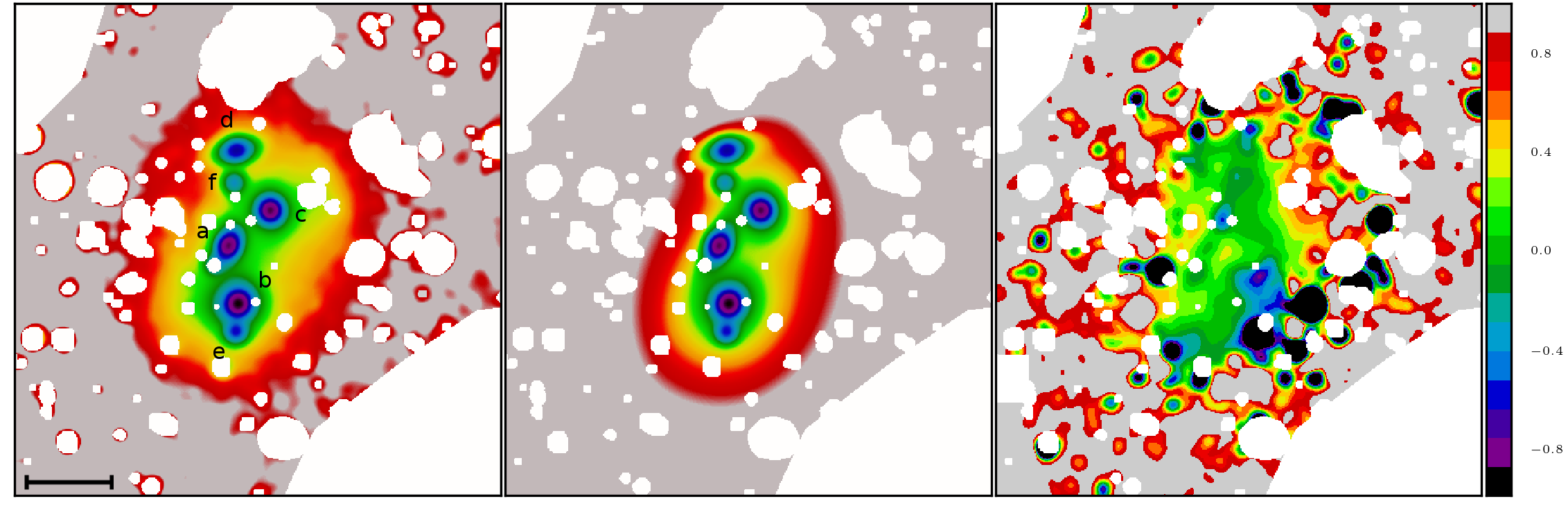}
    \end{center}
    \caption{The same as in Fig.~\ref{fig:hcg8_deco} but for HCG\,35.}
\end{figure*} 

\begin{figure*}
	\label{fig:hcg37_deco}
    \begin{center}
        \includegraphics[width = 0.9\textwidth]{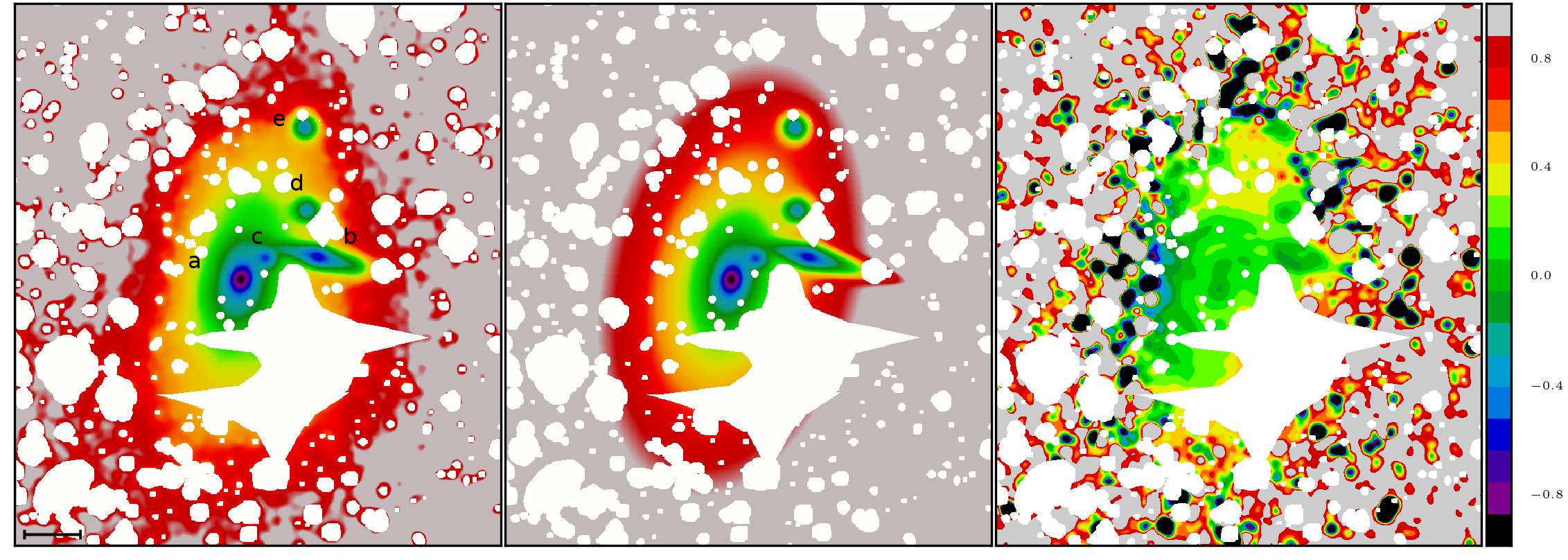}
    \end{center}
    \caption{The same as in Fig.~\ref{fig:hcg8_deco} but for HCG\,37.}
\end{figure*} 

\begin{figure*}
	\label{fig:hcg74_deco}
    \begin{center}
        \includegraphics[width = 0.9\textwidth]{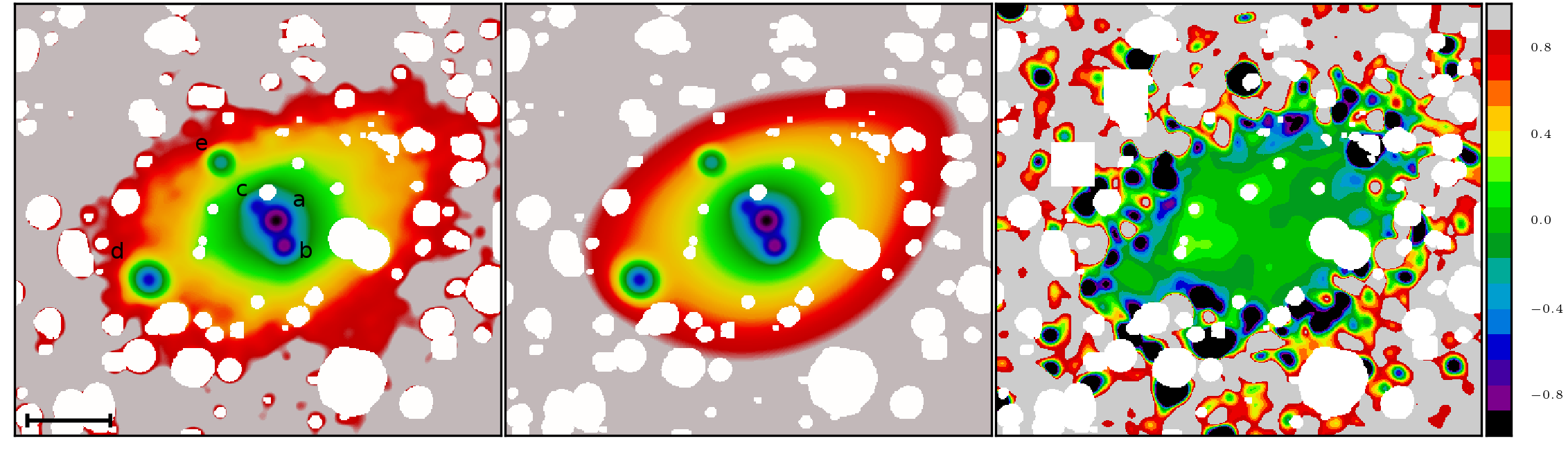}
    \end{center}
    \caption{The same as in Fig.~\ref{fig:hcg8_deco} but for HCG\,74.}
\end{figure*} 

% Don't change these lines
\bsp	% typesetting comment
\label{lastpage}
\end{document}